\documentclass[iop,apj]{emulateapj}
\usepackage{color}
\usepackage{amsmath,graphicx,natbib}
\usepackage{rotating,multirow}

\shorttitle{Non-Gaussian Velocity Distributions in Solar Flares from Extreme Ultraviolet Lines}
\shortauthors{Jeffrey, Fletcher \& Labrosse}

\begin{document}

	\title{Non-Gaussian Velocity Distributions in Solar Flares from Extreme Ultraviolet Lines: A Possible Diagnostic of Ion Acceleration}

	\author{Natasha L. S. Jeffrey, Lyndsay Fletcher \& Nicolas Labrosse}
	\affil{School of Physics \& Astronomy, University of Glasgow, G12 8QQ, Glasgow, UK}

\begin{abstract}
In a solar flare, a large fraction of the magnetic energy released is converted rapidly to the kinetic energy of non-thermal particles and bulk plasma motion. This will likely result in non-equilibrium particle distributions and turbulent plasma conditions. We investigate this by analysing the profiles of high-temperature extreme ultraviolet emission lines from a major flare (SOL2014-03-29T17:44) observed by the {\em EUV Imaging Spectrometer} (EIS) on {\em Hinode}. We find that in many locations the line profiles are non-Gaussian, consistent with a kappa-distribution of emitting ions with properties that vary in space and time. At the flare footpoints, close to sites of hard X-ray emission from non-thermal electrons, the $\kappa$-index for the Fe XVI 262.976 \AA~line at 3 MK takes values of 3-5. In the corona, close to a low-energy HXR source, the Fe XXIII 263.760 \AA~line at 15 MK shows $\kappa$ values of typically 4-7. The observed trends in the $\kappa$ parameter show that we are most likely detecting the properties of the ion population rather than any instrumental effects.  We calculate that a non-thermal ion population could exist if locally accelerated on timescales $\leq$ 0.1 s. However, observations of net redshifts in the lines also imply the presence of plasma downflows which could lead to bulk turbulence, with increased non-Gaussianity in cooler regions. Both interpretations have important implications for theories of solar flare particle acceleration.
\end{abstract}


\section{Introduction}\label{intro}
Solar flare extreme ultraviolet (EUV) spectral line observations with the {\it Hinode} \citep{2007SoPh..243....3K} {\it EUV Imaging Spectrometer} \citep[EIS;][]{2007SoPh..243...19C} provide information on ion line emissions, plasma temperatures, mass flows, ion abundances and electron densities \citep[cf.][]{2015SoPh..290.3399M}. For most purposes, Gaussian fitting is an excellent approximation for the low moments of the spectral line: integrated intensity (zero moment) and line centroid position (first moment), even if the line profile is non-Gaussian. But the shape of the EUV line profile can be used to infer more about the velocity distribution of the emitting ions. \cite{2016A&A...590A..99J} observed non-Gaussian spectral lines in flare EUV emission, showing that many unblended \ion{Fe}{16} lines were consistent with a line shape produced by a kappa rather than a Maxwellian velocity distribution, in different flare regions. Megakelvin flare temperatures produce spectral lines dominated by Doppler broadening, and physically, such a line shape could be produced by (1) non-thermal ions of $\lesssim$ 1 MeV or (2) non-Gaussian turbulent velocity fluctuations, providing a new EUV diagnostic tool.

	\begin{figure*}[t!]
	\centering
	\includegraphics[width=1.0\textwidth]{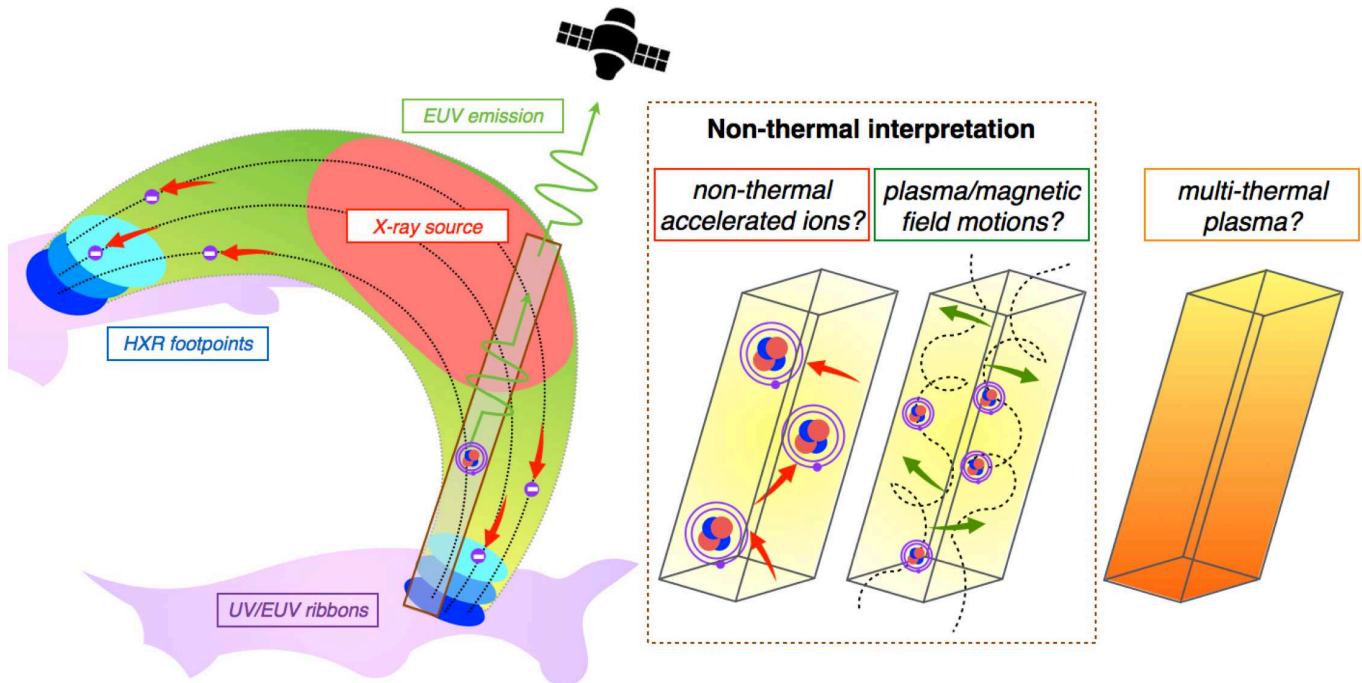}
	\caption{{\it Left:} A cartoon of the flare and the observations. {\it Right:} If an instrumental cause can be eliminated, then the EUV kappa line profiles could be produced by three physical scenarios: 1. a non-thermal ion velocity distribution from isotropic non-thermal ion motions, 2. turbulent motions due to magnetic fluctuations or possibly a superposition of unresolved flows or 3. a multi-thermal plasma distribution (not discussed in this paper).}
	\label{fig1}
	\end{figure*}

Non-thermal flare particles are usually detected by X-ray and gamma ray observations. Most flares have X-ray bremsstrahlung emission from keV electrons, currently detected with the {\it Ramaty High Energy Solar Spectroscopic Imager} \citep[{\it RHESSI};][]{2002SoPh..210....3L}. Only a small minority of (typically) large flares, e.g. SOL2002-07-23 \citep{2003ApJ...595L.103K}, have detectable gamma-ray line emission, produced by interactions between MeV protons and heavier ions \citep[cf.][]{2011SSRv..159..167V}, and hence the properties and occurrence of such ions remain uncertain. Accelerated ions with energies less than a few MeV are almost impossible to detect with methods such as impact polarization or charge-exchange \citep[e.g.][]{1990ApJS...73..303H,1998A&A...334.1136B}, that also require the presence of anisotropic ion beams, remaining inconclusive. But to assess the non-thermal ion energy content requires knowledge of this accelerated but low-energy component. Ion kappa velocity distributions \citep[cf.][]{2010SoPh..267..153P,2009JGRA..11411105L} are routinely detected in space physics e.g. \citet[][]{1998SSRv...86..127G}, but the high density flare environment ($n_e >10^{9}$ cm$^{-3}$) with thermalizing Coulomb collisions is very different to the collisionless solar wind. If such distributions can exist in flare conditions, they could provide a novel diagnostic technique of solar flare ion acceleration unavailable using other methods.

The presence of plasma turbulence might be an alternative explanation of observed solar flare non-Gaussian spectral lines. Excess line broadening, or the presence of broadening larger than expected from isothermal ion motion, is often detected during a flare e.g. \citet[][]{1995ApJ...438..480A,1993SoPh..144..217D,1979ApJ...233L.157D,1980ApJ...239..725D,1990A&A...236L...9A,1986ApJ...301..975A}, and likely produced by either turbulent magnetic fluctuations (magnetohydrodyamic (MHD) turbulence) or possibly by the superposition of unresolved flows. Although, recent EIS studies in active regions and cooler lines in flares showed some correlation between excess line width and directed Doppler shifts \citep[e.g.][]{2011ApJ...740...70M}, other notable observations: larger broadening of hotter lines and isotropy (line broadening is seen for flares located at all heliocentric angles), might be consistent with magnetic fluctuations. Other independent observations using X-ray imaging e.g. \citet{2011ApJ...730L..22K} also show additional and independent evidence for MHD turbulence in the corona. Further, a recent study in preparation (Kontar et al., submitted PRL) shows that MHD turbulence can act as a crucial intermediary in the transfer of large amounts of energy from stressed magnetic fields to accelerated particles. However, irrespective of the cause, this excess turbulent motion is usually assumed to produce a Gaussian line profile. Indeed, plasma motions in a stochastic turbulent system and described by Brownian motion will produce a velocity probability distribution function (PDF) that is normally distributed. However, large, sporadic motions far exceeding the mean, may lead to a velocity PDF with larger and heavier tails than that of a Gaussian, and lead to EUV line profiles better described by a kappa or Lorentzian profile. For example, non-Gaussian magnetic fluctuations are measured in space plasmas \citep[][]{1999GeoRL..26.1801S,2002GeoRL..29.1446H,2016MNRAS.459.3395P}, with this intermittency likely to exist on smaller scales in particular. Therefore, any evidence of non-Gaussian line profiles connected to solar flare turbulence could provide an important observational constraint regarding the nature of the turbulence, vital for MHD and kinetic modelling, which is not available via other techniques. Some possible causes of non-Gaussian spectral line profiles, including turbulence and accelerated ions are shown in Figure \ref{fig1}.

In this paper, we analyse flare SOL2014-03-29T17:44, that shows the presence of non-Gaussian EUV spectral lines. To date, SOL2014-03-29T17:44 is one of the best observed flares in history. As well as observations with {\em RHESSI\,} and {\em Hinode\,} EIS, the flare was also observed by the {\em Interface Region Imaging Spectrograph} \citep[IRIS; ][]{2014SoPh..289.2733D}, instruments onboard the {\em Solar Dynamics Observatory} \citep{2012SoPh..275....3P} and the {\em Dunn Solar Telescope (DST)}. Hence, it has generated a number of papers studying flare energy \citep{2015ApJ...804L..20A}, chromospheric evaporation and white light flare emission \citep[e.g.][]{2014ApJ...794L..23H,2015ApJ...813..113B,2015ApJ...806....9K,2015ApJ...811....7L,2015SoPh..290.3525L,2015ApJ...799..218Y,2016arXiv160200016H,2016ApJ...816...88K,2016arXiv160907390K,2016ApJ...827...38R}, spectropolarimetric data \citep{2015ApJ...814..100J}, sunquakes \citep{2014ApJ...796...85J,2015ApJ...812...35M}, Moreton waves \citep{2016SoPh..291.3217F} and soft X-ray pulsations \citep{2015SoPh..290.3625S}. Here, we show that many \ion{Fe}{16} and \ion{Fe}{23} lines, produced at electron temperatures of $\sim3$ MK and $\sim15$ MK respectively, have a line shape consistent with a k,appa velocity distribution. We discuss whether the observed non-Gaussian line profiles could be produced by the EIS instrumental profile. We create maps showing the spatial distribution of fitted line properties such as the $\kappa$ index and characteristic width that describe the velocity distribution at each location and time. Finally, we weigh the evidence for the line shapes being due to non-Maxwellian flare-accelerated ions or to non-Gaussian turbulent velocity fluctuations, which would be an observational first.

\section{Chosen flare and method}\label{method}

	\begin{figure}[t!]
	\centering
	\includegraphics[width=1.0\linewidth]{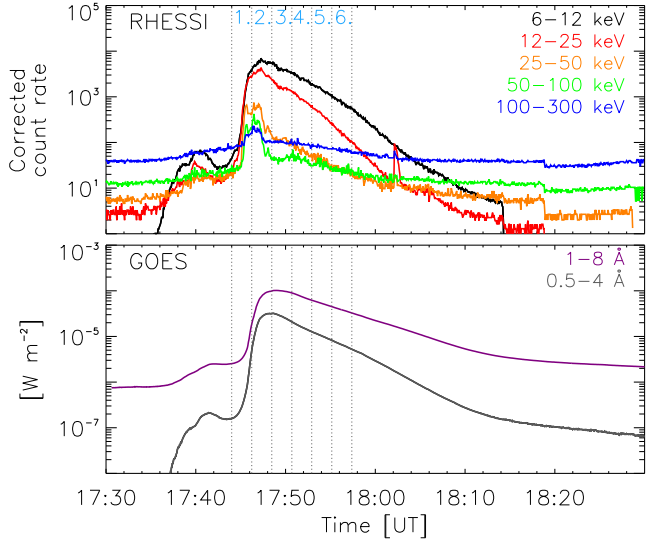}
	\caption{{\em RHESSI\,} (top) and {\em GOES\,} (bottom) light curves for flare SOL2014-03-29T17:44. The grey dashed lines indicate the start and end times of six EIS rasters covering the flare and the times of study.}
	\label{fig2}
	\end{figure}

SOL2014-03-29T17:44 is an X1.0 flare with coordinates [X=510'',Y=265'']. The  X-ray emission starts around 17:44 UT and peaks in soft X-rays (SXR) at $\sim$17:48 UT (in {\em Geostationary Operational Environmental Satellite (GOES)\,} 1-8~\AA). The hard X-ray (HXR, $>$25 keV) emission peaks around 17:46 UT. The flare {\em RHESSI\,} and {\em GOES\,} X-ray light curves are shown in Figure \ref{fig2}. The start and end times of six EIS rasters covering the rise, peak and decay times of SOL2014-03-29T17:44 are indicated by grey dotted lines in Figure \ref{fig2}, denoting the time intervals under study. EIS observes SOL2014-03-29T17:44 in fast-rastering mode. Each raster is two minutes and fourteen seconds long, with slit movements every $\sim12$ seconds. The slit scans in the X direction from solar west to east. The $1''$ slit is used during the observations, moving $3''.99$ every slit jump. The natural binning in the Y direction is $1''$.

\begin{figure*}[ht!]
	\vspace{-30 pt}
	\centering
	\hspace{21 pt}\includegraphics[width=0.47\textwidth]{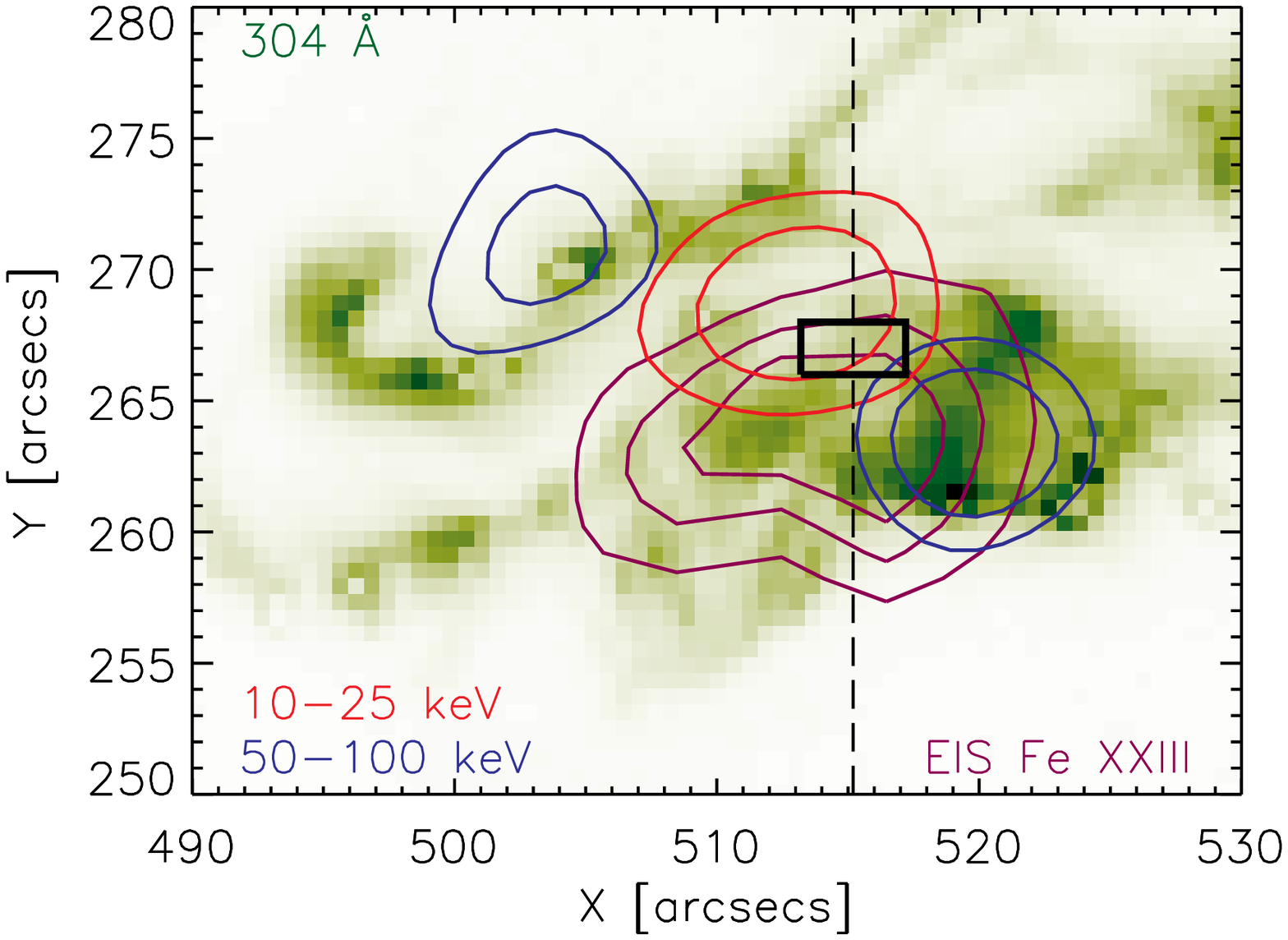}
	\hspace{3 pt}\includegraphics[width=0.47\textwidth]{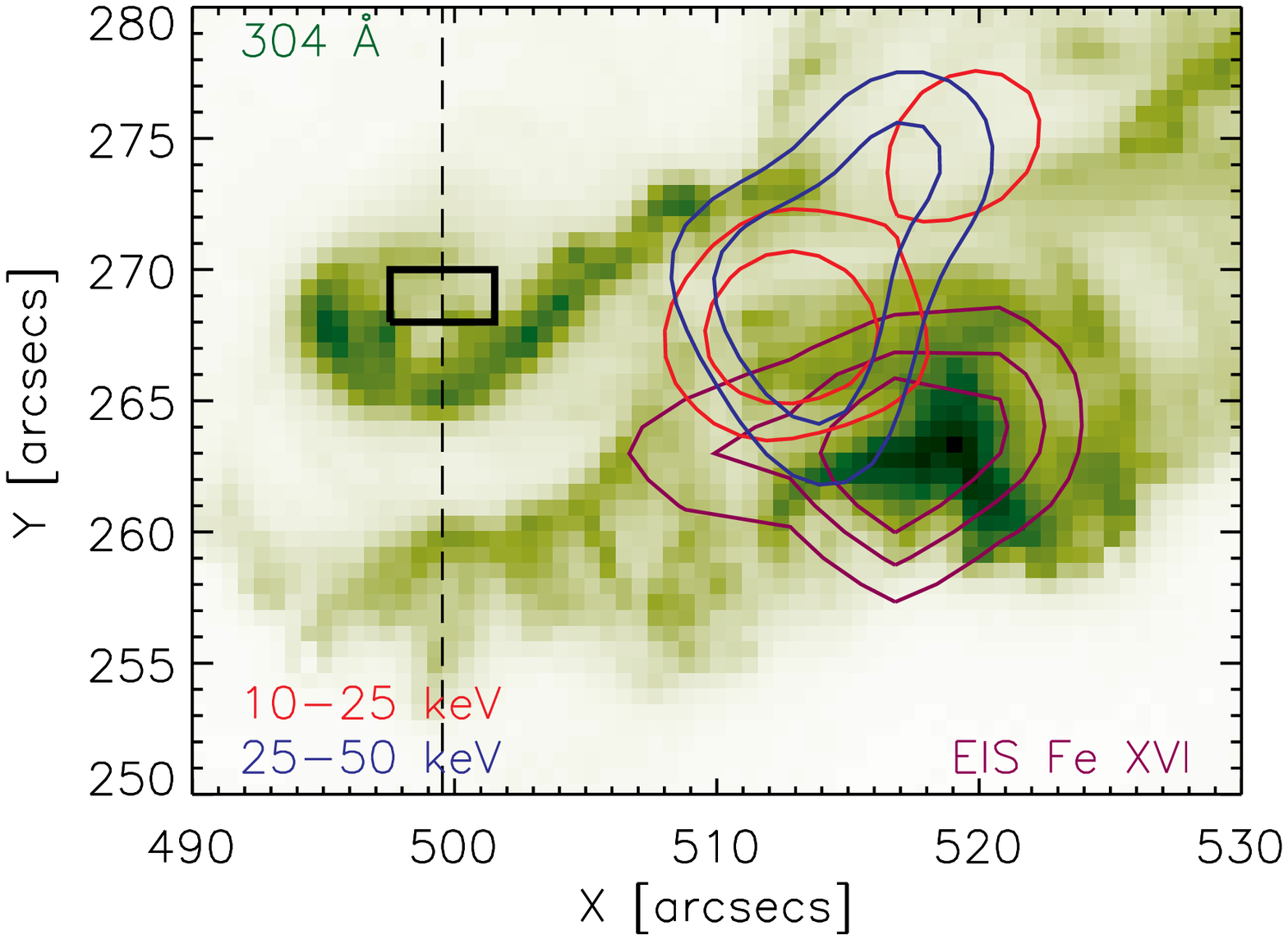}\hspace{-30 pt}
	\includegraphics[width=0.48\textwidth]{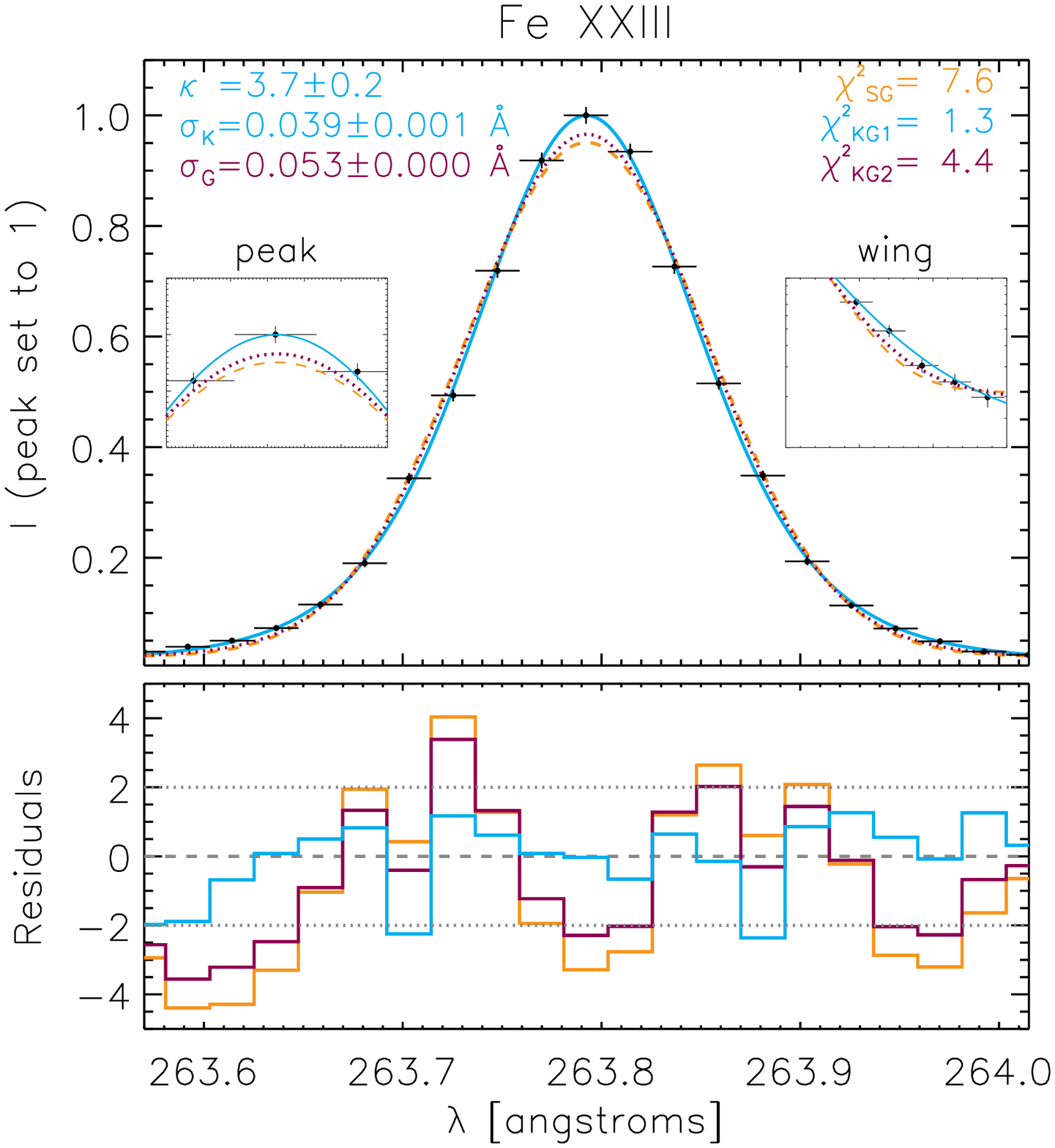}
	\includegraphics[width=0.48\textwidth]{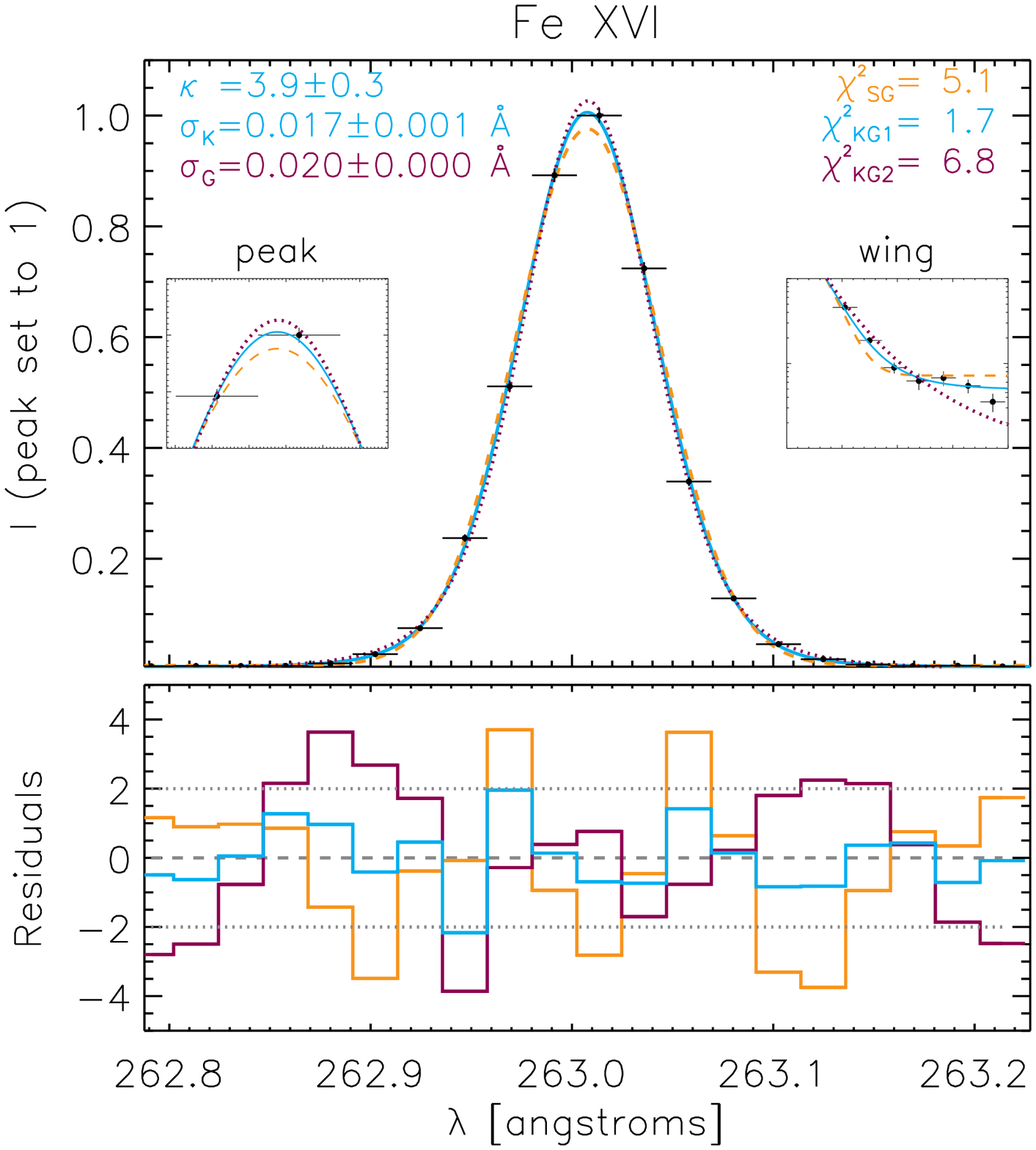}
	\caption{{\it Top row:} Two SDO AIA images of SOL2014-03-29T17:44 using the 304 \AA~passband (green background image) at times of 17:46:59 UT {\it (left)} and 17:49:22 UT {\it (right)}, times within two different EIS rasters. RHESSI contours at 10-25 keV (red) and either 25-50 keV or 50-100 keV (navy blue) are displayed at levels of 50 \% and 70 \% of the maximum. \ion{Fe}{23} {\it (left)} and \ion{Fe}{16} {\it (right)} intensity contours are displayed in purple, at 30 \%, 50 \% and 70 \% of the maximum. {\it Bottom row:} Spectral lines of either \ion{Fe}{23} {\it (left)} or \ion{Fe}{16} {\it (right)} observed at the location of the rectangular box shown in the top images. Each line is fitted with the KG1, KG2 and SG fits (see text for details). The small panels display the line peak and right wings in detail so that the fits can be clearly seen. The reduced $\chi^{2}$, residuals and fit parameters of $\kappa$ and $\sigma$ are also displayed.}
	\label{fig3}
	\end{figure*}

The morphology of SOL2014-03-29T17:44 is shown in Figure \ref{fig3}. The two images in the 304 \AA~passband of SDO {\em Atmospheric Imaging Assembly} \citep[AIA; ][]{2012SoPh..275...17L} at 17:46:58 UT and 17:49:22 UT show the mainly unsaturated flare ribbons. {\em RHESSI\,} X-ray contours at 10-25 keV and either 25-50 keV or 50-100 keV are overlaid. During raster 17:46:14 UT, two HXR footpoints at 50-100 keV are present, at either side of a lower energy 10-25 keV coronal source. At the later time, the 50-100 keV HXR footpoints disappear but we still observe X-rays up to 50 keV. The EIS intensity contours from the \ion{Fe}{16} and \ion{Fe}{23} EIS rasters are also displayed. The EIS data is aligned with AIA using the procedure eis\_aia\_offsets.pro, with a $5''$ error in Y. We assume that AIA and {\em RHESSI\,} are well-aligned for the purposes of our analysis.

The EIS data in the Y direction is binned into $2''$ bins (from $1''$) improving the signal-to-noise ratio and line fitting goodness-of-fit. The EIS instrumental broadening $W_{\rm inst}$ using the $1''$ slit is $W_{\rm inst}=0.059$~\AA~(the full width at half maximum, FWHM) assuming a Gaussian instrumental profile.

\subsection{Non-Gaussian ion and plasma velocity distributions}\label{nonthermal}
The EIS data for SOL2014-03-29T17:44 includes two suitably strong, unblended spectral lines formed at different temperatures: \ion{Fe}{16} ($\approx2.5-4$ MK, $\log{T}=6.4$) and \ion{Fe}{23} ($\approx15-16$ MK, $\log{T}=7.2$). We use the non-Gaussian line profiles to determine the underlying velocity distribution.\footnote{It is also possible that a multi-thermal plasma along the line-of-sight could be responsible, particularly if the ions and electrons have different temperature distributions, but this is not discussed here.}

For the case of an accelerated ion population and following \cite{2014ApJ...796..142B}, a 3-D kappa ion velocity distribution $f(v)$ of the first kind can be written as
\begin{equation}
\begin{split}
f(v)= & \frac{n}{\pi^{3/2}v_{th}^{3}\kappa^{3/2}}\frac{\Gamma(\kappa)}{\Gamma(\kappa-3/2)}\left(1+\frac{v^{2}}{\kappa v_{th}^{2}}\right)^{-\kappa} \\
= & A_{v}\left(1+\frac{v^{2}}{\kappa v_{th}^{2}}\right)^{-\kappa}
\end{split}
\label{fv_3d}
\end{equation}
where $n=\int f(v) d^{3}v$ is the number density associated with an acccelerated ion distribution and $v_{th}=\sqrt{2k_{B}T/M}$ is a Maxwellian thermal velocity at temperature $T$ (for $k_{B}$ the Boltzmann constant and $M$ the ion mass), and $\Gamma(z)=\int_{0}^{\infty}t^{z-1}e^{-t}dt$ is the Gamma function. 

To make the link to observed line profiles, we need to convert Equation \ref{fv_3d} to a 1-D line-of-sight velocity $v_{\parallel}$. The 1-D ion velocity distribution is given by the integral over all perpendicular velocities $v_{\perp}$, so that, assuming isotropy,

\begin{equation}\label{los_kap}
\begin{split}
f(v_{\parallel})= & \int_{0}^{\infty}f(v)2\pi v_{\perp} dv_{\perp} \\
= & A_{v}\int_{0}^{\infty}\left(1+\frac{v_{\parallel}^{2}+v_{\perp}^{2}}{\kappa v_{th}^{2}}\right)^{-\kappa}2\pi v_{\perp} dv_{\perp} \\
= & A_{v}\frac{\pi \kappa v_{th}^{2}}{\kappa-1}\left(1+\frac{v_{\parallel}^{2}}{\kappa v_{th}^{2}}\right)^{-\kappa+1} \\
\rightarrow f(v_{\parallel})=& \frac{n}{\pi^{1/2}\kappa^{1/2}v_{th}}\frac{\Gamma(\kappa-1)}{\Gamma(\kappa-3/2)}\left(1+\frac{v_{\parallel}^{2}}{\kappa v_{th}^{2}}\right)^{-\kappa+1}
\end{split}
\end{equation}

where $n=\int_{-\infty}^{\infty}f(v_{\parallel})dv_{\parallel}$.

As $\kappa\rightarrow\infty$, Equation \ref{los_kap}\footnote{Equation \ref{los_kap} is slightly different to the kappa function used in \cite{2016A&A...590A..99J}, where the index $(-\kappa)$ was used instead of $(-\kappa+1)$.} tends to a 1-D isothermal Maxwellian distribution. In this form, we can think of $v_{th}$ as the thermal speed of a Maxwellian ion population before acceleration, or a characteristic speed of the distribution. For low $\kappa$ and large $v_{\parallel}$, we can approximate the ion velocity distribution as a power law with $f(v_{\parallel})\approx v_{\parallel}^{-2(\kappa-1)}=v_{\parallel}^{-\beta}$. 

The line-of-sight velocity distribution is related to the emitted line profile by $f(v_{\parallel})\propto I(\lambda)\frac{d\lambda}{dv_{\parallel}}=I(\lambda)\frac{\lambda_{0}}{c}$, for wavelength $\lambda$, rest wavelength $\lambda_{0}$ and speed of light $c$, giving

\begin{equation}\label{I_line}
I(\lambda)=A_{\lambda}\left(1+\frac{(\lambda-\lambda_{0})^{2}}{\kappa 2\sigma_{\kappa}^{2}}\right)^{-\kappa+1}
\end{equation}

where $v_{th}^{2}=2k_{B}T/M=2\sigma_{\kappa}^{2} c^{2}/\lambda_{0}^{2}$ and $A_{\lambda} \propto A_{v} c/\lambda_{0}$. As $\kappa\rightarrow\infty$ in Equation \ref{I_line}, the line shape becomes Gaussian. Also, if $\kappa=2$, the line profile is the same as a Lorentzian. Hence, a kappa line profile can be used as a general line fitting form that can cover the specific cases of both Gaussian and Lorentzian line profiles.
A kappa distribution might also be used to describe a spectrum of velocities $F(u_{\parallel})$ produced by plasma turbulence. In this case, the plasma velocity distribution (excluding the ion thermal motions) could be described by,
\begin{equation}\label{pvel}
F(u_{\parallel})=\frac{F_{0}}{\kappa^{1/2}}\frac{\Gamma(\kappa-1)}{\Gamma(\kappa-3/2)}\left(1+\frac{(u_{\parallel}-u_{1})^{2}}{\kappa u_{0}^{2}}\right)^{-\kappa+1}
\end{equation}
where $u_{\parallel}$ is the plasma velocity, $F_{0}$ is a function dependent on plasma properties, $u_{0}$ is a characteristic speed of the turbulence and $u_{1}$ is a bulk flow plasma velocity. The overall velocity distribution would then be a convolution of $F(u_{\parallel})$ with the ion velocity distribution, but the overall line profile and its non-Gaussianity could still be approximated by a kappa line distribution. Hence, regardless of the physical process, a kappa line profile is an excellent starting point for the detection and analysis of non-Gaussian ion or plasma velocities. Even if the kappa distribution does not describe all the underlying physics, it provides a mathematically convenient line profile for the determination of non-thermal/non-Gaussian velocities from {\it Hinode} EIS data (where more detailed fitting is not possible), providing a fitting function that can range from a Gaussian to a Lorentzian.

\subsection{EIS line fitting of \ion{Fe}{16} and \ion{Fe}{23}}\label{lfit}
The \ion{Fe}{16} and \ion{Fe}{23} lines are fitted with a single Gaussian to estimate the Gaussian intensity, centroid and line width. Lines with skewness $|S|>0.08$ indicating lack of symmetry, and probable moving components, are removed from the study. Many of the \ion{Fe}{16} and \ion{Fe}{23} lines fitted with a Gaussian have high reduced $\chi^{2}$ values, greater than 6. From the Gaussian fitting, even after the removal of a Gaussian instrumental profile with FWHM $W_{\rm inst}=0.059$ \AA, the Doppler broadening in most regions is larger than expected from an isothermal plasma. The Gaussian line widths after the removal of $W_{\rm inst}$ for \ion{Fe}{23} can be as large as $0.12$ \AA~~and for \ion{Fe}{16} as large as $0.08$ \AA. The expected isothermal widths for \ion{Fe}{23} and \ion{Fe}{16} are $W_{\rm th}\sim0.1$~\AA~(for $\log{T}=7.2$) and $W_{\rm th}\sim0.04$ \AA~(for $\log{T}=6.4$).

Next, as in \cite{2016A&A...590A..99J}, we re-fit the lines with a convolved kappa - Gaussian distribution, accounting for (1.) a Gaussian EIS instrumental profile with $W_{\rm inst}=0.059$ \AA~and (2.) the possibility of a non-Maxwellian velocity distribution resulting in line profiles with higher peaks and `heavier' wings than a Gaussian. The convolved kappa ($\mathcal{K}$)- Gaussian ($\mathcal{G}$) line profile is given by

	\begin{equation}
	\begin{split}
	&\mathcal{W}(\lambda) = \mathcal{G}(\lambda)*\mathcal{K}(\lambda)= A[0]+A[1]\times\\ 
	& \sum_{\lambda^{'}}\exp{\left(-\frac{(\lambda^{'}-A[2])^{2}}{2\sigma_{I}^{2}}\right)\left(1+\frac{(\lambda-\lambda^{'}-A[2])^{2}}{2A[3]^{2}A[4]}\right)^{-A[4]+1}}
	\end{split}
	\label{fit_IW}	
	\end{equation}

where there are five free fit parameters $A$. For further details see \cite{2016A&A...590A..99J}.  From Equation \ref{fit_IW}, we are interested in determining the values of the kappa index $\kappa$ and characteristic width $\sigma_{\kappa}$ (fit parameters A[4] and A[3] respectively); parameters that provide information about the velocity distribution. We call this fit KG1. As discussed in \cite{2016A&A...590A..99J}, this function is a generalized Voigt function, with the traditional Voigt function, a convolution of a Gaussian and a Lorentzian, being the limiting case when A[4] $=\kappa=2$.

It is possible that all or part of the non-Gaussian line shape results from the instrumental profile. There is no reason for the EIS instrumental profile to be Gaussian, although it may be extremely well-approximated as such. For example, the spectrometer might be expected to have an instrumental response closer to a ${\rm sinc}^{2}\lambda$ function. To account for the possibility of a non-Gaussian instrumental response we fit another convolved kappa - Gaussian (see Equation \ref{IW} in Appendix \ref{app}), where the kappa part is fixed to represent an instrumental profile with chosen $\kappa_{I}$ and $\sigma_{I}$ and the Gaussian parameters are free to vary, representing a physical line profile.  The EIS instrumental profile can be approximated by a Gaussian profile with FWHM $W_{\rm inst}=0.059$ \AA. Therefore the kappa instrumental profile is constrained by the requirement that $\kappa_{I}$ and $\sigma_{I}$ produce $W_{\rm inst}=0.059$~\AA~when approximated by a Gaussian. To obtain this we choose $\kappa_{I}=3$ and $\sigma_{I}=0.0395$ \AA. This parameter choice is not unique and the choice of values are discussed further in Appendix \ref{app}. We call this fit KG2.

The line goodness-of-fits are judged by a combination of ``judgement by eye'', a reduced $\chi^{2}=\frac{1}{\rm DOF} \displaystyle \sum_{i}\frac{(o_{i}-m_{i})^{2}}{\epsilon_{i}^{2}}$ from the weighted least squares fit, where $o_{i}$ are the observed intensity values, $\epsilon_{i}$ are the observed intensity error values, $m_{i}$ are the model values and degree of freedom ${\rm DOF=number\;of\;data\;points} - {\rm number\;of\;fitted\;parameters}$, and by examining the fit residuals $R=\frac{o-m}{\epsilon}$.

The bottom row of Figure \ref{fig3} displays two examples: one \ion{Fe}{16} and one \ion{Fe}{23} profile and fit. Here the lines are fitted with: (1.) a physical kappa  - instrumental Gaussian fit (KG1), (2.) an instrumental kappa  - physical Gaussian fit (KG2) and (3.) a single Gaussian (SG). The corresponding spatial locations are indicated in the images shown in the top row of Figure \ref{fig3} by the rectangular boxes and slit positions (dashed lines). Each image displays the AIA 304~\AA~passband where two north and south ribbons can be clearly seen. {\em RHESSI\,} X-ray contours at 10-25 keV and 25-50 keV or 50-100 keV and \ion{Fe}{16} or \ion{Fe}{23} contours are displayed. Figure \ref{fig3} shows how the kappa part of the KG1 fit is able to account for the higher peaks and broader wings of the observed spectral lines. For both profiles in Figure \ref{fig3}, the single Gaussian fits produce the large reduced $\chi^{2}$ values of $\chi_{G}^{2}=7.6$ (\ion{Fe}{23}) and $\chi_{G}^{2}=5.1$ (\ion{Fe}{16}). The KG1 fits give the lowest reduced $\chi^{2}$ values of $\chi_{KG}^{2}=1.3$ (\ion{Fe}{23}) and $\chi_{KG}^{2}=1.7$ (\ion{Fe}{16}). The KG2 fits produce higher reduced $\chi^{2}$ values than the KG1 fits with $\chi_{KG2}^{2}=4.4$ (\ion{Fe}{23}) and $\chi_{KG2}^{2}=6.8$ (\ion{Fe}{16}). The lines displayed in Figure \ref{fig3} are two examples where the KG1 fit (physical kappa profile) gives a lower goodness-of-fit than the KG2 fit (instrumental kappa profile). The example line fits in Figure~\ref{fig3} support a physical rather than an instrumental origin since the lines are best fitted with different kappa parameters, and not the single, fixed $\kappa_{I}$ and $\sigma_{I}$ values of the constraint, as we might expect if the non-Gaussian part of the profile was wholly instrumental. We discuss this in greater detail in Appendix \ref{app} and later in subsection \ref{inst_comp}.

	\begin{figure*}[t]
	\centering
	\includegraphics[width=0.95\linewidth]{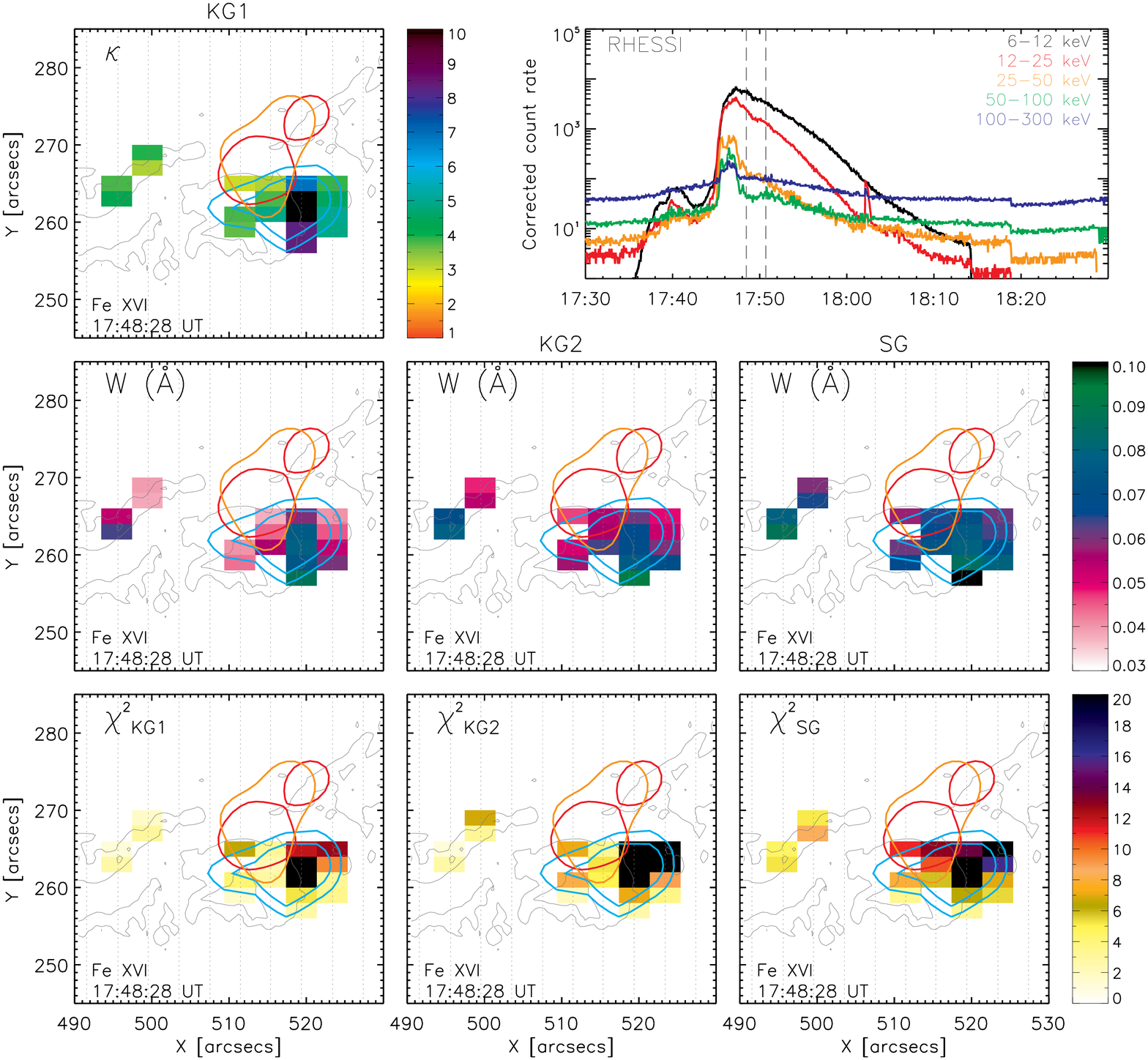}
	\caption{Gaussian (SG) and kappa-Gaussian (KG1 and KG2) fits for \ion{Fe}{16}. {\it Left column:} kappa-Gaussian fit using a Gaussian instrumental profile (KG1), {\it middle column:} kappa-Gaussian using a kappa instrumental profile (KG2) and {\it right column:} single Gaussian fit (SG). {\it Row 1:} $\kappa$ index, {\it row 2:} $2\sqrt{2\ln{2}}\times\sigma$ of each fit ($\sigma_{\kappa}$ or $\sigma_{G}$) and {\it row 3:} reduced $\chi^{2}$ values for each fit. The {\em RHESSI\,} light curves are also displayed, with grey dashed lines showing the time of observation (time $t_3$). The parameters from lines shown in this figure satisfy criteria (1.) and (2.) only (inital study, see text for details).}
	\label{fig4a}
	\end{figure*}
	\begin{figure*}[t]
	\centering
	\includegraphics[width=0.95\linewidth]{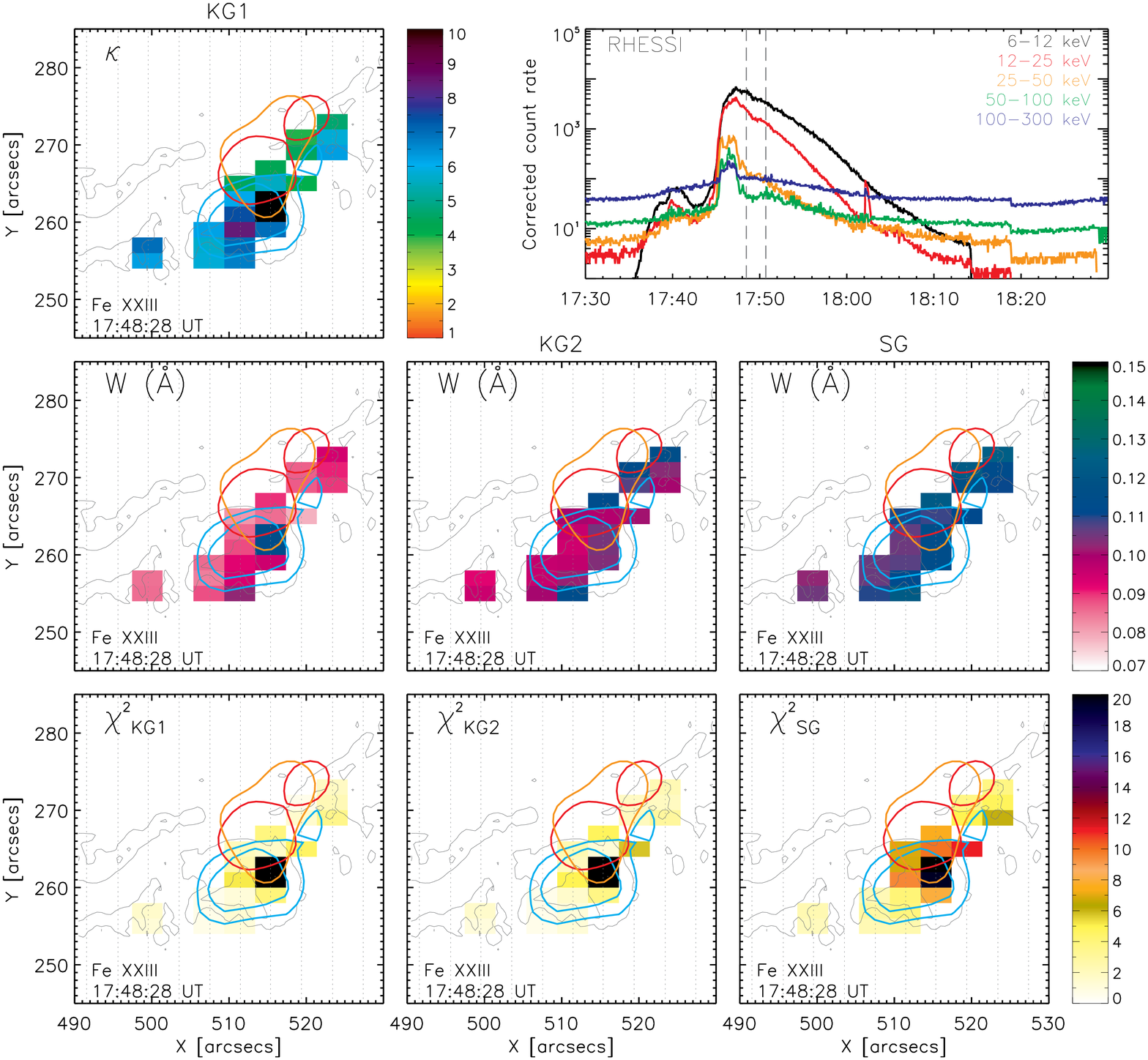}
	\caption{Gaussian (SG) and kappa-Gaussian (KG1 and KG2) fits for \ion{Fe}{23}. {\it Left column:} kappa-Gaussian fit using a Gaussian instrumental profile (KG1), {\it middle column:} kappa-Gaussian using a kappa instrumental profile (KG2) and {\it right column:} single Gaussian fit (SG). {\it Row 1:} $\kappa$ index, {\it row 2:} $2\sqrt{2\ln{2}}\times\sigma$ of each fit ($\sigma_{\kappa}$ or $\sigma_{G}$) and {\it row 3:} reduced $\chi^{2}$ values for each fit. The {\em RHESSI\,} light curves are also displayed, with grey dashed lines showing the time of observation (time $t_3$). The parameters from lines shown in this figure satisfy criteria (1.) and (2.) only (inital study, see text for details).}
	\label{fig5}
	\end{figure*}

In Figure \ref{fig3}, the residuals for each spectral line are also shown. For the chosen \ion{Fe}{23} line, the residuals clearly show that the KG1 model is a better fit for the line, as indicated by the low $\chi^{2}_{KG1}$ value. This is particularly noticable around the peak and the wings of the line, where the KG1 residuals are very close to zero (values within $\pm$2), compared to the fixed KG2 and SG residuals (values within $\pm$4). Again, for the chosen \ion{Fe}{16} line, the KG1 residuals show that this model is a better description of the line than a Gaussian (SG), for all wavelengths covering the line profile (again the KG1 residual values are within $\pm$2).

We perform two line profile studies. The initial study fits, with KG1, KG2 and SG functions, lines that satisfy the following two criteria:
\begin{enumerate}
\item Lines must have an absolute value of skewness less than 0.08 \citep[to remove lines with moving components as discussed. Also see][]{2016A&A...590A..99J}.
\item The estimated noise level (calculated as the standard deviation of the ratio of the intensity errors to intensity for each line) for the line must be below 9\% and the ratio of the integrated intensity error to integrated intensity less than 0.9\%.
\end{enumerate}

Following the line fitting with KG1, KG2 and SG profiles, we identify those fits where we are confident that KG1 is the best fit, according to the following extra criteria:

\begin{enumerate}
\setcounter{enumi}{2}
\item The reduced $\chi^{2}$ values of the kappa - Gaussian fits must be less than 5.0.
\item The reduced $\chi^{2}$ values of the Gaussian fits must be greater than 3.0.
\item The ratio $\chi_{G}^{2}/\chi_{KG}^{2}$ must be greater than 2.0 (for both KG1 and KG2).
\end{enumerate}

Criterion 2 is used as a ``noise value'', which we define as 100\%$\times{\rm STD}(\epsilon/o)$ (for STD=standard deviation). From the work in \citet{2016A&A...590A..99J} and by testing model lines with different levels of Gaussian noise, we found that lines with a noise value less than $\sim$10\% were usually suitable (i.e. small intensity error values) for a line model comparison. The integrated intensity error to integrated intensity ratio of 0.9\% was chosen by trial and error and by examining how this value changed for lines found to be either suitable or unsuitable for study. Criterion 2 allows us to quickly remove a large fraction of unsuitable lines in each raster without examining each line in detail, since each map has a total of $\sim$660 lines. Further, criteria 3-5 help to find non-Gaussian line shapes and remove lines with larger errors that can be well-fitted by all models (i.e. all producing low $\chi^{2}$ values), helping to pinpoint and only examine lines that have a definite non-Gaussian shape. In particular, criteria 4 and 5 are used to remove lines where the Gaussian model has low $\chi^{2}<3$ since we want to look at (a) non-Gaussian lines and (b) remove noisy lines well-fitted by any model.

None of the \ion{Fe}{16} regions contained warm pixels \citep[as discussed in][]{2016A&A...590A..99J} but four \ion{Fe}{23} regions did contain warm pixels. In the initial analysis of Section \ref{results} warm pixels are included but they are removed in the further analysis of KG1. We also varied the EIS line intensities using the codes of \cite{2016SoPh..291...55K} that account for finite binning in wavelength before the lines are fitted.

\section{Results}\label{results}

\subsection{Initial comparison of the KG1, KG2 and SG fits}
In Figures \ref{fig4a} and \ref{fig5}, maps of line fit parameters $\kappa$ and $W=2\sqrt{2\ln{2}}\times\sigma_{\kappa}$, and the goodness-of-fit $\chi^{2}$ are displayed. These are shown for a single EIS raster time of 17:48:23 UT (start time, $t_{3}$) and for each of the three fits: KG1 (first column, three panels), KG2 (second column, two panels) and SG (third column, two panels), for \ion{Fe}{16} (Figure \ref{fig4a}) and \ion{Fe}{23} (Figure \ref{fig5}). Line widths $W$ are displayed as a `Gaussian FWHM' for easy comparison with line widths found from Gaussian line fitting. At this time, fits satisfying criteria 1 and 2 are located $\sim $within the \ion{Fe}{16} 30\% intensity contour and along the northern ribbon, and for \ion{Fe}{23},  mainly within the \ion{Fe}{23} 30\% contour and close to the 10-20 keV and 25-50 keV X-ray contours.

For the \ion{Fe}{16} KG1 fit, the lowest values of $\chi_{\rm KG1}^{2}$ ($<3$) are located at the edges of the \ion{Fe}{16} source and along the northern ribbon. Closer inspection of the actual line fitting for all fits shows that the high $\chi_{\rm KG1}^{2}$ ($\sim$10) values close to the centre of the \ion{Fe}{16} source are due to the line having a moving component not removed by the skewness condition (the shape of some line profiles with a large moving component can lead to the line shape having a lower skewness than 0.08). The KG1 $W$ values in regions of low $\chi_{\rm KG1}^{2}$ are between $0.04$ \AA~and $0.07$ \AA. For the KG2 fit, the $\chi_{\rm KG2}^{2}$ values are low in a number of locations, but with higher values than the KG1 fit. The $\kappa$ index and $\sigma_{\kappa}$ values for the KG2 fit are kept constant at 3 and 0.0395 \AA~ respectively, and the KG2 Gaussian widths ($2\sqrt{2\ln{2}}\times \sigma_{G}$) are found to be $>0.05$ \AA. The SG $\chi_{\rm SG}^{2}$ values are higher (often greater than 3) and the SG widths are $>0.06$ \AA. For the \ion{Fe}{23} KG1 fit, the majority of $\chi_{\rm KG1}^{2}$ values are again very low (mainly $<$3), apart from two points that have very high $\chi_{\rm KG1}^{2}$ values (greater than 16). Again, on closer inspection, these lines appear to include blue-shifted moving components (for all fits). For KG1, the $\kappa$ index values are found to be between 4 and 10 and $W$ between $\sim$0.08 \AA~and 0.10 \AA. For KG2, the $W$ values are greater than $\sim$0.10 \AA~ but the $\chi_{\rm KG2}^{2}$ values are low ($\le4$). The $\chi_{\rm SG}^{2}$ for the \ion{Fe}{23} SG fits are again higher, just as for the \ion{Fe}{16} fits, with values above 6. The uncertainties associated with $\sigma$ inferred from each of the KG1, KG2 and SG fits are small, of the order $10^{-3}$~\AA~or less. The errors for the KG1 $\kappa$ values are of the order $10^{-1}$ for both \ion{Fe}{16} and \ion{Fe}{23}. The initial analysis and Figures \ref{fig4a} and \ref{fig5} show three main results:
\begin{enumerate}
\item{Spatial patterns for $\kappa$ index and characteristic width $\sigma_{\kappa}$ (KG1) emerge and this is discussed further in subsection \ref{FA_KG1}.}
\item{The KG1 and KG2 $W$ values are smaller than those found from the SG fit, which requires the presence of larger excess line broadening to explain the observed values.}
\item{Overall, the KG1 $\chi_{\rm KG1}^{2}$ values are smaller than the KG2 and SG values, for both \ion{Fe}{16} and \ion{Fe}{23} (with most reduced $\chi_{\rm KG1}^{2}$ values less than two).}
\end{enumerate} 
	\begin{figure*}[t]
	\centering
	\hspace{87pt}\includegraphics[width=0.21\linewidth]{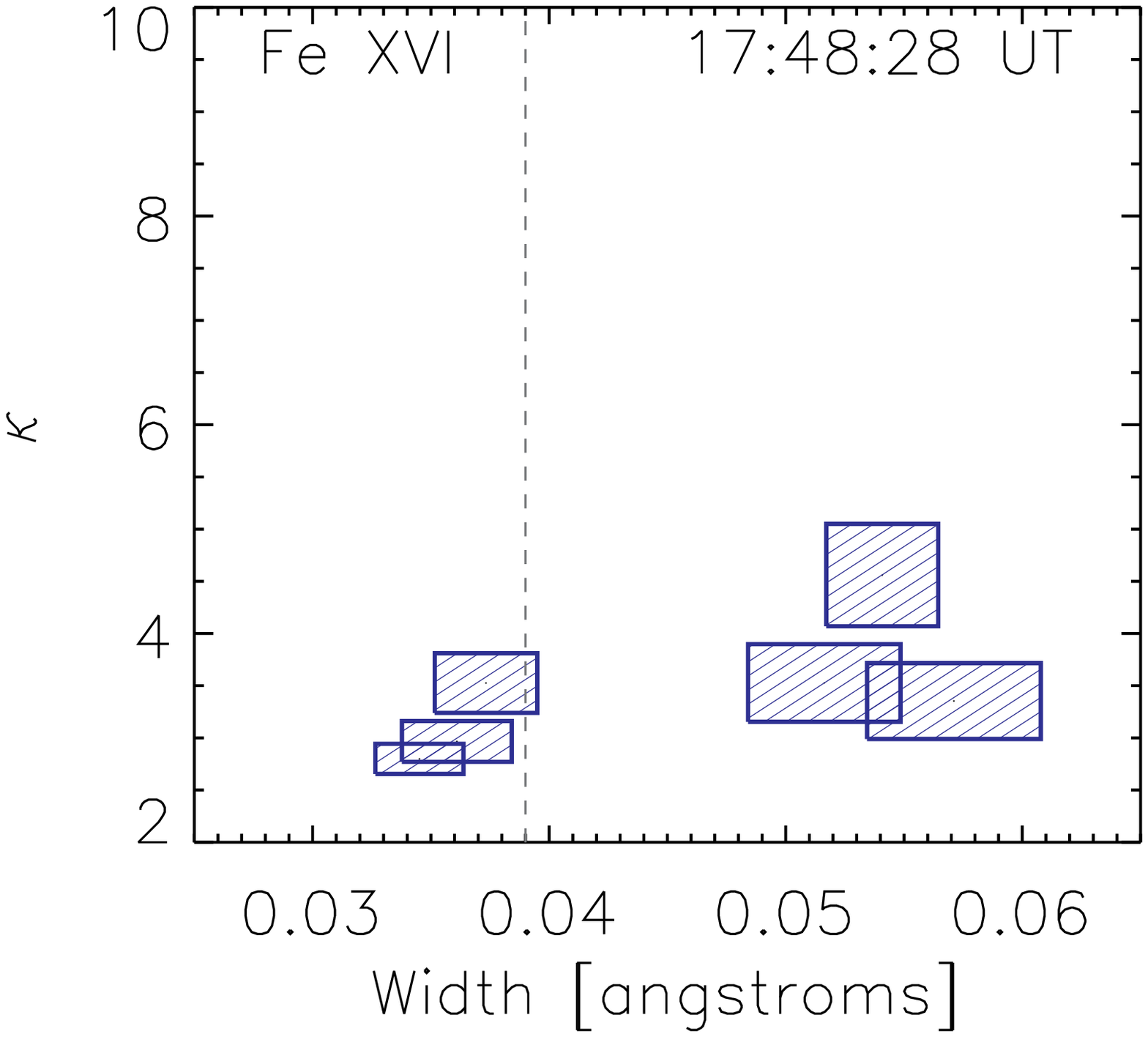}\hspace{-23pt}
	\includegraphics[width=0.21\linewidth]{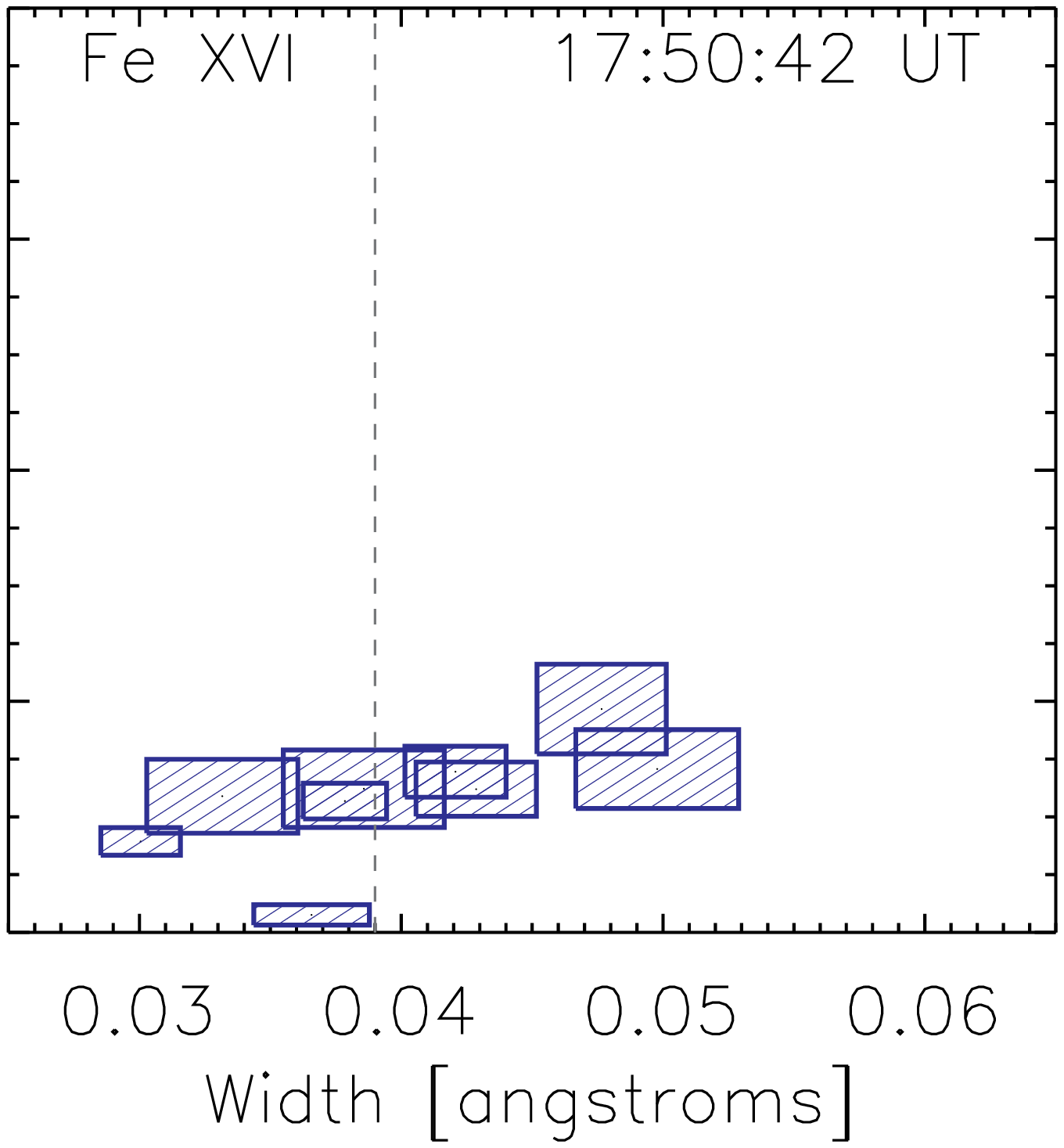}\hspace{-23pt}
	\includegraphics[width=0.21\linewidth]{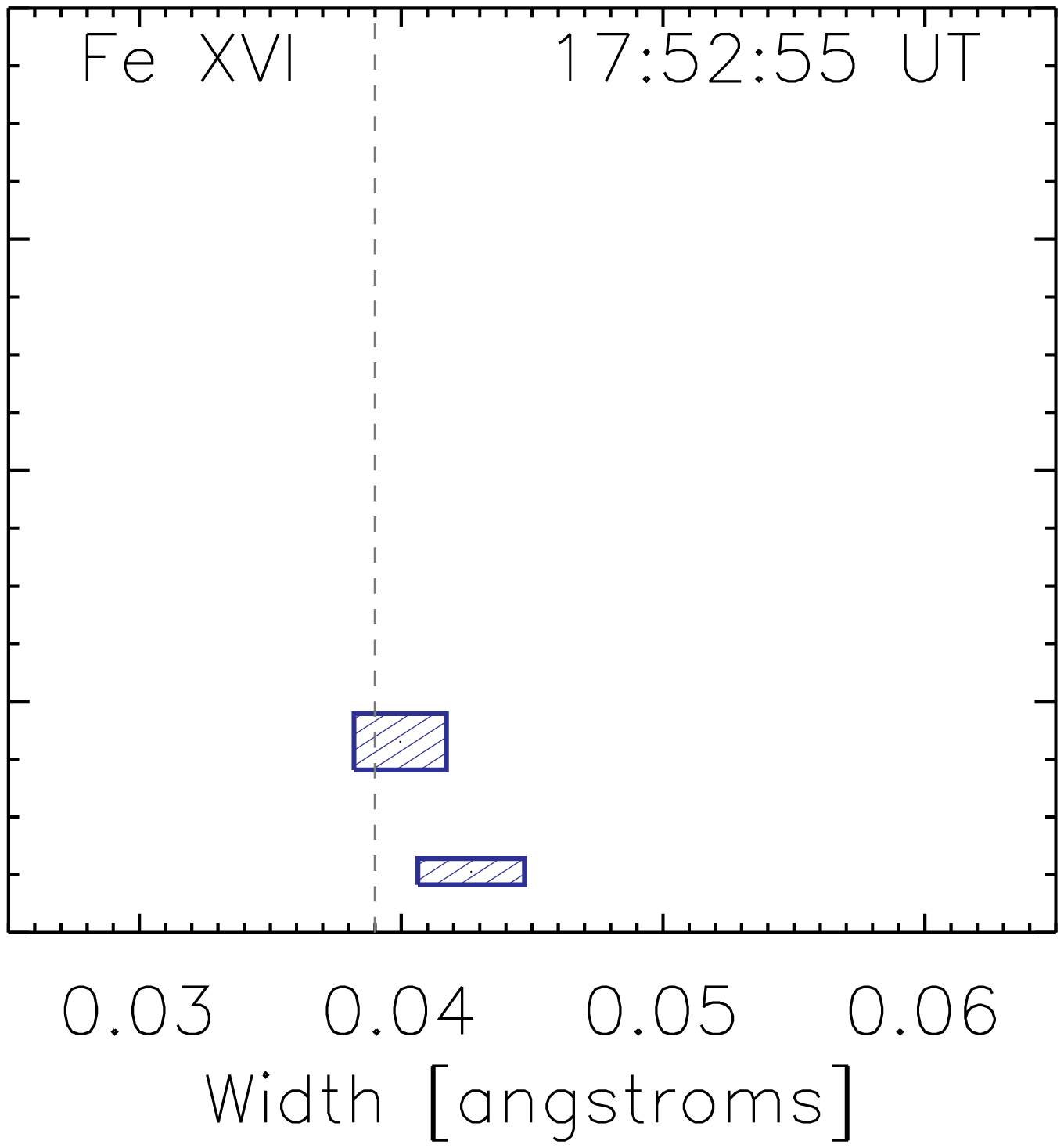}\hspace{-23pt}
	\includegraphics[width=0.21\linewidth]{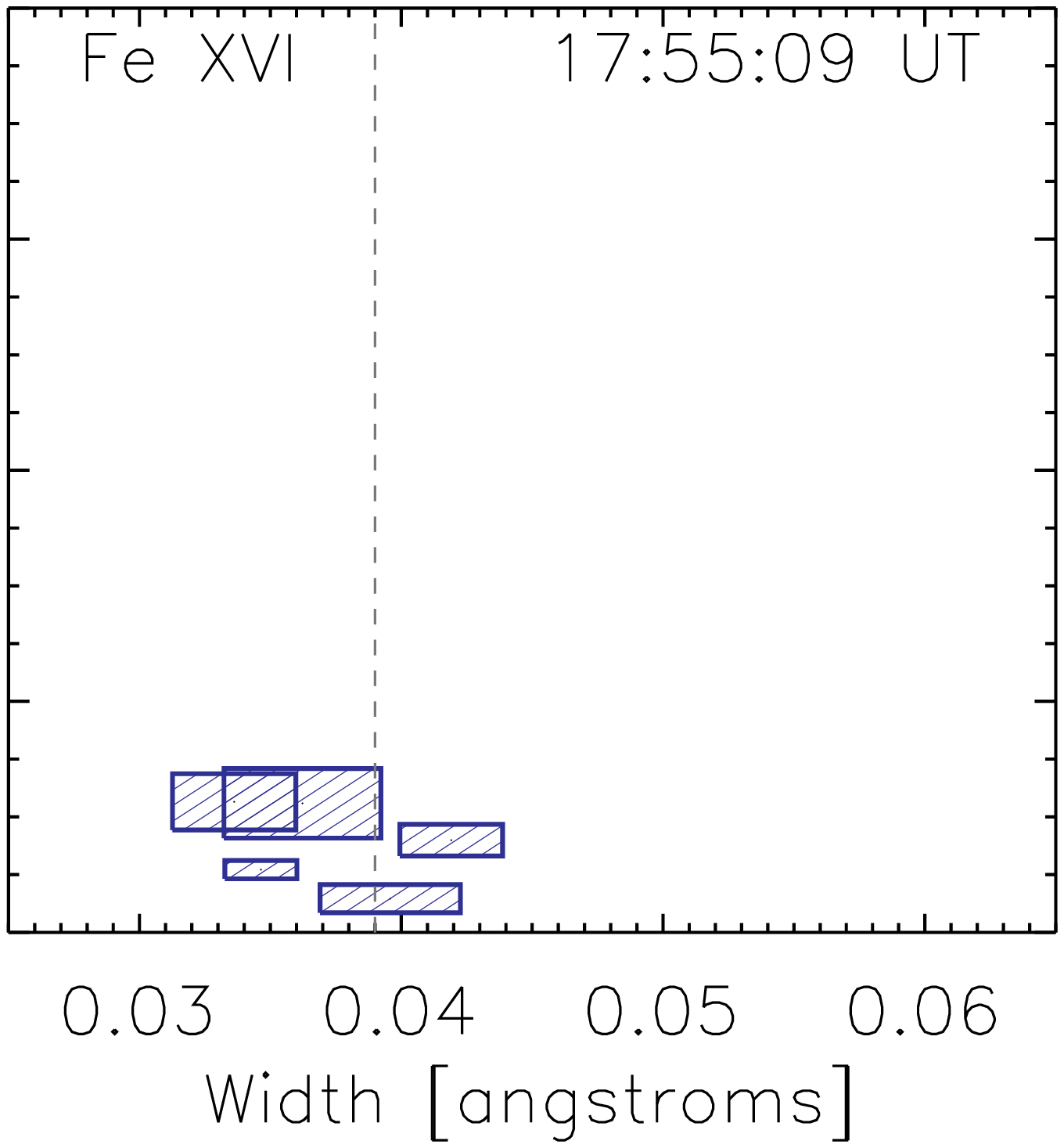}\hspace{-23pt}
	\includegraphics[width=0.21\linewidth]{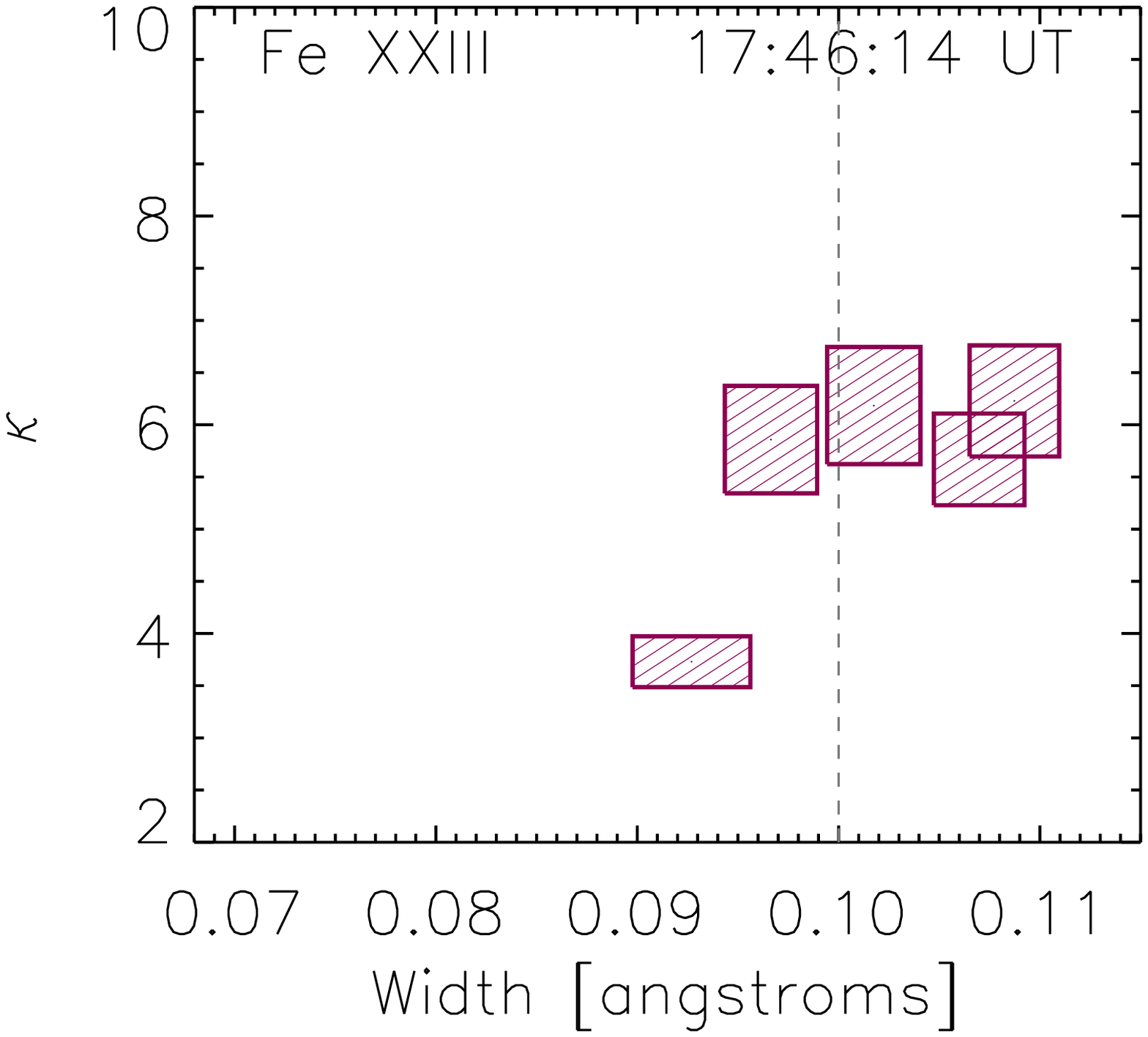}\hspace{-23pt}
	\includegraphics[width=0.21\linewidth]{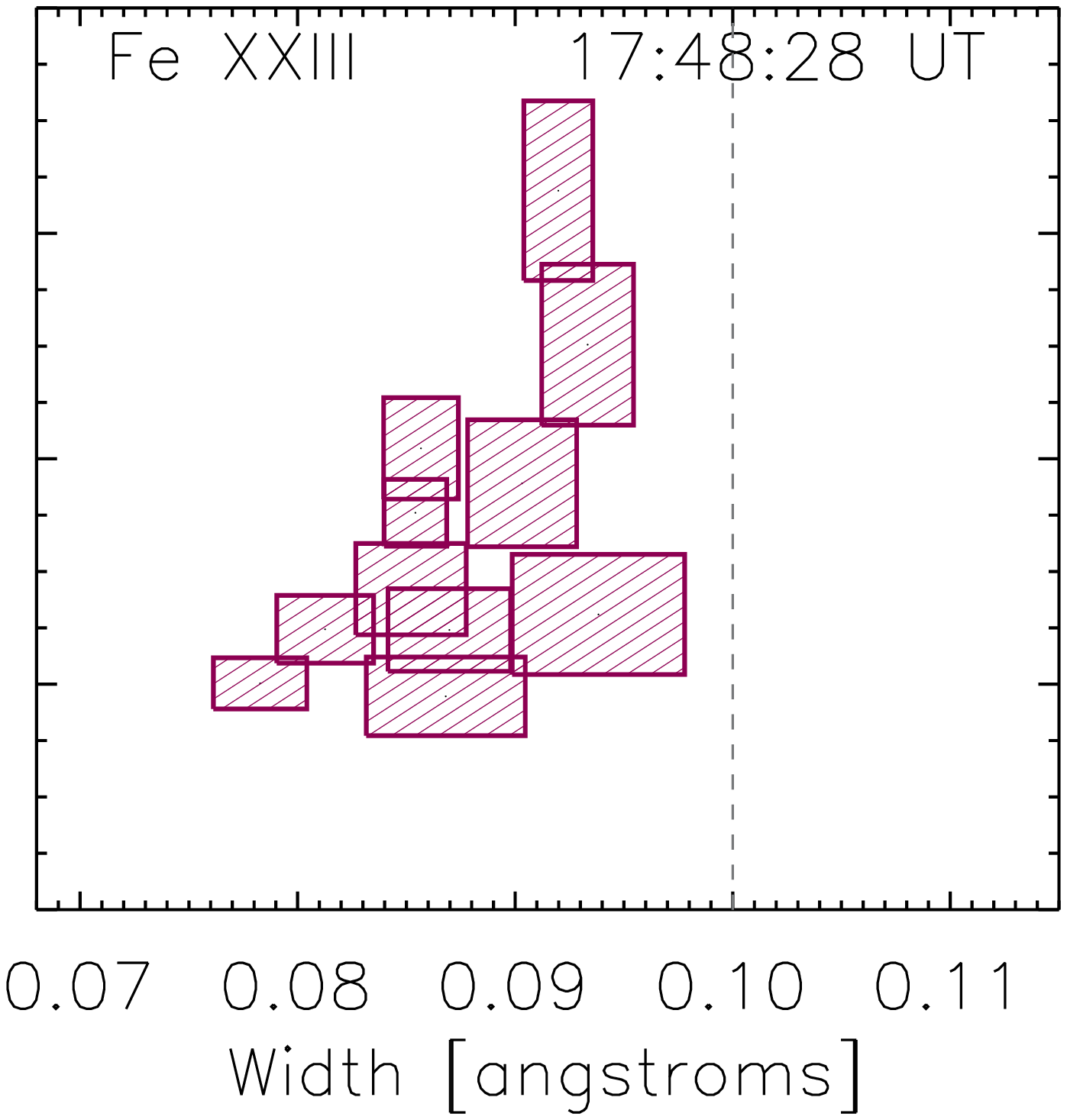}\hspace{-23pt}
	\includegraphics[width=0.21\linewidth]{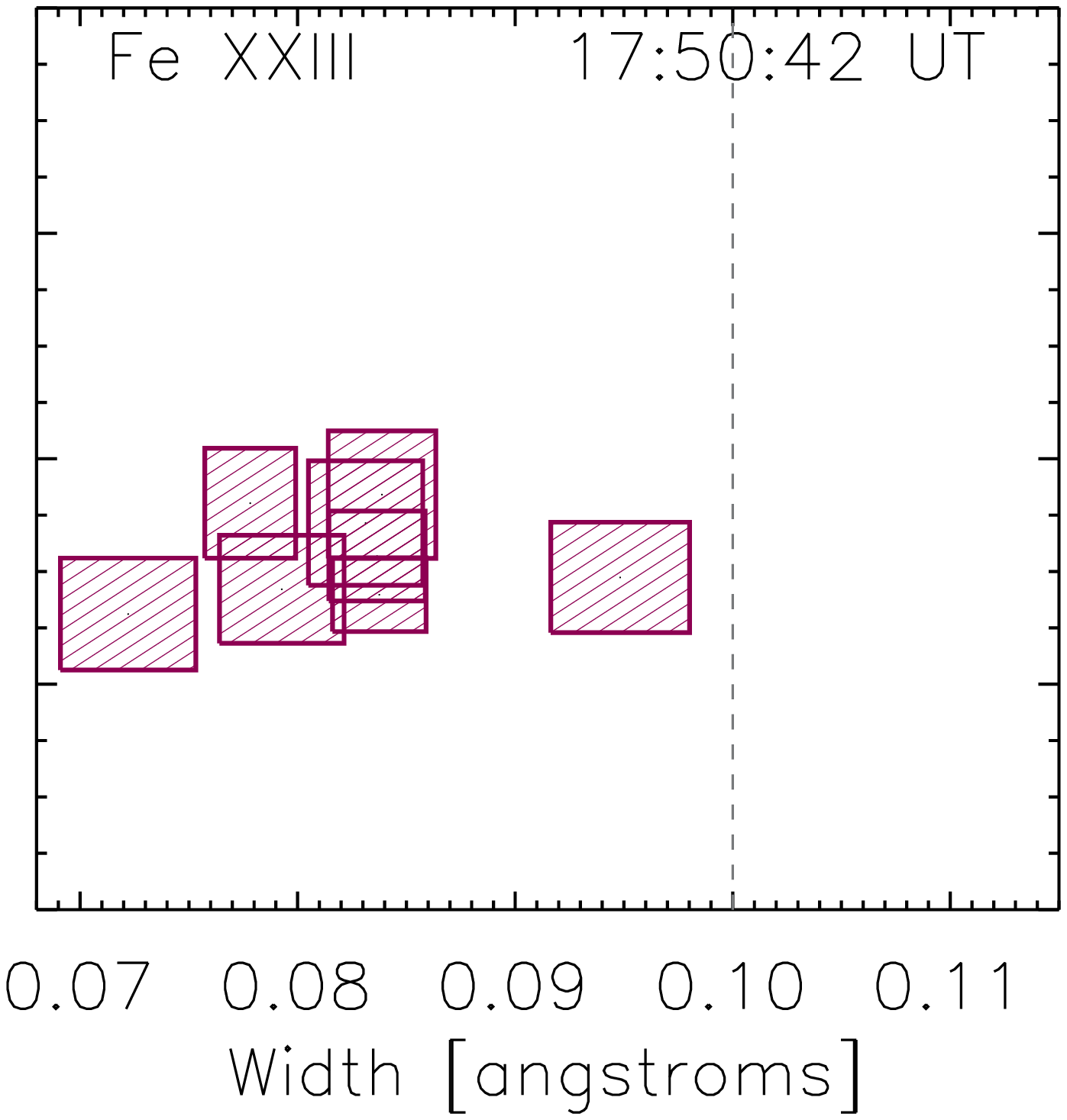}\hspace{-23pt}
	\includegraphics[width=0.21\linewidth]{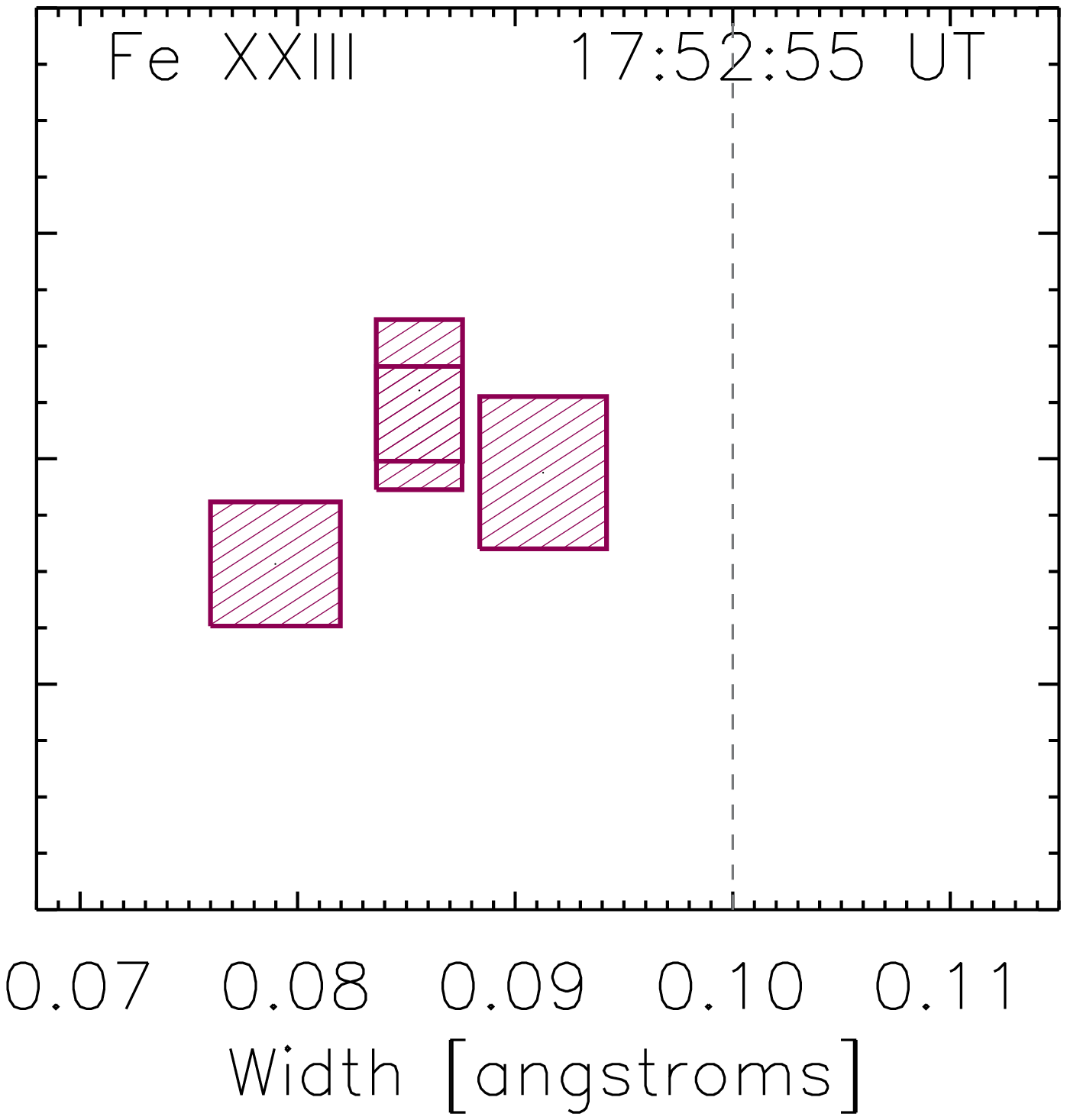}\hspace{-23pt}
	\includegraphics[width=0.21\linewidth]{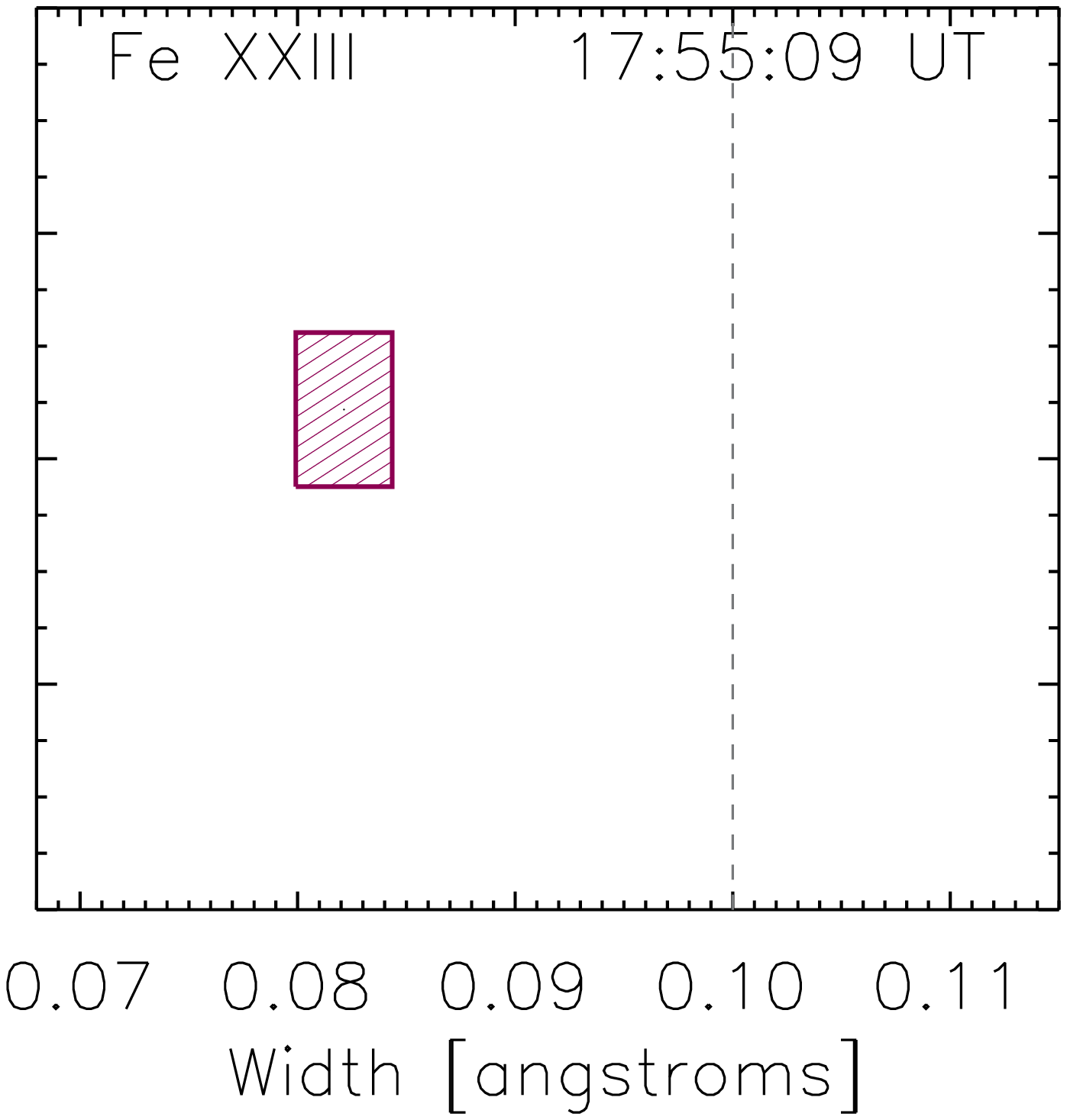}
	\caption{The KG1 fit values of $\kappa$ versus $W=2\sqrt{2\ln{2}}\times\sigma_{\kappa}$ at each time for \ion{Fe}{16} ({\it top row}) and \ion{Fe}{23} ({\it bottom row}). By plotting $\kappa$ versus $2\sqrt{2\ln{2}}\times\sigma_{\kappa}$ we look for trends that might indicate that non-Gaussian line profiles are due to the instrumental response instead of a physical process. This figure is compared with Figure \ref{app3} in Appendix \ref{app}. The multiple values of $\kappa$ for a given $W$ (and vice-versa) is one observation that supports a physical cause.}
	\label{fig6}
	\end{figure*}

\subsection{Further evidence against an instrumental origin for the non-Gaussian property}\label{inst_comp}
Before analysing the KG1 fits in detail, we present evidence that the non-Gaussian component of the line profiles are more consistent with a physical rather than an instrumental cause. Further details are provided in Appendix \ref{app}. In Figure \ref{fig6}, we plot the KG1 $\kappa$ values versus the characteristic widths $W=2\sqrt{2\ln(2)}\sigma_{\kappa}$ to observe if there is a trend between changes in $\kappa$ index and $W$, for both \ion{Fe}{16} and \ion{Fe}{23} lines. Only lines that satisfy all five criteria listed in Section \ref{method} are shown in Figure \ref{fig6}. We look for common trends that might suggest that the kappa line profiles are due to an instrumental process instead of a physical one. Importantly, we compare the observed KG1 values in Figure \ref{fig6} with Figure \ref{app3} in Appendix \ref{app}. Figure \ref{app3} displays the results of two modelled lines closely representing \ion{Fe}{16} and \ion{Fe}{23}. Each modelled line is chosen to have an instrumental response either represented by: (1.) ${\rm sinc}^{2}{\lambda}$ function (as discussed in Appendix \ref{app}) or by (2.) a kappa function with the chosen parameters $\kappa_{I}=3$ and $\sigma_{I}=0.0395$~\AA~(the same as fitting function KG2). Then, each modelled instrumental response is convolved with a Gaussian line representative of a physical line profile and the line width of this Gaussian is varied between sensible values for both \ion{Fe}{16} and \ion{Fe}{23} (see Appendix \ref{app}). Each resulting modelled line is fitted with the KG1 fitting function, and the KG1 fitted values of $\kappa$ index versus $W$ values are then plotted in Figure \ref{app3}. Figure \ref{app3} shows that as the (physical) Gaussian width of the modelled line increases, so do the resulting KG1 fit values of $\kappa$ index and $W$, for all modelled lines. The KG1 parameters found from actual fitting to the observed lines and originally shown in Figure \ref{fig6} are then re-plotted in Figure \ref{app3} for comparison with the model line results (lines only satisying criteria (1.) and (2.) in Section \ref{method} are also shown). Both Figure \ref{fig6} and Figure \ref{app3} show that the observed values show a range of different $W$ values for a given $\kappa$ value (and vice-versa), which is not suggested by the model line results. The results for \ion{Fe}{23} do not match the expected curves at all, while there is a much better match for \ion{Fe}{16}, although again we see different values of $W$ for a given $\kappa$ index. Therefore, this test is suggestive (but not conclusive) that the observed non-Gaussian line profiles are physical instead of instrumental and we interpret the KG1 fitting results as such in the next subsection\footnote{Although we have provided evidence of why the non-Gaussian line profiles are more likely to be physical, it is difficult to rule out an instrumental cause completely. Therefore, if we wish to perform more detailed flare spectroscopy studies and use line shape as a reliable diagnostic tool in the future, then the exact instrumental profile must be laboratory tested before launch.}.

\subsection{Further analysis of the KG1 lines}\label{FA_KG1}
\begin{figure*}[hbp!]
	\centering
	
	\hspace{-60pt}\includegraphics[width=0.30\textwidth,angle=0]{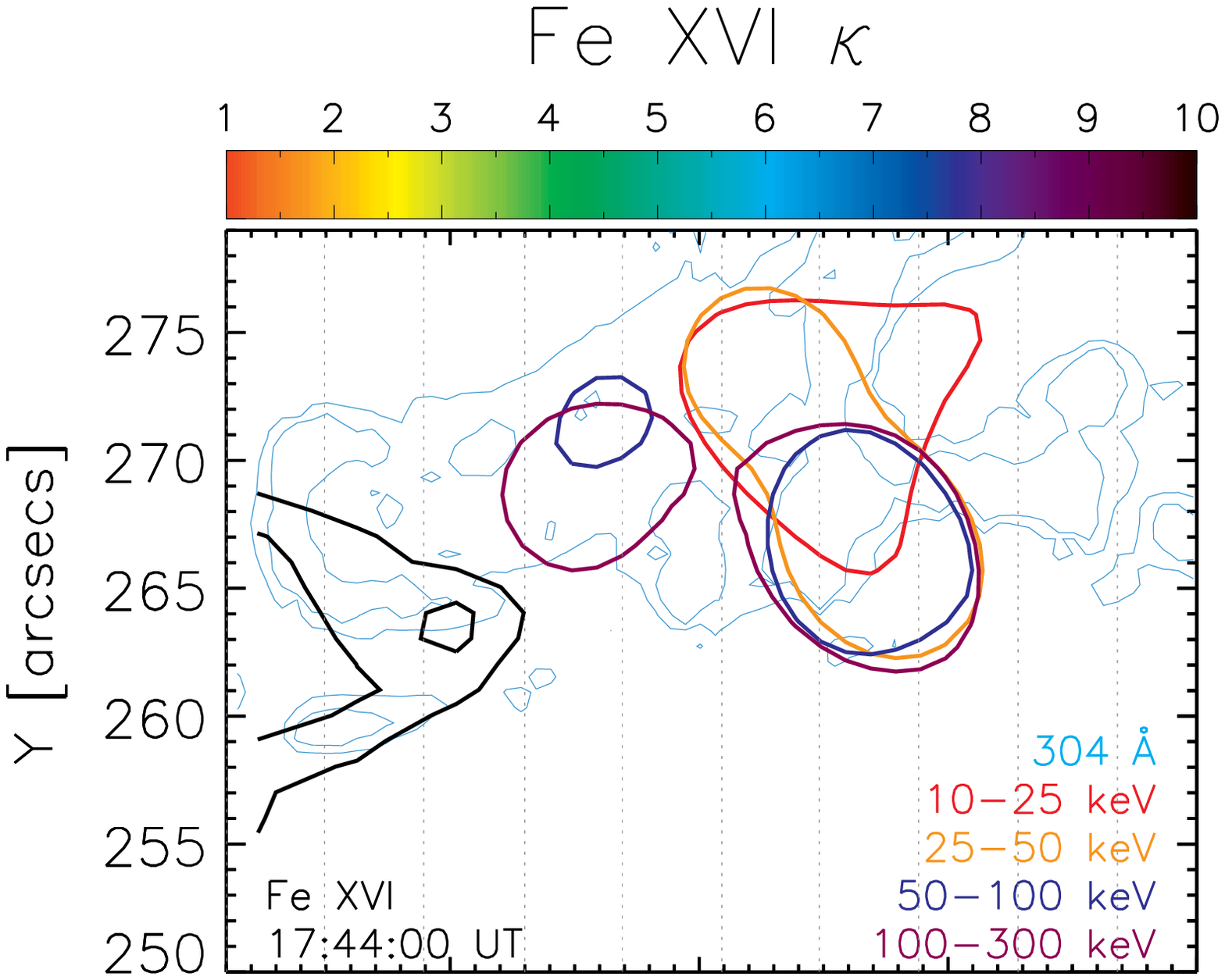}\hspace{-0.6cm}
	\includegraphics[width=0.30\textwidth,angle=0]{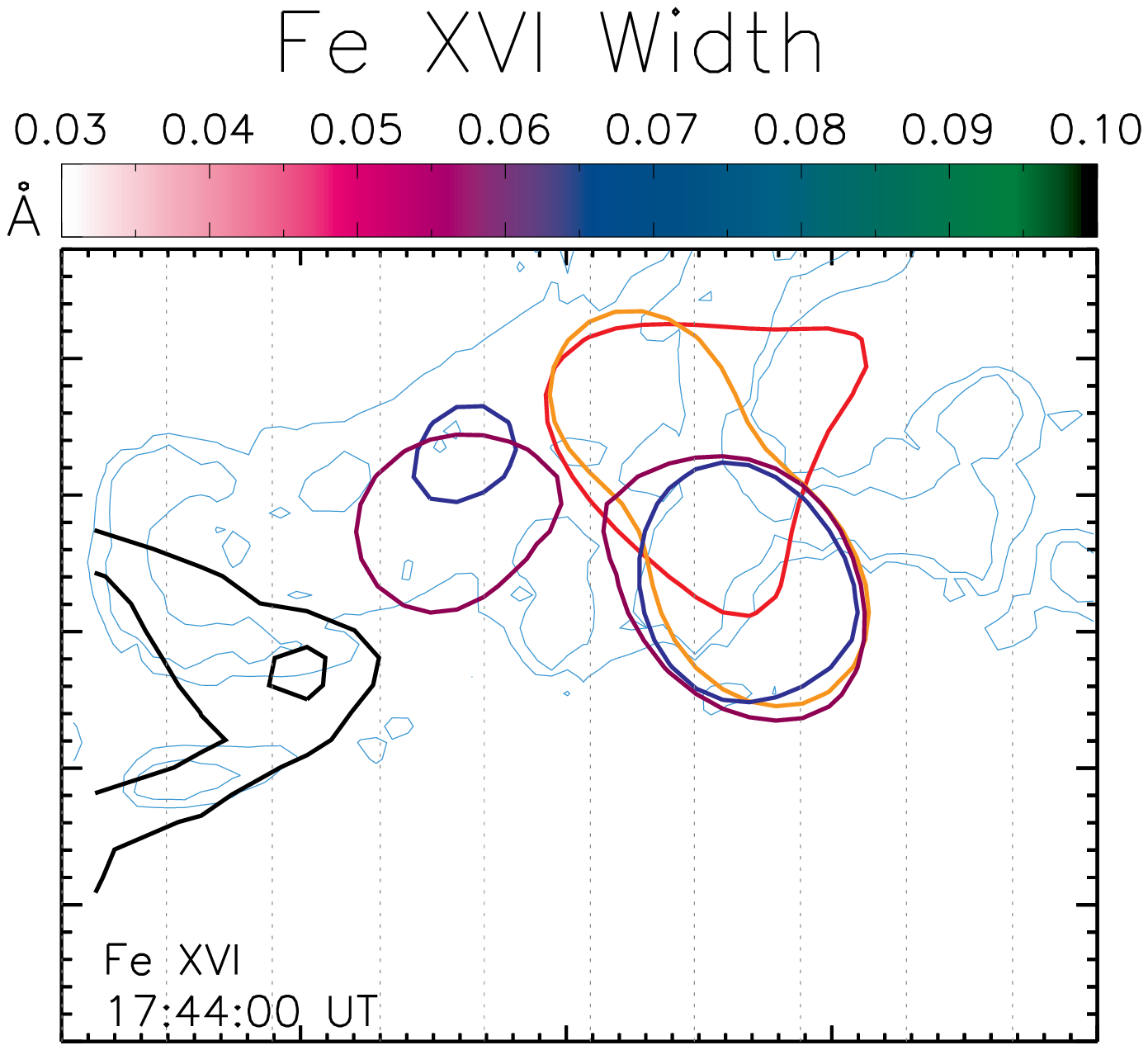}\hspace{-0.6cm}
	\includegraphics[width=0.30\textwidth,angle=0]{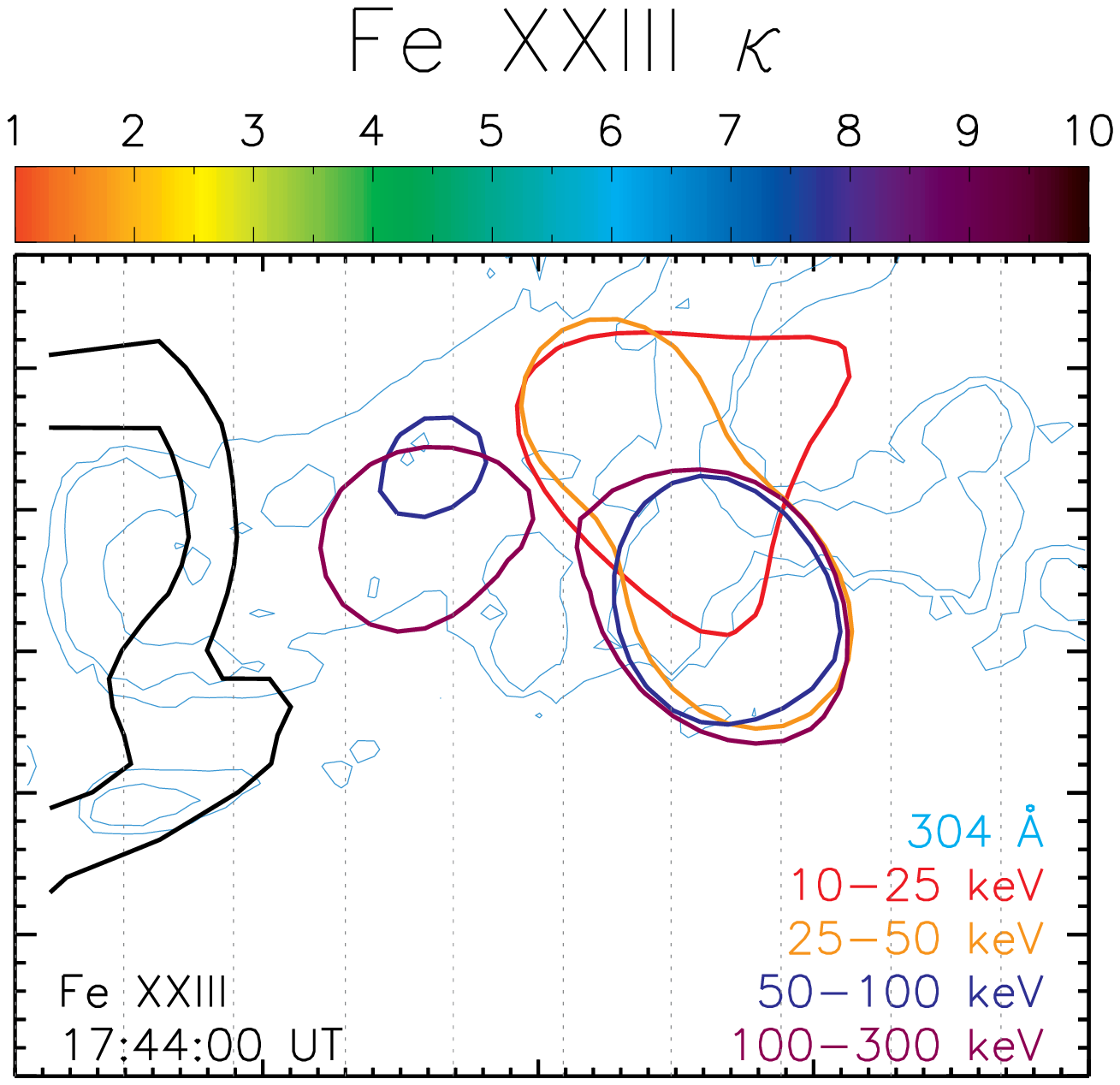}\hspace{-0.6cm}
	\includegraphics[width=0.30\textwidth,angle=0]{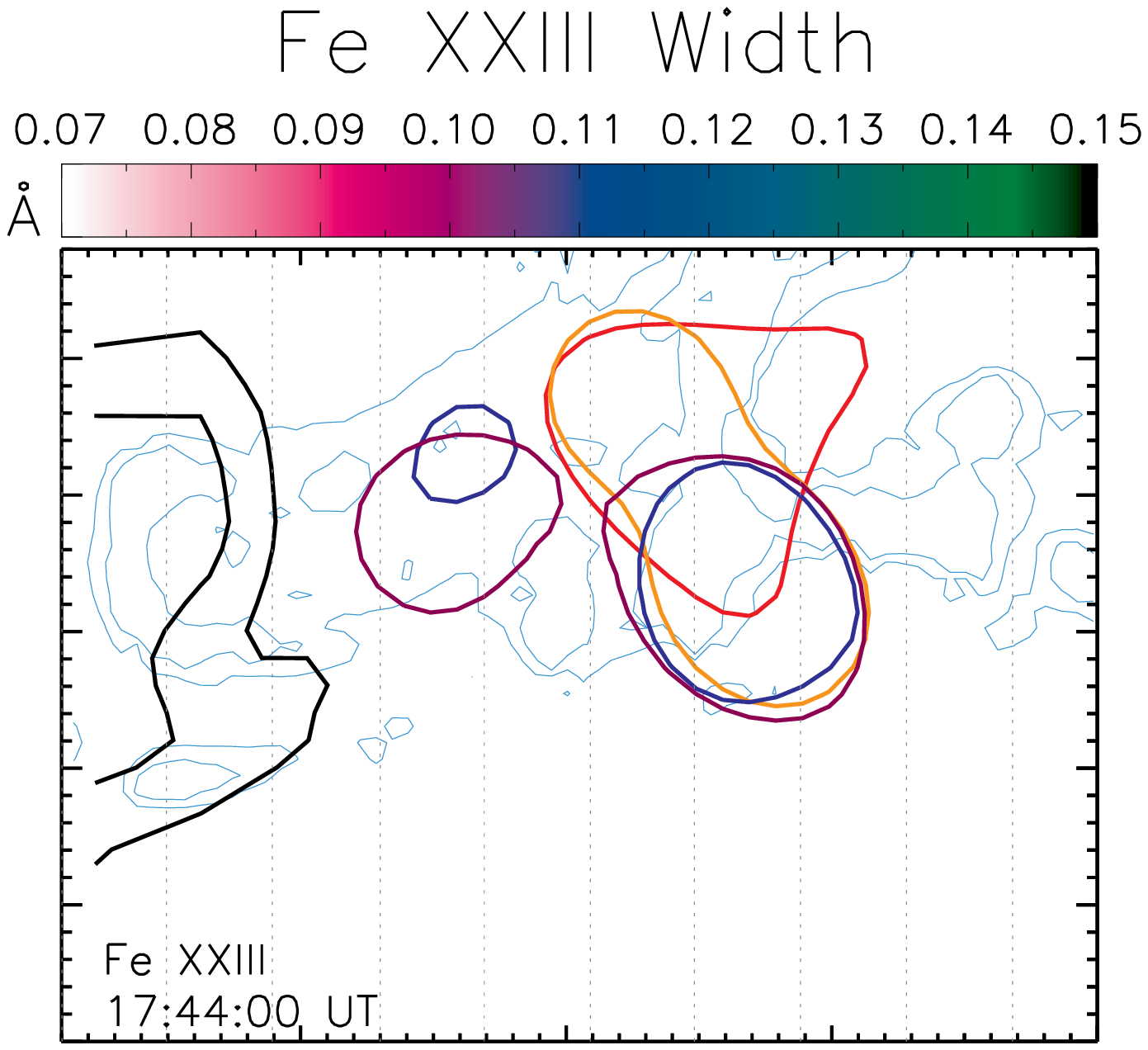}\hspace{-60pt}

\vspace{-41pt}

	\hspace{-60pt}\includegraphics[width=0.30\textwidth,angle=0]{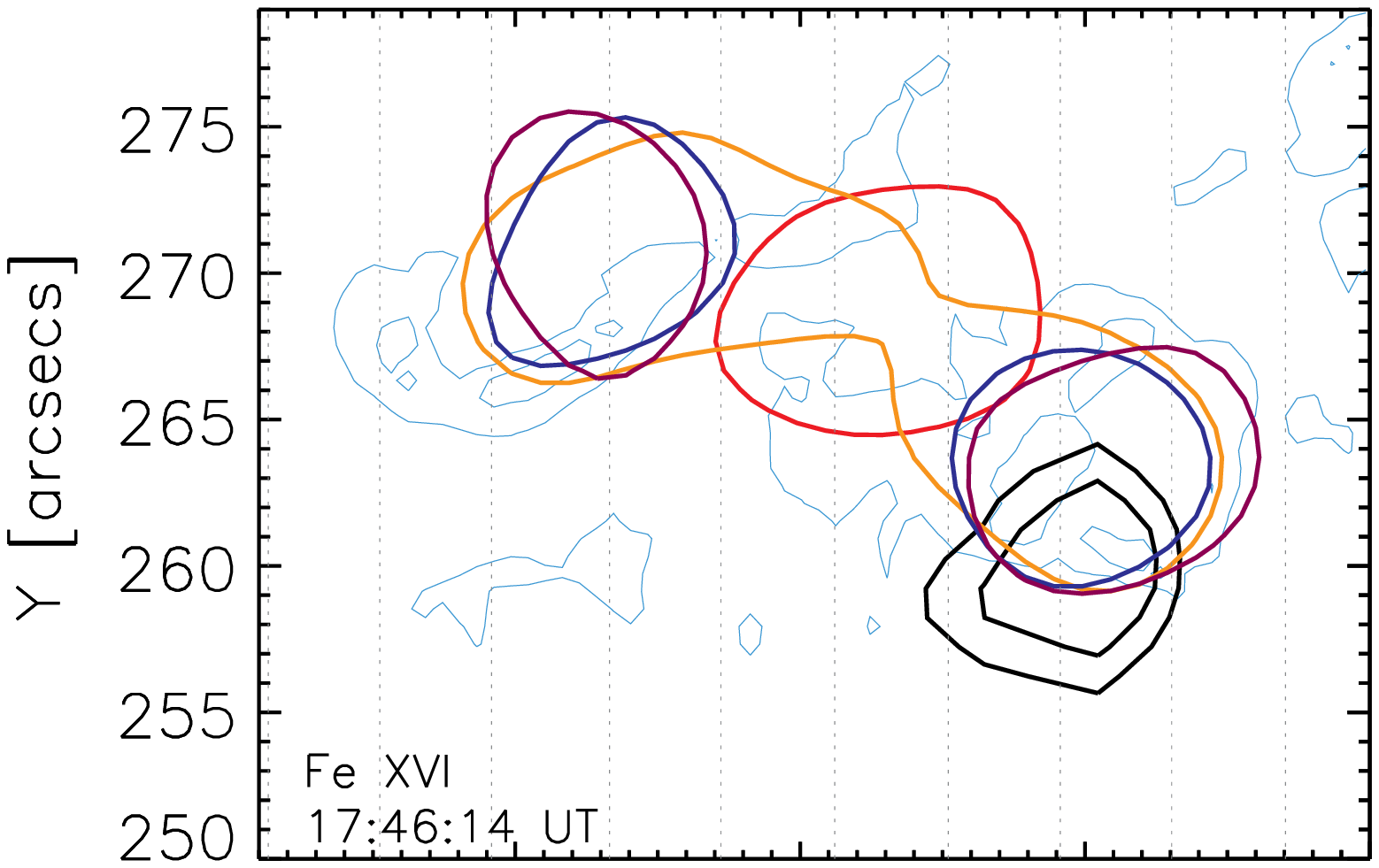}\hspace{-0.6cm}
	\includegraphics[width=0.30\textwidth,angle=0]{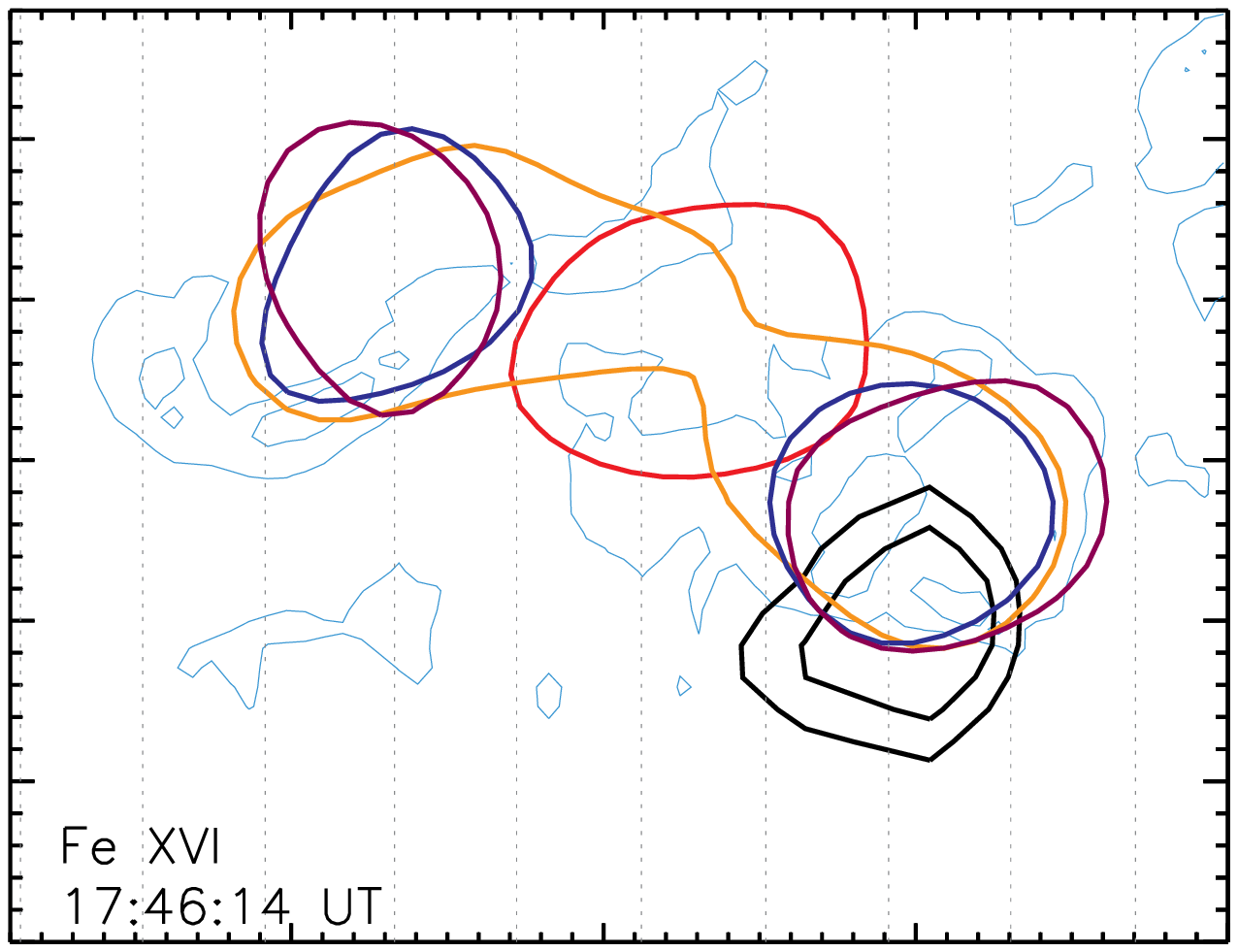}\hspace{-0.6cm}
	\includegraphics[width=0.30\textwidth,angle=0]{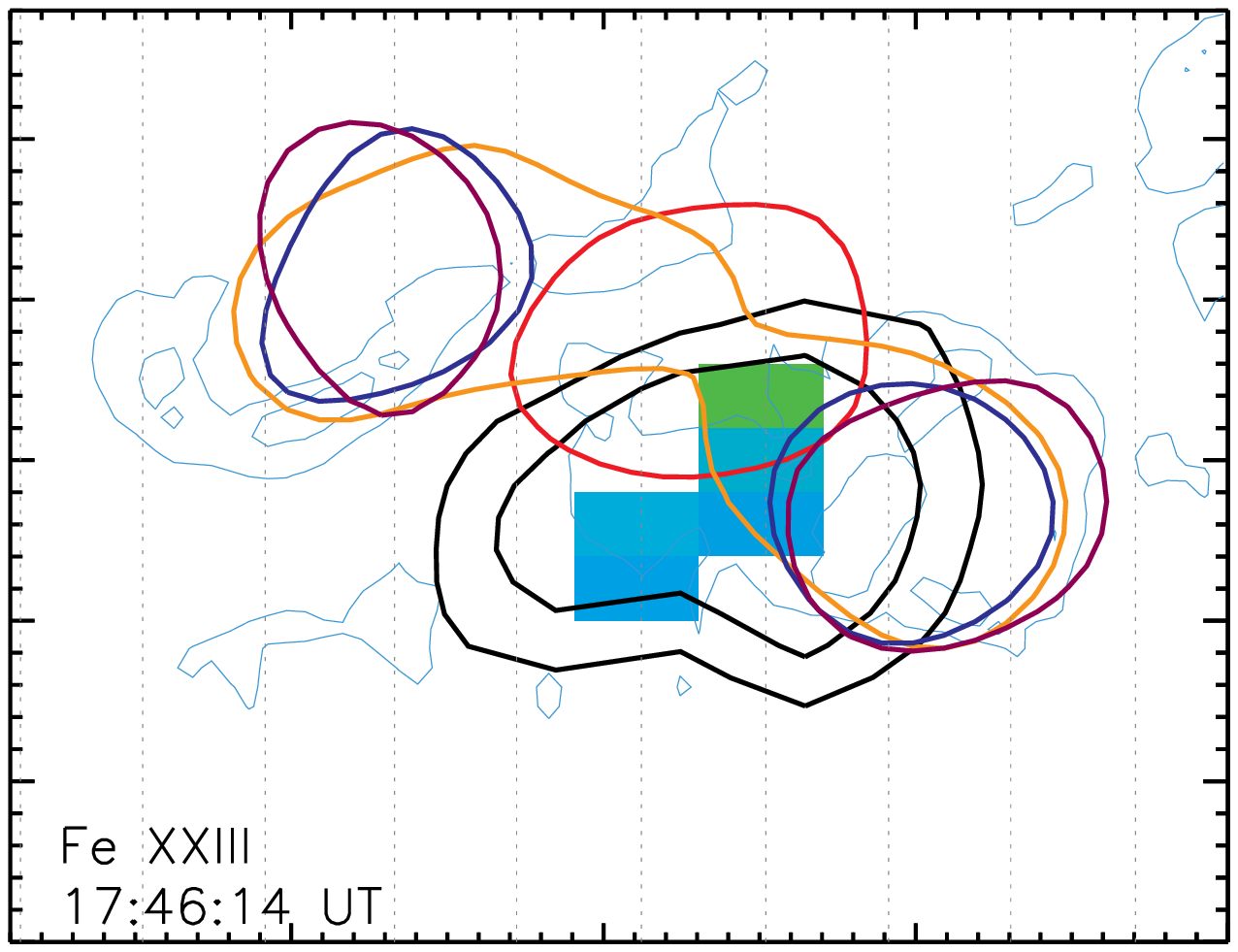}\hspace{-0.6cm}
	\includegraphics[width=0.30\textwidth,angle=0]{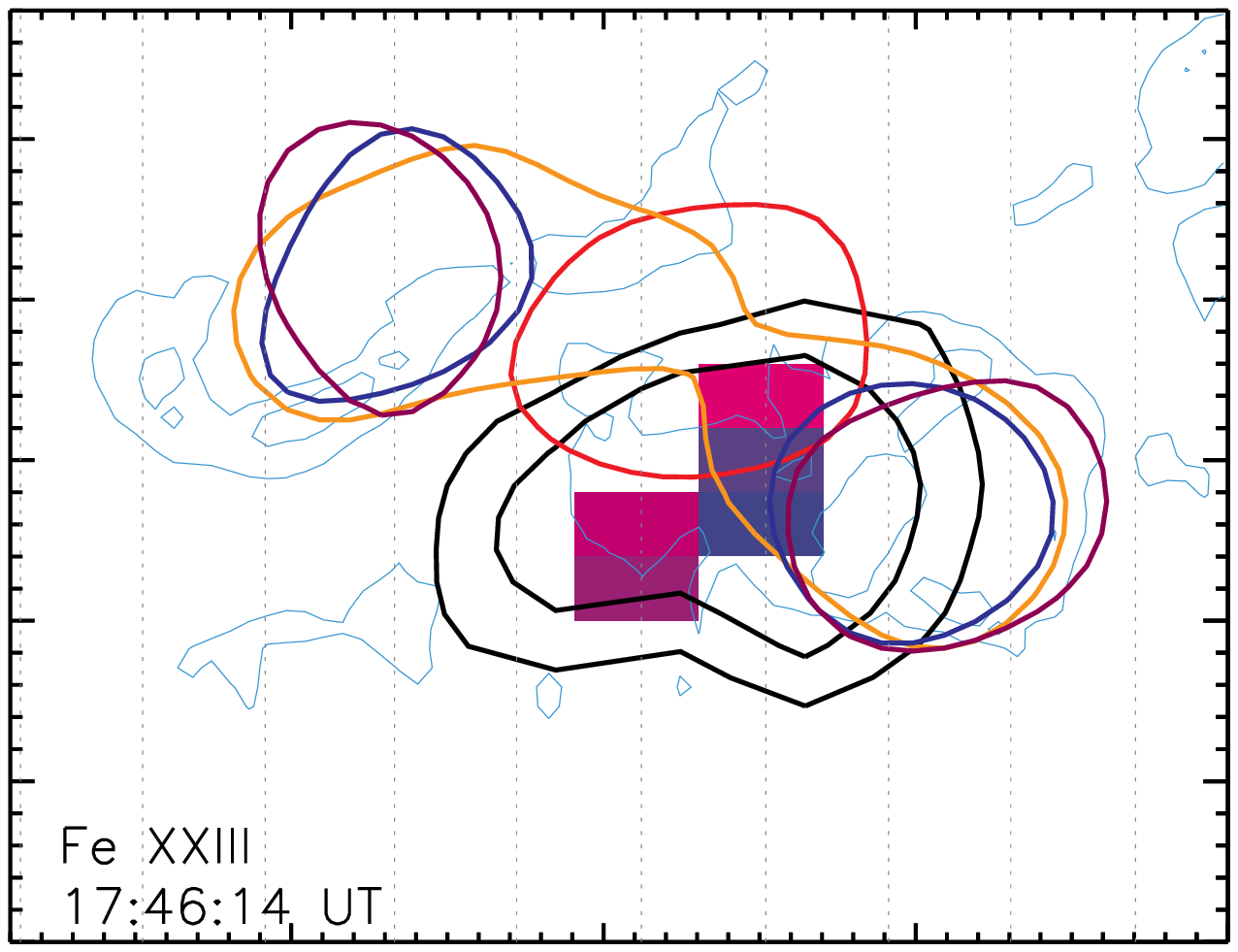}\hspace{-60pt}

\vspace{-41pt}

	\hspace{-60pt}\includegraphics[width=0.30\textwidth,angle=0]{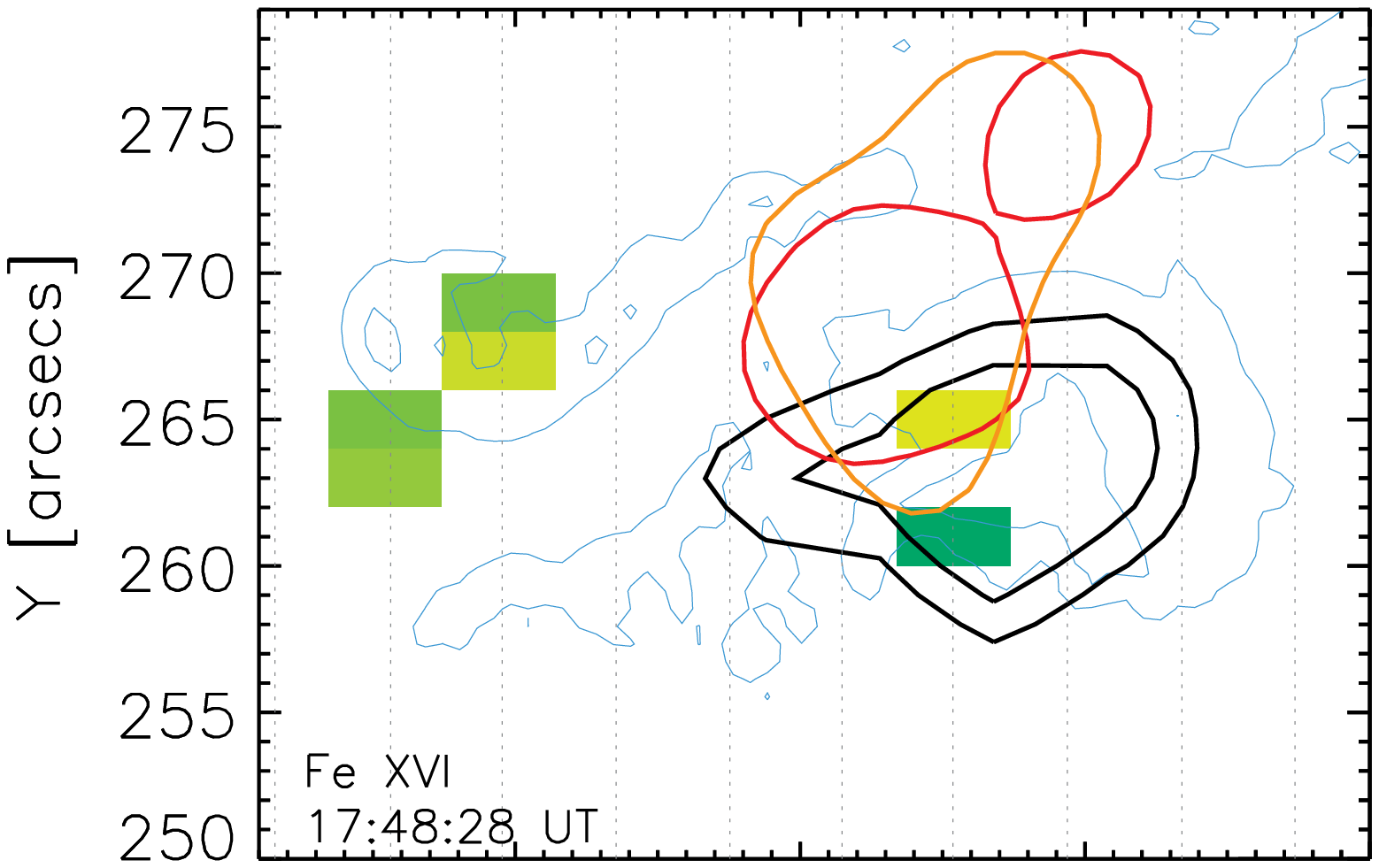}\hspace{-0.6cm}
	\includegraphics[width=0.30\textwidth,angle=0]{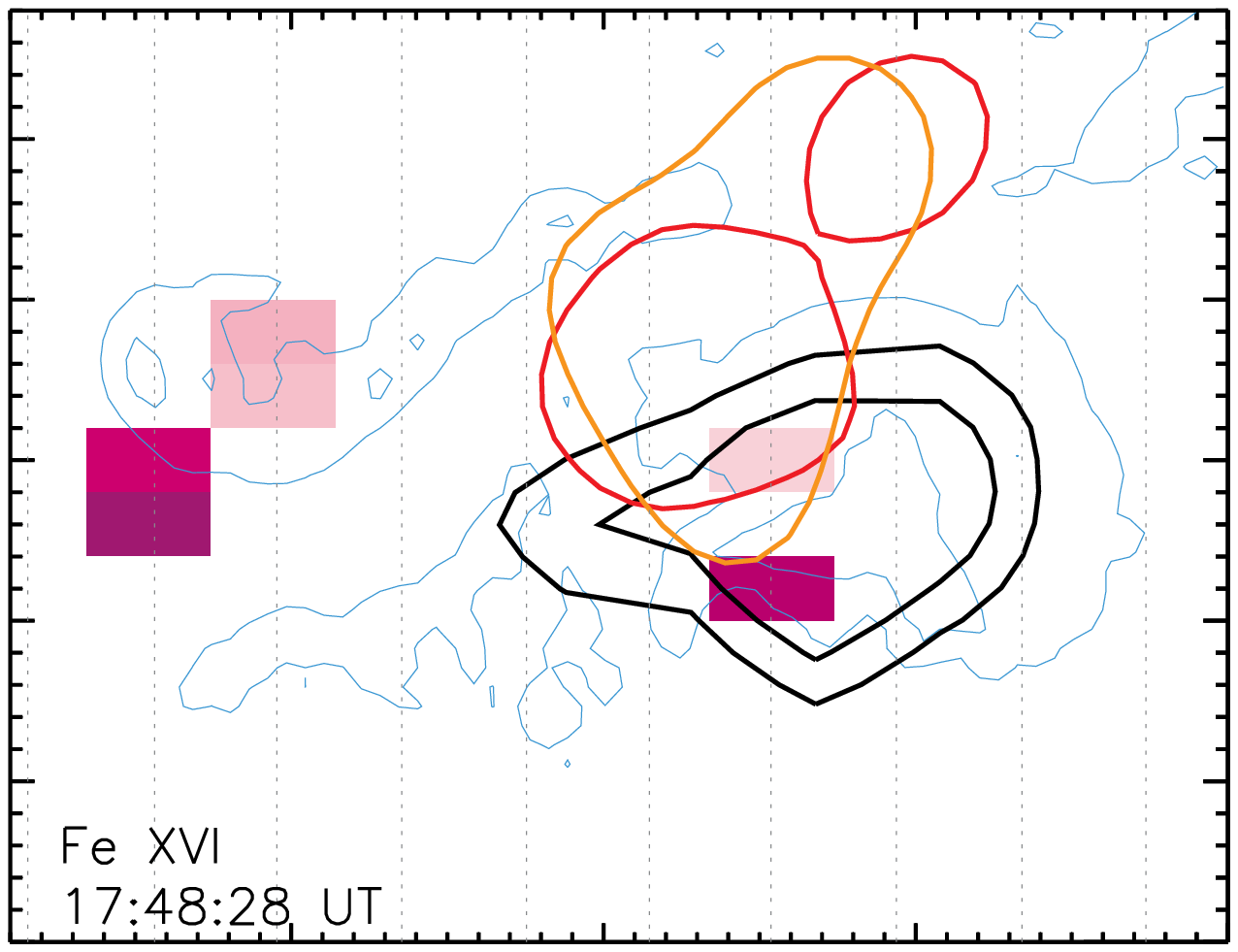}\hspace{-0.6cm}
	\includegraphics[width=0.30\textwidth,angle=0]{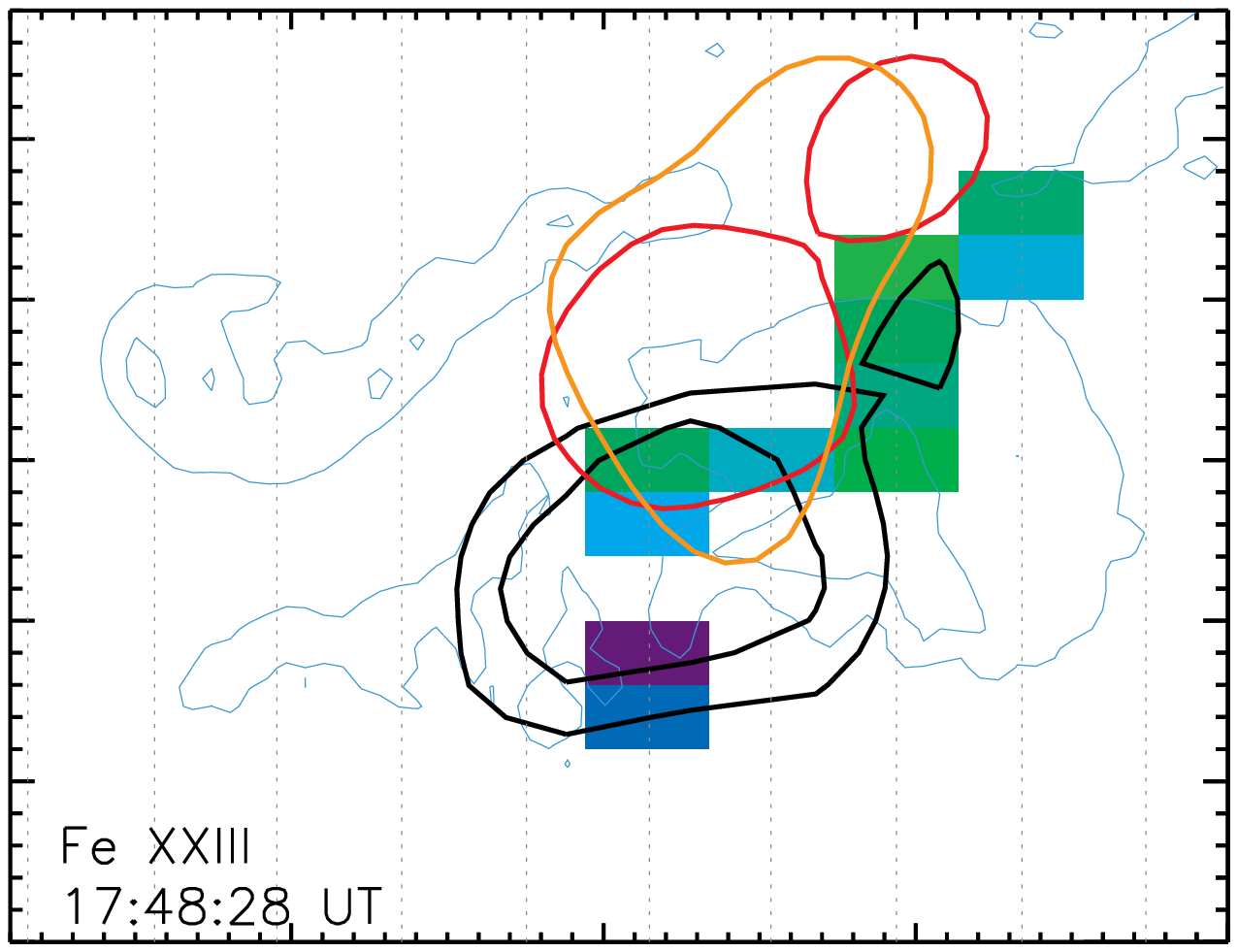}\hspace{-0.6cm}
	\includegraphics[width=0.30\textwidth,angle=0]{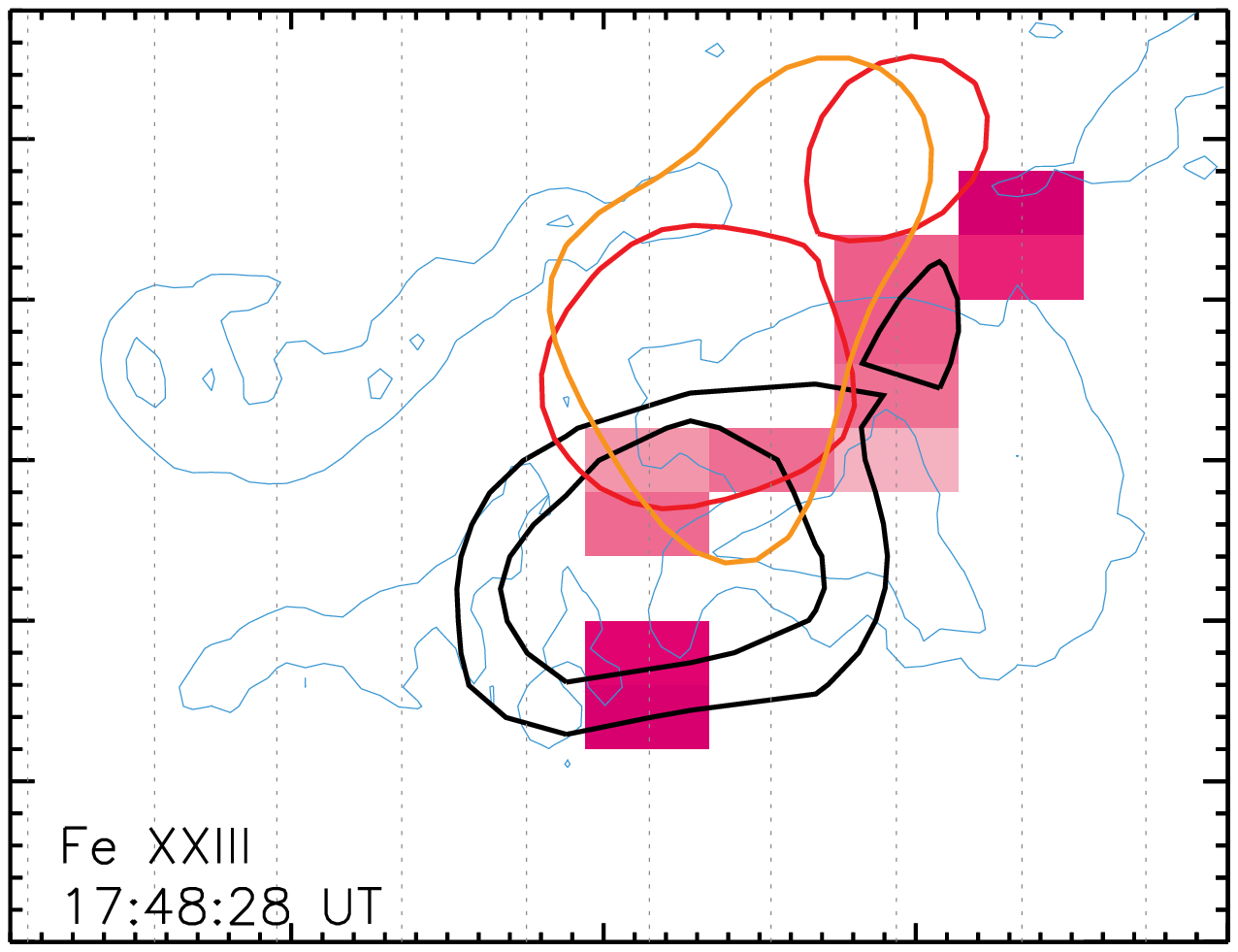}\hspace{-60pt}

\vspace{-41pt}

	\hspace{-60pt}\includegraphics[width=0.30\textwidth,angle=0]{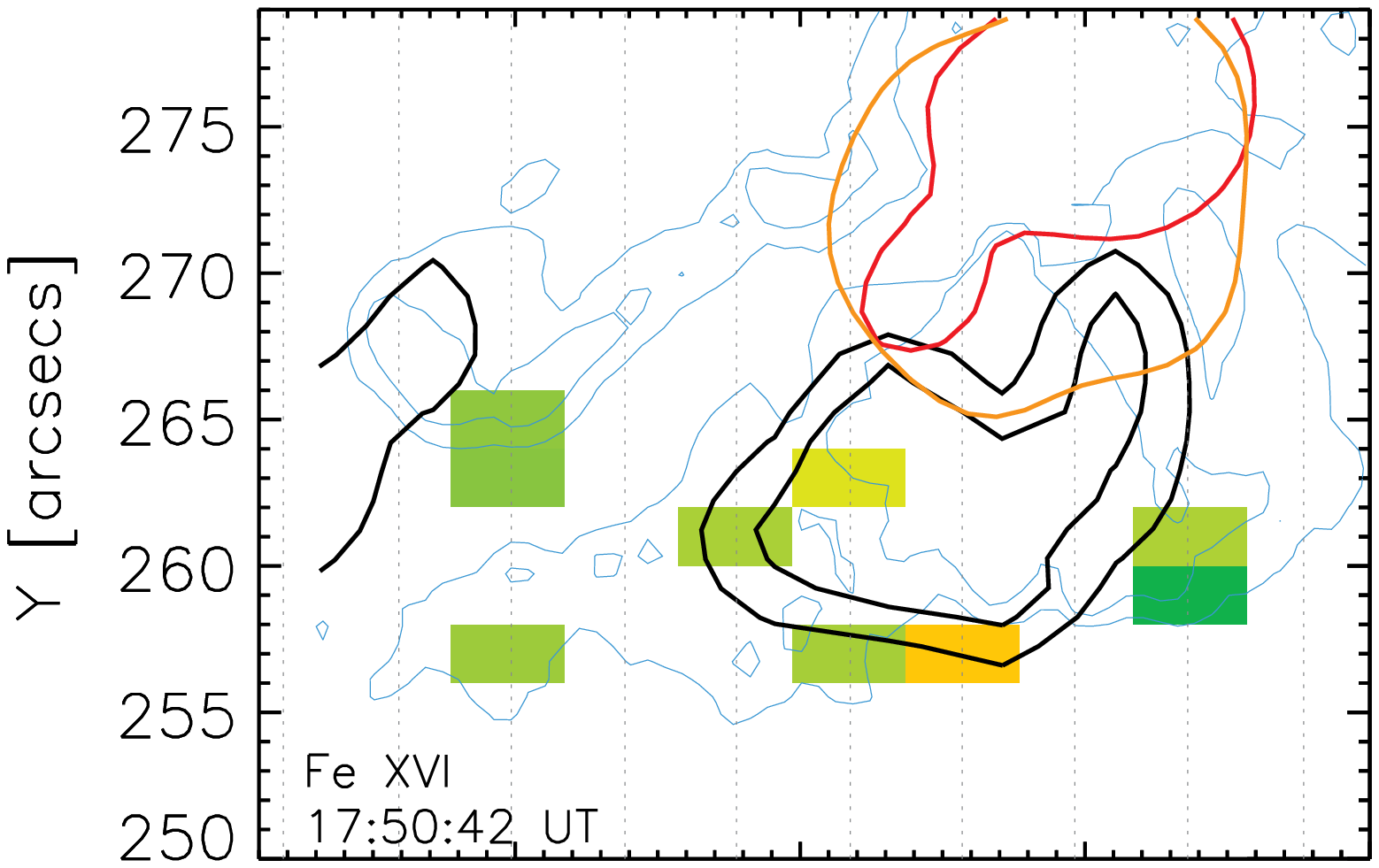}\hspace{-0.6cm}
	\includegraphics[width=0.30\textwidth,angle=0]{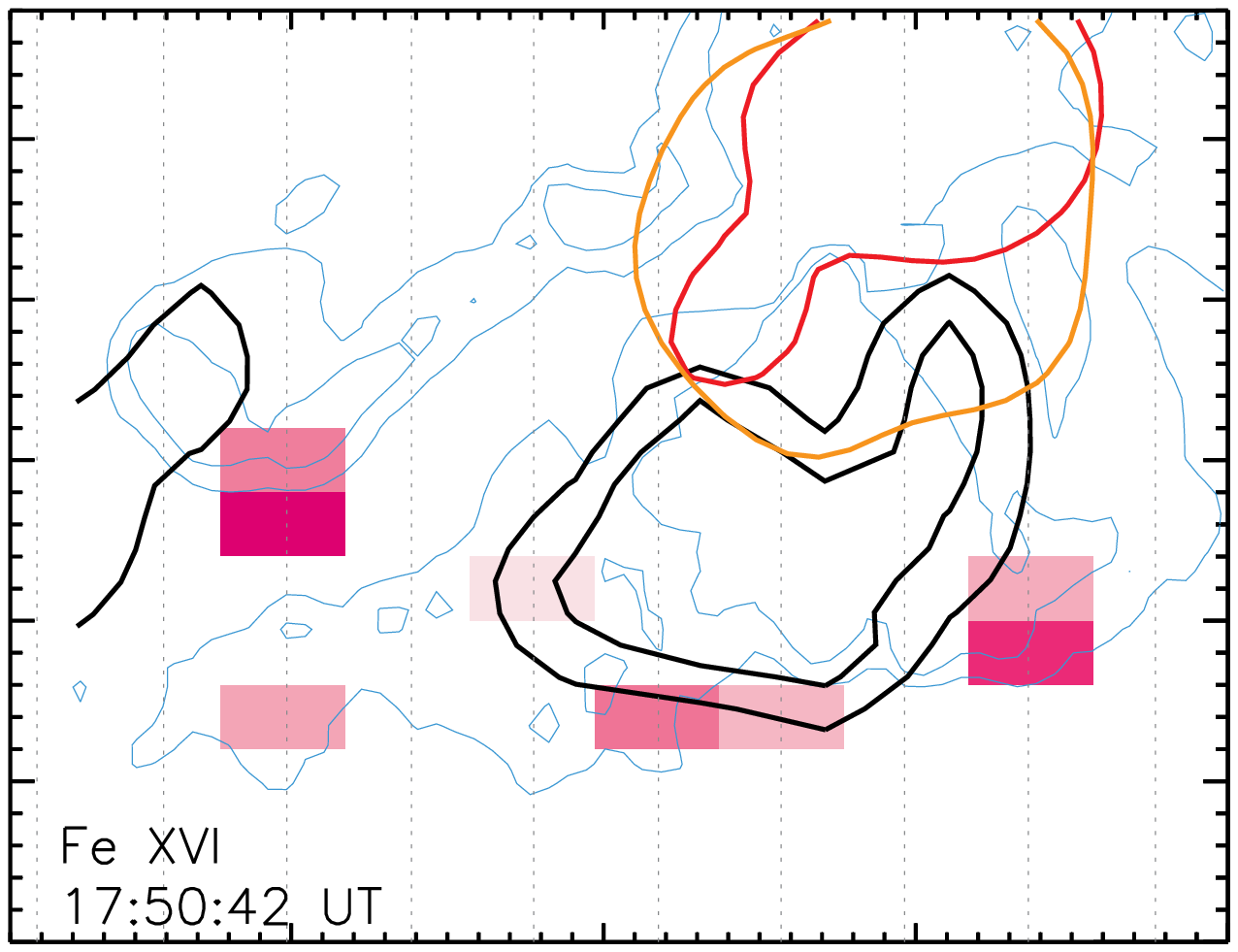}\hspace{-0.6cm}
	\includegraphics[width=0.30\textwidth,angle=0]{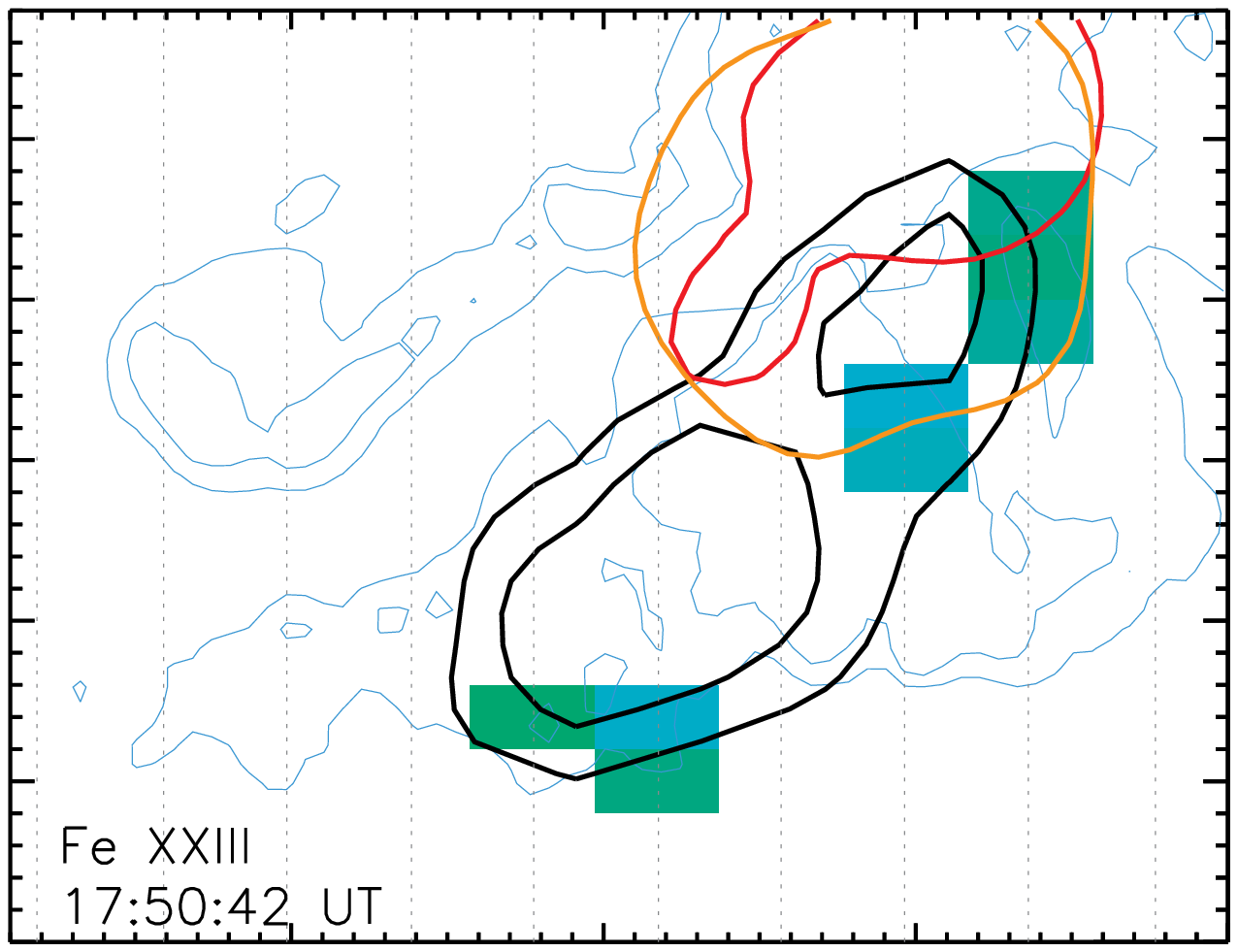}\hspace{-0.6cm}
	\includegraphics[width=0.30\textwidth,angle=0]{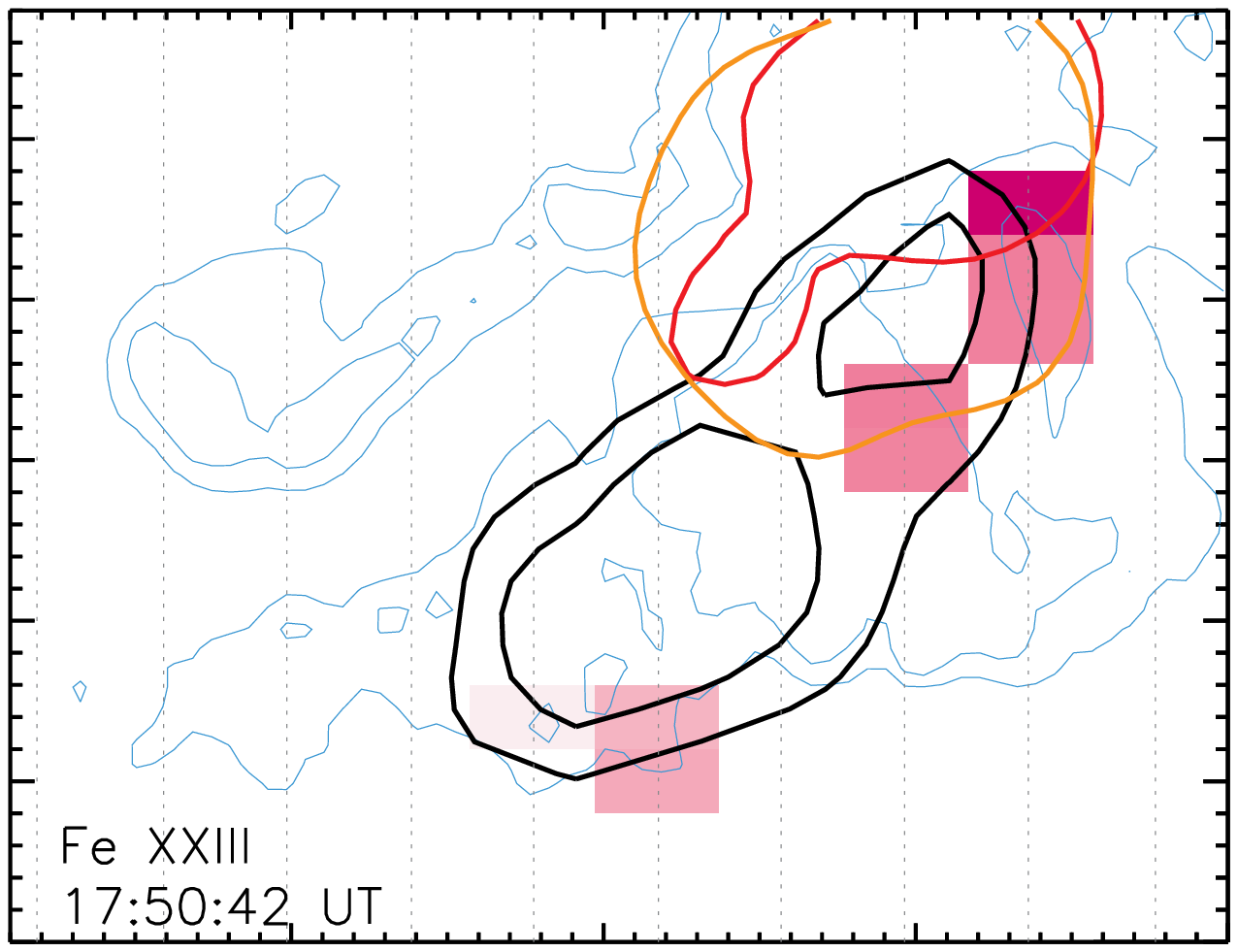}\hspace{-60pt}

\vspace{-41pt}

	\hspace{-60pt}\includegraphics[width=0.30\textwidth,angle=0]{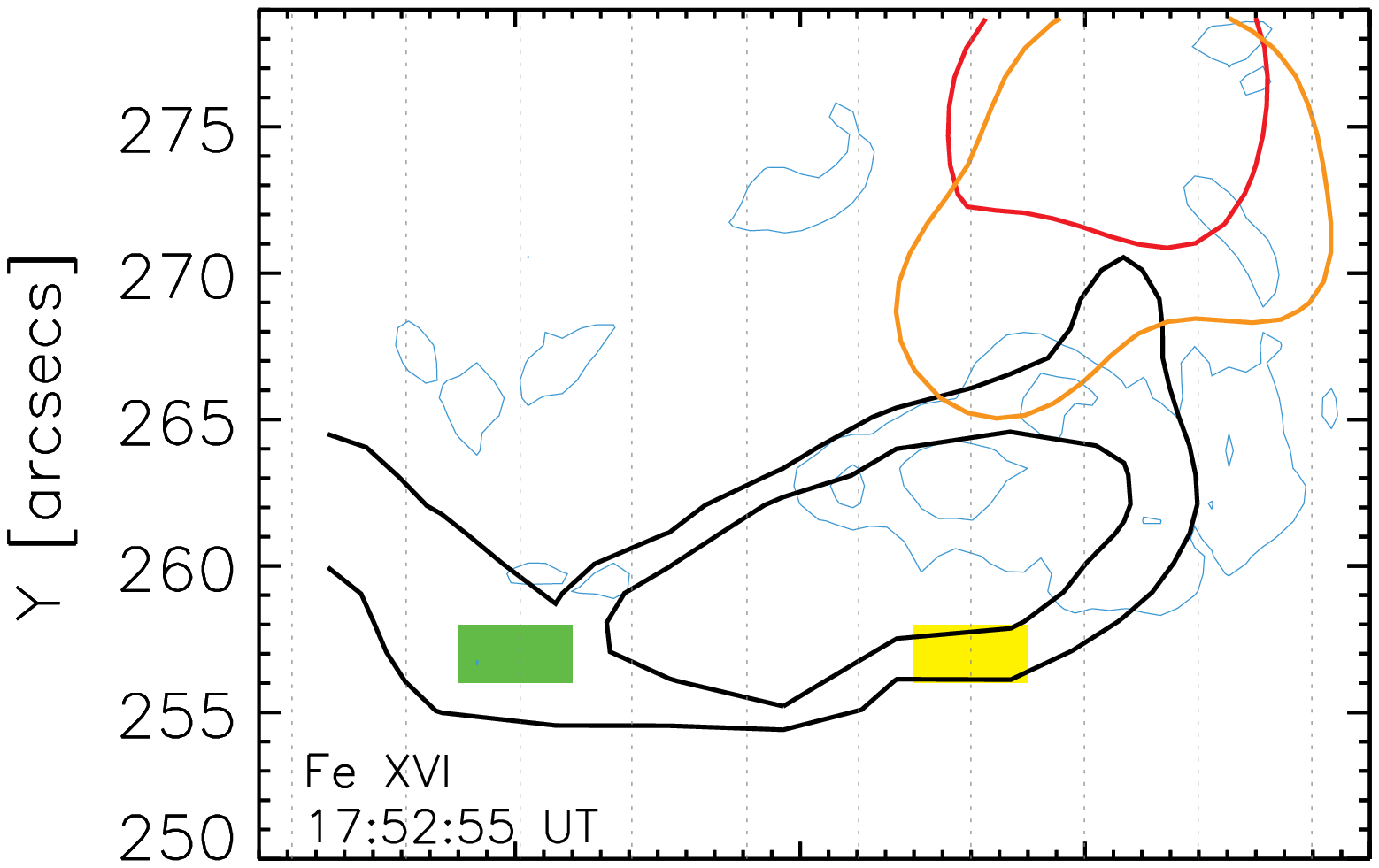}\hspace{-0.6cm}
	\includegraphics[width=0.30\textwidth,angle=0]{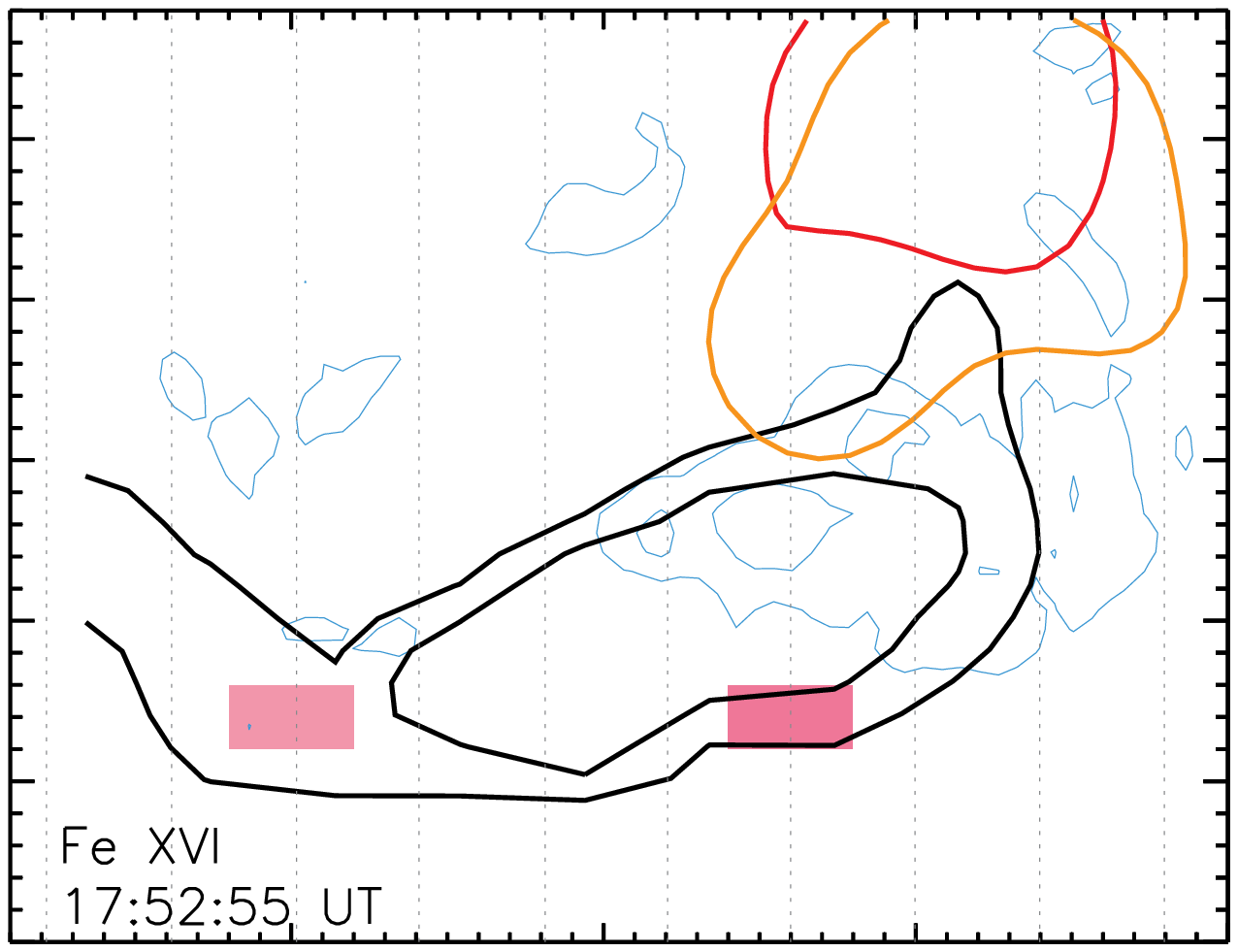}\hspace{-0.6cm}
	\includegraphics[width=0.30\textwidth,angle=0]{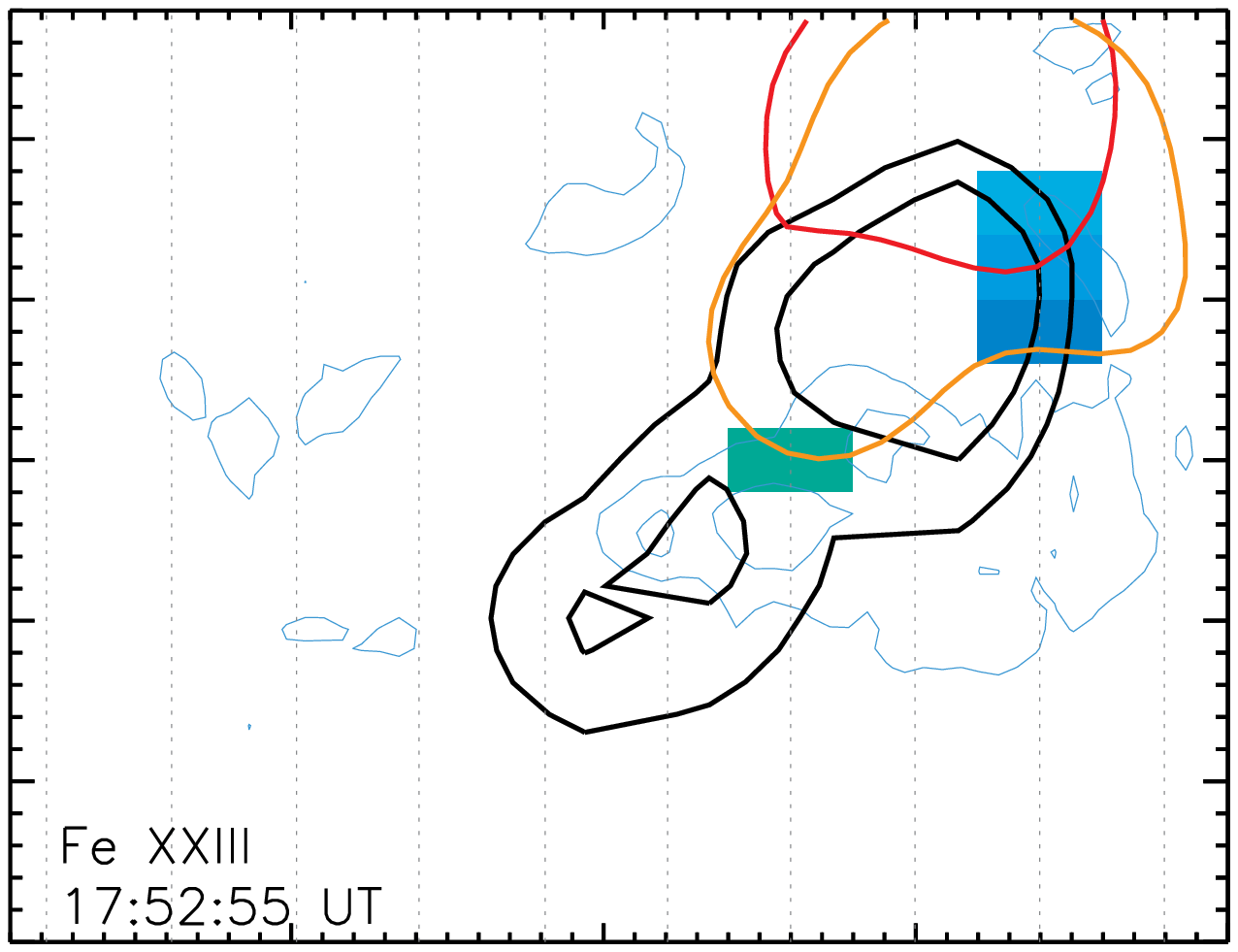}\hspace{-0.6cm}
	\includegraphics[width=0.30\textwidth,angle=0]{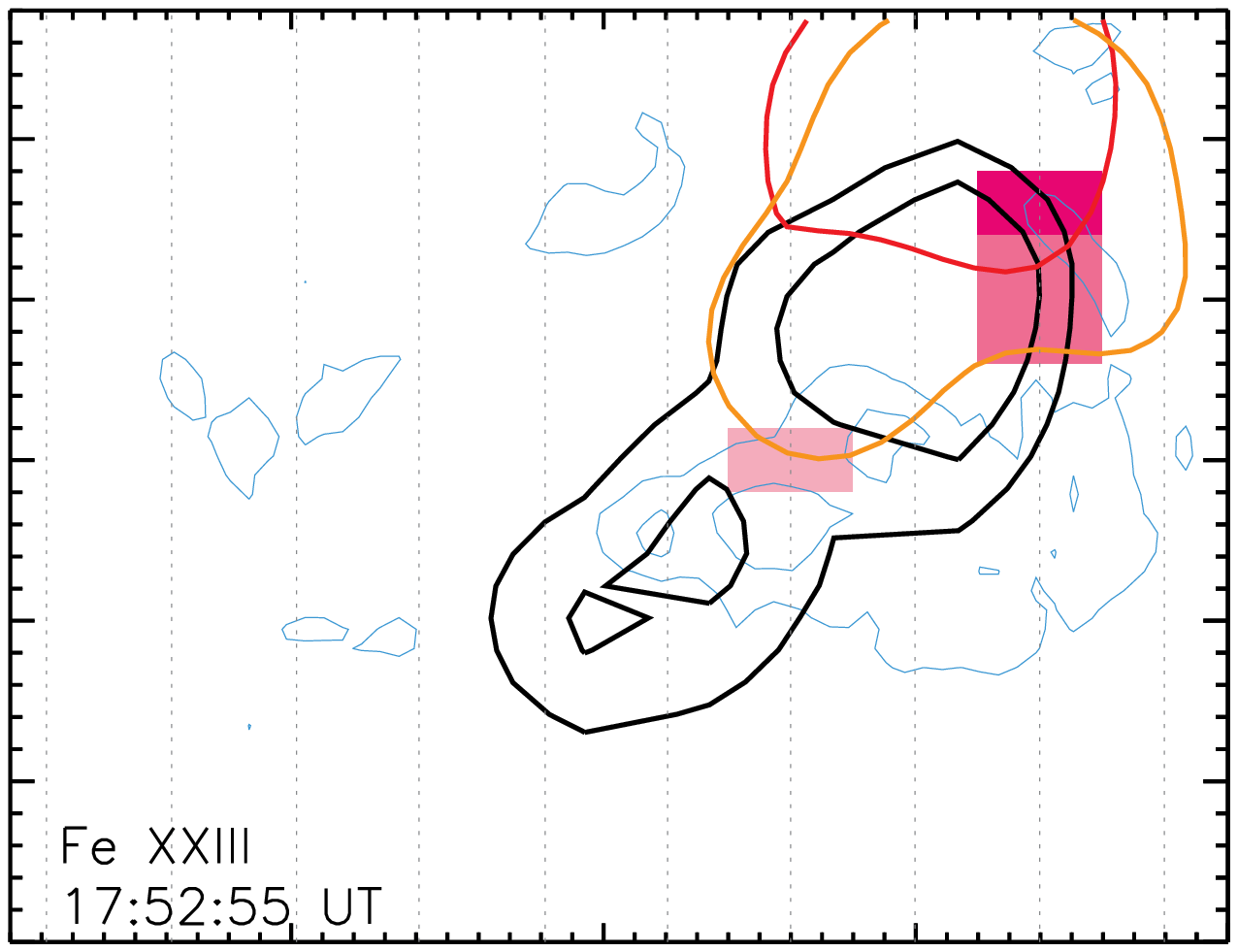}\hspace{-60pt}

\vspace{-41pt}

	\hspace{-60pt}\includegraphics[width=0.30\textwidth,angle=0]{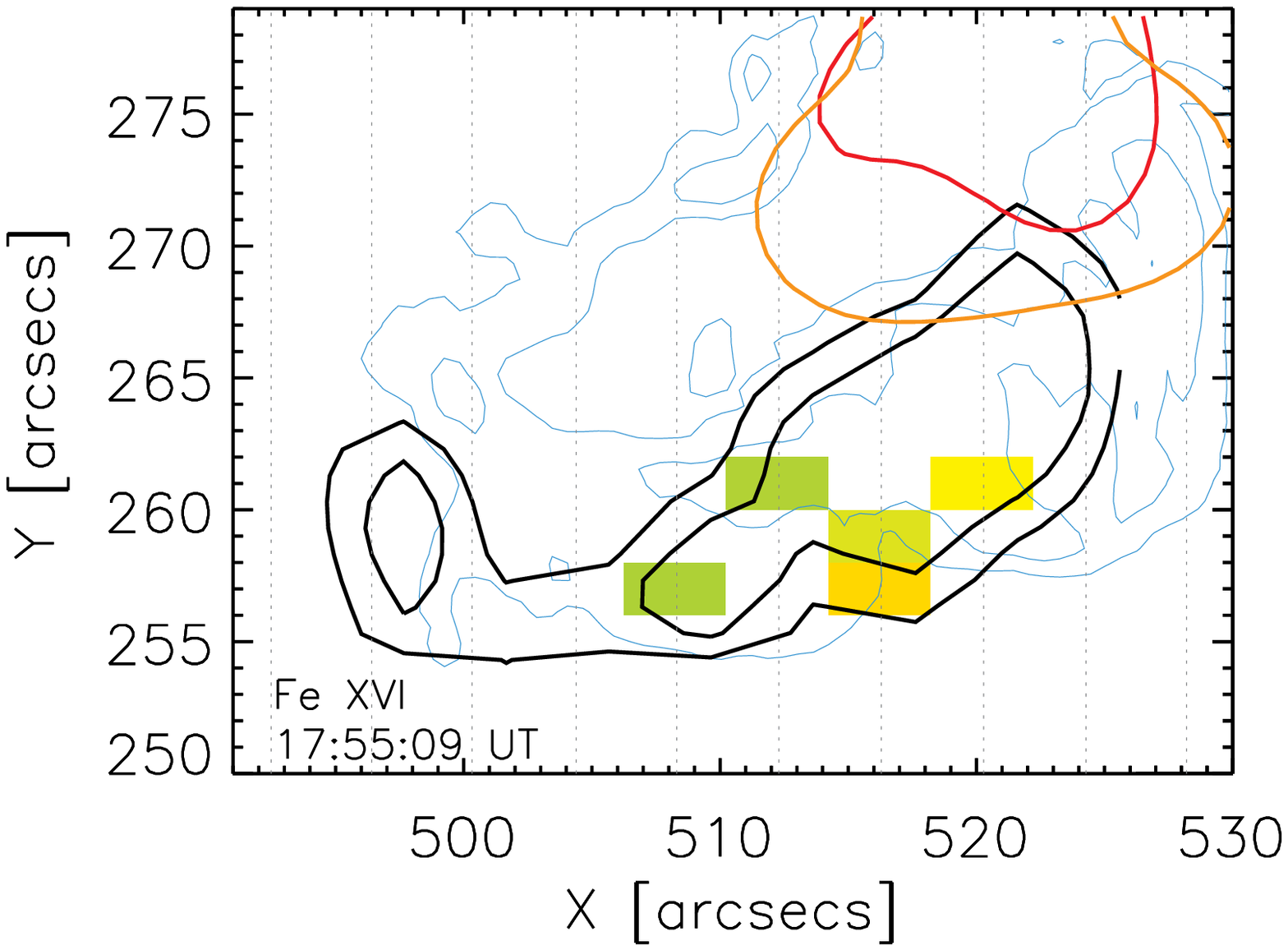}\hspace{-0.6cm}
	\includegraphics[width=0.30\textwidth,angle=0]{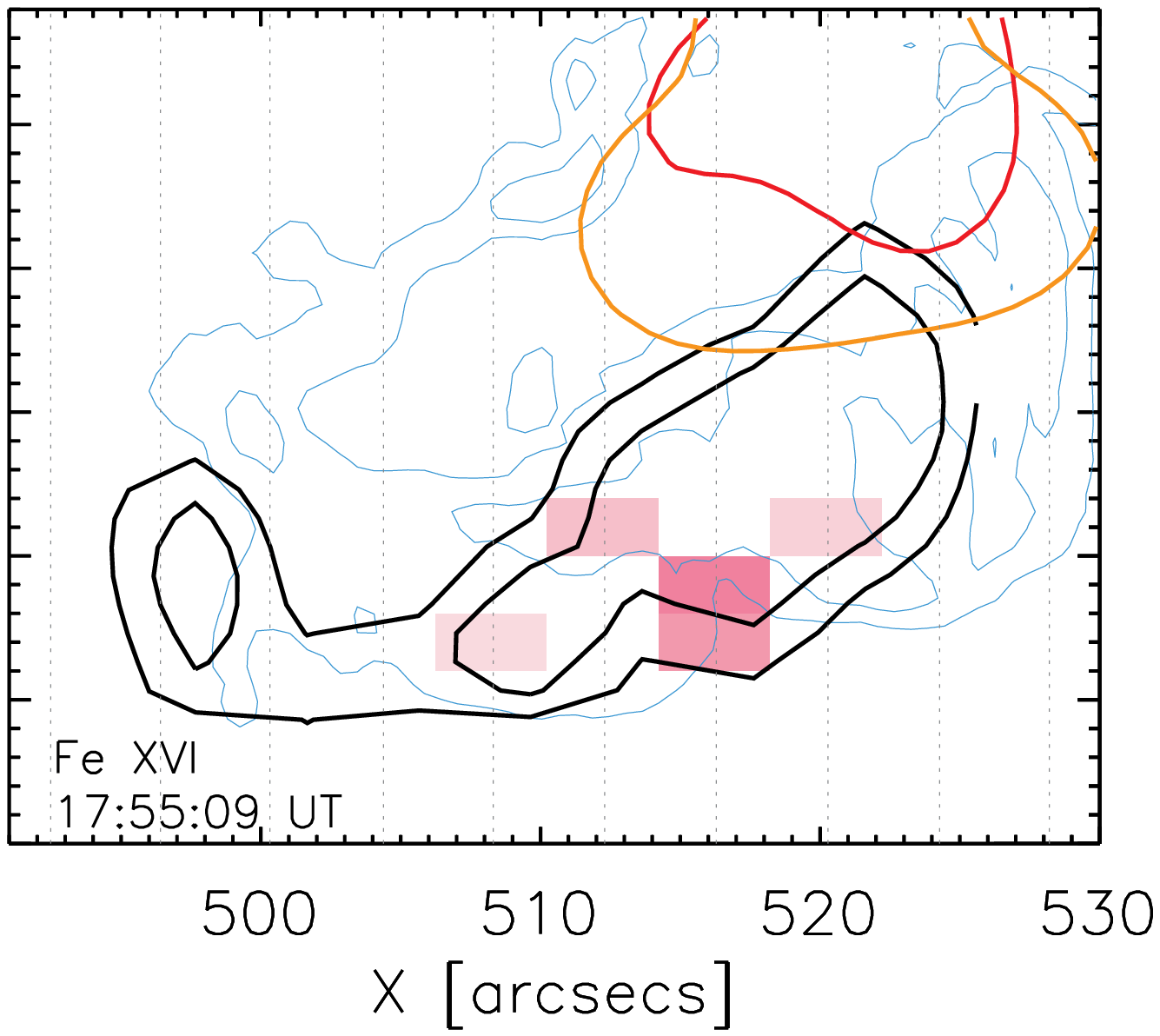}\hspace{-0.6cm}
	\includegraphics[width=0.30\textwidth,angle=0]{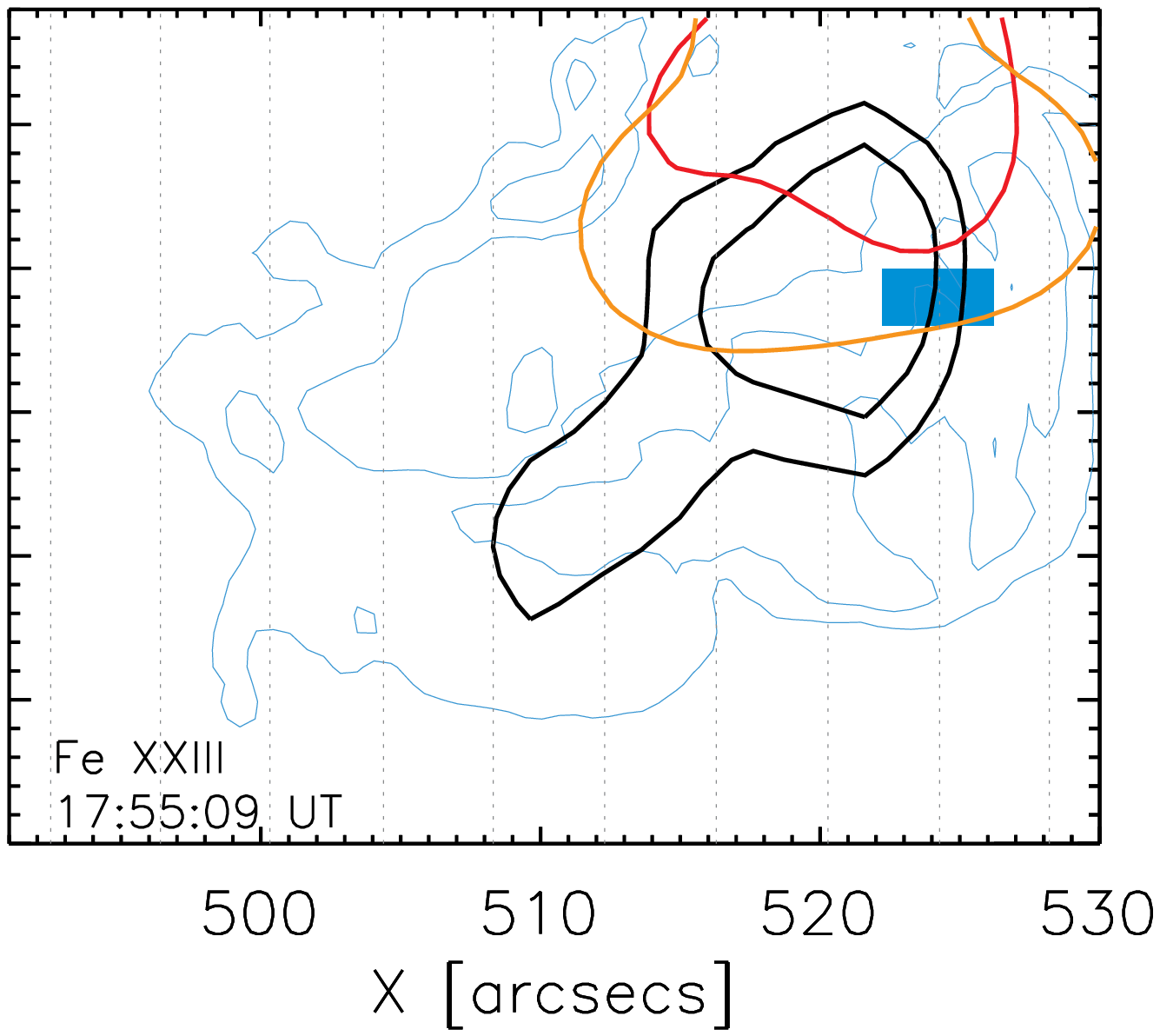}\hspace{-0.6cm}
	\includegraphics[width=0.30\textwidth,angle=0]{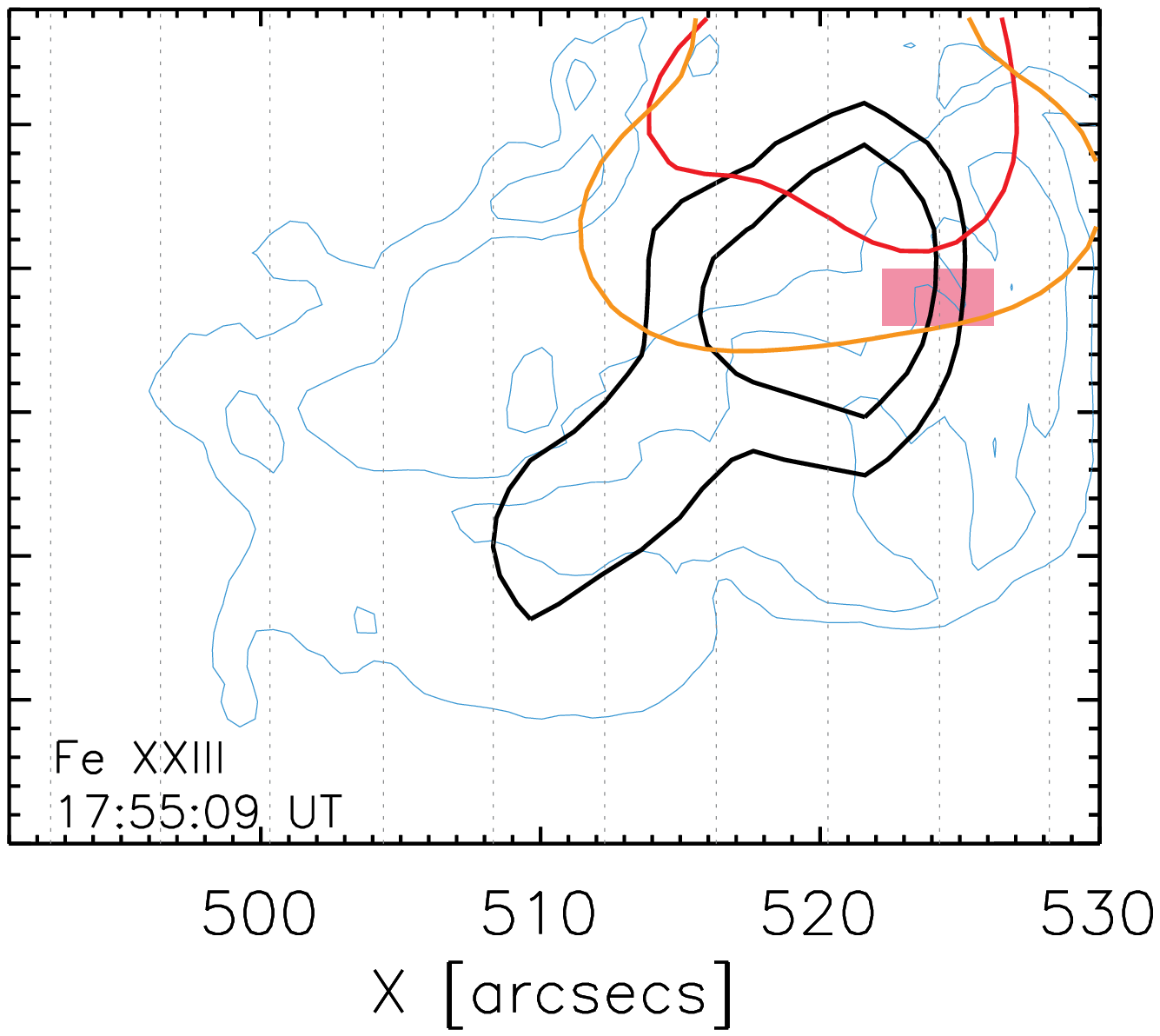}\hspace{-60pt}

	\vspace{0.5cm}
	\caption{Maps of SOL2014-03-29T17:44 during six different EIS raster times (time $t_{1} - t_{6}$ increases from top to bottom), showing the results of KG1 fits satisying all five criteria (Section \ref{method}). {\it Columns one and two:} \ion{Fe}{16} $\kappa$ (1) and $W=2\sqrt{2\ln{2}}\times\sigma_{\kappa}$ (2). {\it Columns three and four:} \ion{Fe}{23} $\kappa$ (3) and $W=2\sqrt{2\ln{2}}\times\sigma_{\kappa}$ (4). The values are also shown in Table \ref{table1}.}

	\label{fig7}
	\end{figure*}

In Figure \ref{fig7}, the flare is shown at the six different (EIS start) times $t_1$ = 17:44:00 UT, $t_2$ = 17:46:14 UT, $t_3$ = 17:48:28 UT, $t_4$ = 17:50:42 UT, $t_5$ =17:52:55 UT and $t_6$ = 17:55:09 UT. Each map shows AIA 304~\AA, {\em RHESSI\,} and EIS contours. Maps of the \ion{Fe}{16} and \ion{Fe}{23} KG1 fit parameters, $\kappa$ and $W=2\sqrt{2\ln{2}}\sigma_{\kappa}$, are displayed for regions that satisfy all the criteria listed in Section \ref{method} in Figure \ref{fig7}. 
At times $t_1$  and $t_2$, two HXR footpoints (energies $>$ 50 keV) and an X-ray coronal source are present. At times $t_3 - t_6$, the HXR footpoints disappear but the X-ray coronal source can still be observed. At time $t_1$ there is no \ion{Fe}{23} emission suitable for analysis due to a low signal-to-noise ratio and high skewness,  likewise for \ion{Fe}{16} at $t_1$ and $t_2$. The \ion{Fe}{16} and \ion{Fe}{23} KG1 fitting parameters ($\kappa$ and $W$) plus errors and reduced $\chi^{2}$ values are also shown in Table \ref{table1}. For comparision, the KG2 and SG $W$ and $\chi^{2}$ values are also shown for each line. Overall, over 60 \% of the lines shown in Figure \ref{fig7} and Table \ref{table1} have KG1 $\chi^{2}_{KG1}\le2.0$.

\subsubsection{\ion{Fe}{23}}
During the interval starting at $t_2$ (covering the HXR peak), five \ion{Fe}{23} regions satisfy the five criteria in Section~\ref{lfit}. These cover part of the coronal 10-25 keV X-ray source and lie within the \ion{Fe}{23} 50\% contour line. We can see that the $\kappa$ index increases from north to south, with the lowest values of $\kappa$ close to the centre of the coronal X-ray source increasing from $\kappa\sim3.8$ to $\kappa\sim6.5$. Similarly, the largest values of $W$ occur closer to the centre of the \ion{Fe}{23} source, with values ranging from $W=0.09$ \AA~to $W=0.11$~\AA. At $t_3$ the HXR footpoints disappear and there is an X-ray coronal source located close to $X=520''$, $Y=275''$. We fit with KG1 \ion{Fe}{23}, lines from eleven locations along the southern edge of the coronal X-ray source, finding $\kappa$ between 4 and 9. The $W$ values are lower than $t_2$, ranging between $W\sim0.08$ \AA~and $W\sim0.095$ \AA. At $t_4$ there are eight regions suitable for study with $\kappa$ ranging between 4.5 and 6. $W$ ranges between 0.07 and 0.095 \AA, with all values lower than the expected Gaussian thermal width of 0.1 \AA. At $t_5$ and $t_6$, the \ion{Fe}{23} $\kappa$ values are $\approx6-7$ with $W\sim0.08-0.095$ \AA. 

\subsubsection{\ion{Fe}{16}}
Overall, the $\kappa$ values found for \ion{Fe}{16} are smaller than those for \ion{Fe}{23}.
At time $t_3$ there are six locations with \ion{Fe}{16} lines satisfying our criteria. Close to the centre of the \ion{Fe}{16} source and overlapping slightly with the edge of the coronal X-ray source we find two locations with $\kappa$ values of 3 and 4. 
At $X=495''$, $Y=265''$, close to the eastern footpoint, the $\kappa$ values are between 3.5 and 4. Overall, the $\kappa$ values for \ion{Fe}{16} are lower than for \ion{Fe}{23} at this time. At time $t_4$ there are nine suitable \ion{Fe}{16} pixels, with $\kappa$ between 2.5 and 4. These are scattered, mostly located at the periphery of the main \ion{Fe}{16} source and at some distance from the coronal X-ray source. 

At $t_4$, $t_5$ and $t_6$, the \ion{Fe}{16} $W$ values are between $0.03$ \AA~and $0.05$~\AA, slightly lower than the \ion{Fe}{16} values at $t_2$ with the majority between 0.05-0.06~\AA. Overall, the largest KG1 $W$ values occur at early times for \ion{Fe}{23} and \ion{Fe}{16}.

\begin{table}[ht!]
\caption{Table showing the \ion{Fe}{16} \textit{(top)} \ion{Fe}{23} \textit{(bottom)} KG1 fitting parameters (width $W$ and kappa index $\kappa$) and reduced $\chi^{2}$ values for all the lines displayed in Figure \ref{fig7}. For comparison with the KG1 fits shown in Figure \ref{fig7}, the KG2 and SG fitting parameters (width $W$) and $\chi^{2}$ are also displayed. The W error values for KG2 and SG are small, all of the order $\sim10^{-4}$~\AA, and are not shown.}
\begin{center}
\scriptsize
\begin{tabular}{l||lll|ll|ll}
\hline
\multicolumn{8}{l}{\textbf{\small{\ion{Fe}{16}}}}                                                                                                            \\ \hline
                        & \multicolumn{3}{l|}{KG1}                  & \multicolumn{2}{l|}{KG2}         & \multicolumn{2}{l}{SG}          \\ \hline
                        & $\kappa$ & W (\AA) & $\chi^{2}$ & W (\AA) & $\chi^{2}$ & W (\AA) & $\chi^{2}$ \\ \hline
\multirow{6}{*}{t$_{3}$} & 4.7$\pm$0.5   &   0.054$\pm$0.002     &        2.7                &    0.070    &2.4                       &   0.074     &       6.1                \\ \cline{2-8} 
                        & 3.4$\pm$0.4   &     0.057$\pm$0.004   &          1.2             &     0.084   &2.6                        &    0.088    &  5.0 \\ \cline{2-8} 
                        & 3.5$\pm$0.4   &     0.052$\pm$0.003   &    1.0                    &    0.075    &1.7                        &   0.079     &     4.6                   \\ \cline{2-8} 
                        & 2.8$\pm$0.1   &   0.035$\pm$0.002     &     3.2                   &    0.061    &4.4                        &    0.065    &    14.6                    \\ \cline{2-8} 
                        & 3.1$\pm$0.2   &   0.038$\pm$0.002    &     3.3                   &     0.054   &3.5                        &     0.065   &    12.9                   \\ \cline{2-8} 
                        & 3.9$\pm$0.3   &    0.040$\pm$0.002    &     1.7                   &     0.047   &6.8                        &    0.060    &    5.1          
                                         \\ \hline
\multirow{9}{*}{t$_{4}$} & 3.2$\pm$0.3   &     0.039$\pm$0.003   &      1.6                  &    0.060    &        1.6               &   0.064     &      4.7                 \\ \cline{2-8} 
                        & 3.2$\pm$0.2   &   0.043$\pm$0.003     &      1.1                  &    0.067    &        1.9                &   0.070     &8.0                        \\ \cline{2-8} 
                        & 2.2$\pm$0.1   &    0.037$\pm$0.002    &     3.5                   &   0.080     &      17.7                  &    0.084    &27.3                        \\ \cline{2-8} 
                        & 3.9$\pm$0.4   &    0.048$\pm$0.002    &     1.1                   &   0.067     &       1.5                 &      0.070  &5.0                        \\ \cline{2-8} 
                        & 3.2$\pm$0.3   &     0.033$\pm$0.003   &      1.8                  &     0.053   &      1.6                  &    0.058    &4.6                       \\ \cline{2-8} 
                        & 3.1$\pm$0.2   &    0.038$\pm$0.002    &     2.5                   &    0.061    &        3.0                &   0.065     &15.1                        \\ \cline{2-8} 
                        & 3.4$\pm$0.3   &   0.050$\pm$0.003     &     1.2                   &     0.074   &            2.2            &   0.077     &5.2
                        \\ \cline{2-8} 
                        & 2.8$\pm$0.2   &   0.030$\pm$0.002     &    0.7                    &    0.055    &           1.0             &   0.059     &15.3
                          \\ \cline{2-8} 
                        & 3.4$\pm$0.2   &    0.042$\pm$0.002    &    4.3                    &    0.064    &          4.4              &   0.068     &12.4

                                                \\ \hline
\multirow{2}{*}{t$_{5}$} & 3.6$\pm$0.2   &   0.034$\pm$0.002     &   0.6                     &    0.058    &         0.6               &    0.063    &      6.0                  \\ \cline{2-8} 
                        & 2.5$\pm$0.1   &    0.040$\pm$0.002    &   3.5                     &    0.079    &       13.6                 &    0.083    &16.4            
                                                \\ \hline
\multirow{5}{*}{t$_{6}$} & 3.1$\pm$0.2   &   0.034$\pm$0.002     &      1.2                  &    0.055    &           1.0             &     0.059   &5.6                                                \\ \cline{2-8} 
                        & 2.3$\pm$0.1   &    0.040$\pm$0.002    &     1.6                   &    0.080    &    9.9                    &     0.085   &17.5                        \\ \cline{2-8} 
                        & 2.8$\pm$0.1   &   0.042$\pm$0.002     &    1.8                   &   0.072     &         6.8               &     0.077   &4.7
                                                \\ \cline{2-8} 
                        & 3.1$\pm$0.3   &    0.036$\pm$0.003    &       1.7                 &   0.059     &         1.6               &     0.062   &5.1
                               \\ \cline{2-8} 
                        & 2.5$\pm$0.1   &   0.035$\pm$0.001     &     4.2                   &    0.067    &          15.1              &    0.072    &38.7

                                                \\ \hline
\end{tabular}
\end{center}

\vspace{10pt}

\begin{center}
\scriptsize
\begin{tabular}{l||lll|ll|ll}
\hline
\multicolumn{8}{l}{\textbf{\small{\ion{Fe}{23}}}}                                                                                                            \\ \hline
                        & \multicolumn{3}{l|}{KG1}                  & \multicolumn{2}{l|}{KG2}         & \multicolumn{2}{l}{SG}          \\ \hline
                        & $\kappa$ & W (\AA) & $\chi^{2}$ & W (\AA) & $\chi^{2}$ & W (\AA) & $\chi^{2}$ \\ \hline
\multirow{5}{*}{t$_{2}$} & 6.2$\pm$0.6   &   0.102$\pm$0.002     &2.7                        & 0.115       &        2.8                &    0.125    &     5.9                   \\ \cline{2-8} 
                        & 5.9$\pm$0.5   &     0.097$\pm$0.002   &2.3                        &0.110        &    2.5                    &    0.125    & 5.8                        \\ \cline{2-8} 
                        & 6.2$\pm$0.5   &   0.109$\pm$0.002     &2.9                        &0.123        &     3.4                   &    0.120    &6.9                        \\ \cline{2-8} 
                        & 5.7$\pm$0.4   &    0.107$\pm$0.002    &1.6                        &0.124       &     2.8                  &   0.132     &6.5                        \\ \cline{2-8} 
                        & 3.7$\pm$0.2   &   0.092$\pm$0.003     &1.3                        &0.125        &     4.4                   &    0.133    &7.6                        \\ \hline
\multirow{11}{*}{t$_{3}$} & 7.0$\pm$0.7   &    0.093$\pm$0.002    &    1.4                    &    0.102    &          1.1              &    0.112    &     4.0                   \\ \cline{2-8} 
                        & 8.4$\pm$0.7   &    0.092$\pm$0.002    &    1.1                    &   0.096     &   1.6                     &     0.107   &4.1                        \\ \cline{2-8} 
                        & 6.1$\pm$0.8   &    0.086$\pm$0.002    &    2.5                    &    0.096    &      3.1                  &   0.106     &6.9                        \\ \cline{2-8} 
                        & 4.5$\pm$0.4   &    0.081$\pm$0.002    &     2.0                   &   0.100     &     2.4                   &    0.110    &6.91                        \\ \cline{2-8} 
                        & 5.5$\pm$0.3   &    0.085$\pm$0.002    &    1.7                    &  0.099      &    1.3                    &   0.109     &10.1                        \\ \cline{2-8} 
                        & 4.0$\pm$0.3   &    0.078$\pm$0.002    &     4.6                   &   0.100     &      5.8                  &    0.111    &11.4                        \\ \cline{2-8} 
                        & 4.8$\pm$0.4   &    0.085$\pm$0.003    &    1.9                    &  0.102      &     2.1                   &   0.112     &5.5
                                                \\ \cline{2-8} 
                        & 4.5$\pm$0.4   &    0.087$\pm$0.003    &     1.5                   &   0.107     &       3.2                 &    0.117    &5.0
                        \\ \cline{2-8} 
                        & 3.9$\pm$0.3   &    0.087$\pm$0.004    &     0.6                   &   0.113     &     2.6                   &    0.123    &3.8
                                                \\ \cline{2-8} 
                        & 5.8$\pm$0.6   &    0.090$\pm$0.003    &      2.2                  &   0.103     &2.9 & 0.113 &4.8
                                                    \\ \cline{2-8} 
						& 4.6$\pm$0.5   &    0.094$\pm$0.004    &      1.8                  &   0.115     &2.9 & 0.125 &3.8
                                                \\ \hline
\multirow{8}{*}{t$_{4}$} & 4.8$\pm$0.5   &    0.079$\pm$0.003    &    0.9                    &  0.094      &       1.0                 &    0.105    &    3.3                    \\ \cline{2-8} 
                        & 4.6$\pm$0.5   &    0.072$\pm$0.003    &    1.2                    &    0.087    &     1.2                   &   0.098     &3.1                        \\ \cline{2-8} 
                        & 5.6$\pm$0.5   &    0.078$\pm$0.002    &     2.8                   &   0.088     &     3.0                   &    0.099    &5.7                        \\ \cline{2-8} 
                        & 5.4$\pm$0.6   &     0.083$\pm$0.003   &    2.1                    &   0.096     &     2.6                   &   0.106     &4.4                        \\ \cline{2-8} 
                        & 5.7$\pm$0.6   &    0.084$\pm$0.002    &    1.4                    &   0.095     &     2.0                   &    0.106    &3.9                        \\ \cline{2-8} 
                        & 5.1$\pm$0.4   &     0.084$\pm$0.002   &   3.1                     &     0.098   &      4.2                  &    0.109    &6.9                        \\ \cline{2-8} 
                        & 4.8$\pm$0.3   &    0.084$\pm$0.002    &   1.5                     &   0.101     &   1.5                    &   0.111     &6.8
                                                \\ \cline{2-8} 
                        & 5.0$\pm$0.5   &    0.095$\pm$0.003    &   0.8                     &   0.114     &   1.4                     &   0.124     &3.6

                                                \\ \hline
\multirow{4}{*}{t$_{5}$} & 5.1$\pm$0.6   &    0.079$\pm$0.003    &  1.8                      &  0.093      &           1.6             &    0.103    &3.6                                                \\ \cline{2-8} 
                        & 6.6$\pm$0.6   &   0.086$\pm$0.002     &  2.4                      &   0.094     &         2.1               &    0.104    &5.2                        \\ \cline{2-8} 
                        & 6.3$\pm$0.5   &    0.086$\pm$0.002    &     2.1                   &     0.095   &        1.8                &    0.106    &5.2
                                                \\ \cline{2-8} 
                        & 5.9$\pm$0.7   &   0.091$\pm$0.003     &     1.1                   &    0.104    &      1.7                  &   0.114     &3.0

                                                \\ \hline
\multirow{1}{*}{t$_{6}$} & 6.4$\pm$0.7   &    0.082$\pm$0.002    &  1.4                      &  0.090      &           1.4             &    0.100    &3.5             
                                                \\ \hline
\end{tabular}
\end{center}
\label{table1}
\end{table}

\section{Physical interpretation and discussion}

Our analysis can be summarised as follows:

\begin{enumerate}
\item Non-Gaussian line profiles consistent with kappa distributions of emitting ions were found during the flare, close to the flare loop-top, HXR footpoints and ribbons \citep[similar to SOL2013-05-15T01:45 analysed in][]{2016A&A...590A..99J}.
\item \ion{Fe}{16} lines exhibiting kappa profiles were situated further from the coronal source than the \ion{Fe}{23} lines, and often in regions where HXR sources were previously observed.
\item \ion{Fe}{23} lines exhibiting kappa profiles were situated close to the coronal source and appeared to move with the coronal source over time.
\item The $\kappa$ index values of the \ion{Fe}{16} lines were smaller than those of \ion{Fe}{23} and not so systematic in terms of position and value.
\item \ion{Fe}{23} showed interesting spatial variations close to the coronal X-ray sources with smaller values of $\kappa$ index located closer to the X-ray coronal sources early in the flare.
\end{enumerate}

We considered the possibility that the observed non-Gaussian line profiles result from the EIS instrumental response. Although we cannot rule this out completely, we find that parameter trends for $\kappa$ index and $\sigma_{\kappa}$ (particularly for \ion{Fe}{23}) do not behave as we would expect if the instrumental response were non-Gaussian. We will now discuss the possible origins of the results.

\subsection{The possible origin of non-Gaussian spectral lines}
\begin{figure*}[htp]
	\centering
	\includegraphics[width=0.45\textwidth,angle=0]{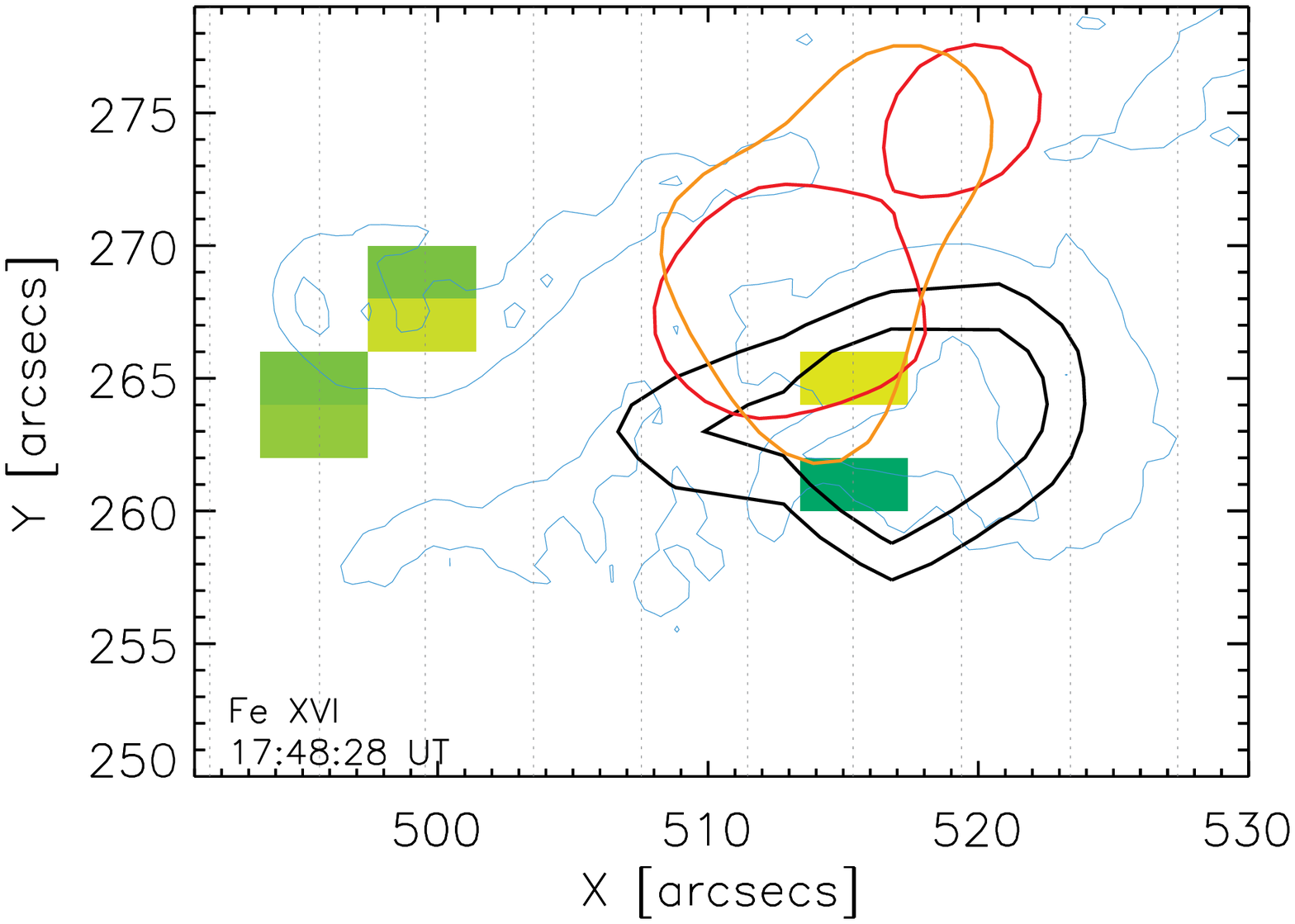}
	\includegraphics[width=0.45\textwidth,angle=0]{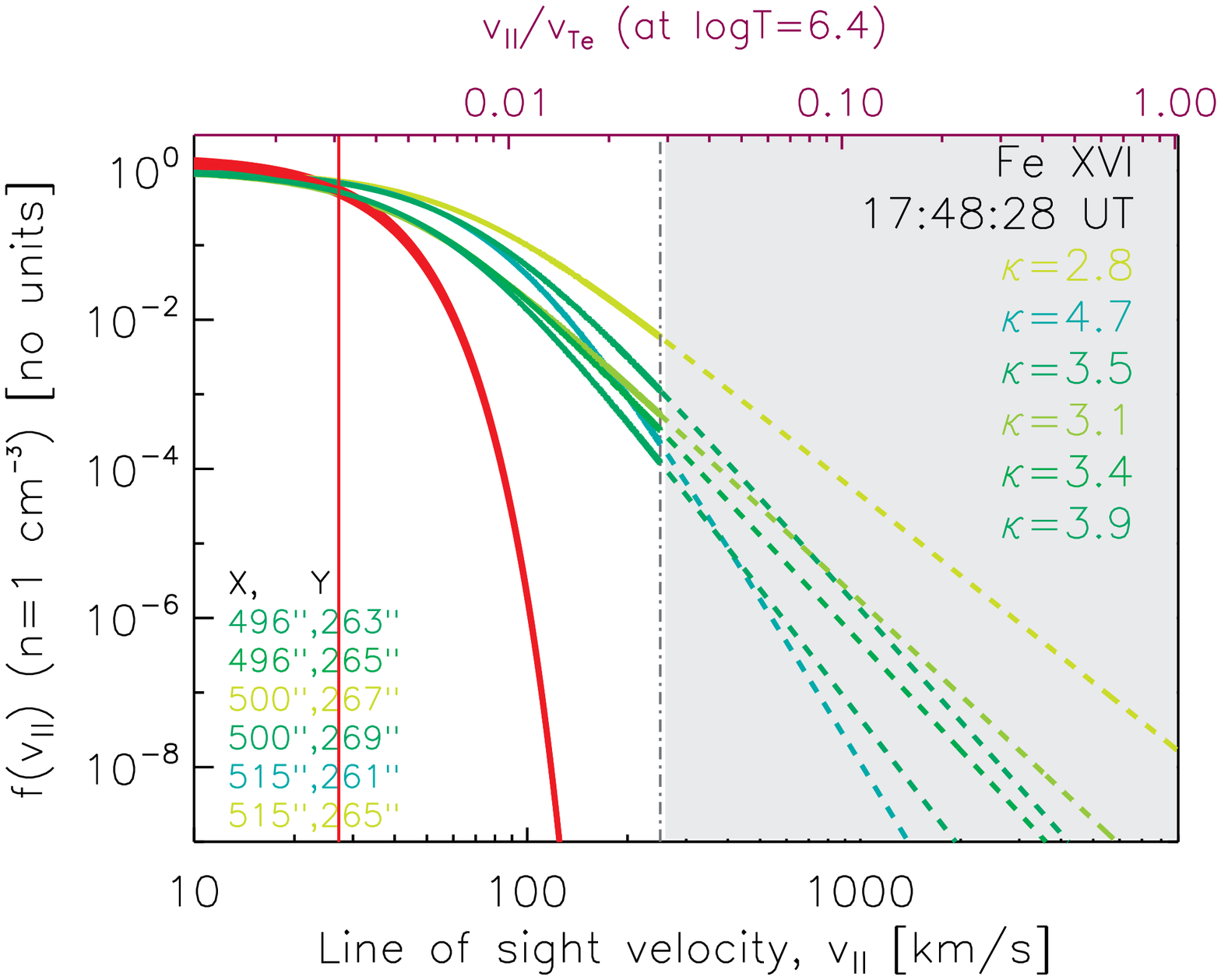}
	\includegraphics[width=0.45\textwidth,angle=0]{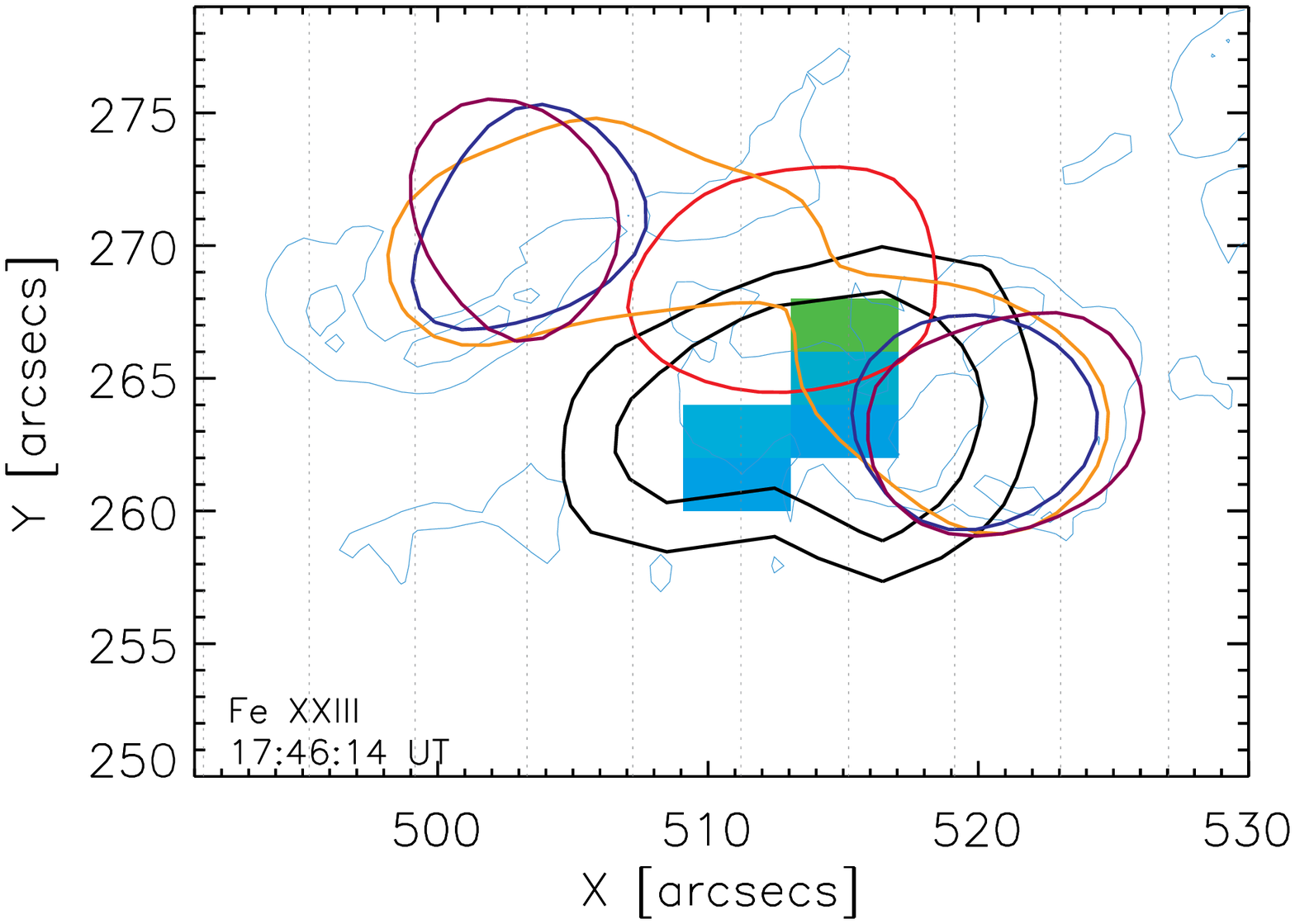}
	\includegraphics[width=0.45\textwidth,angle=0]{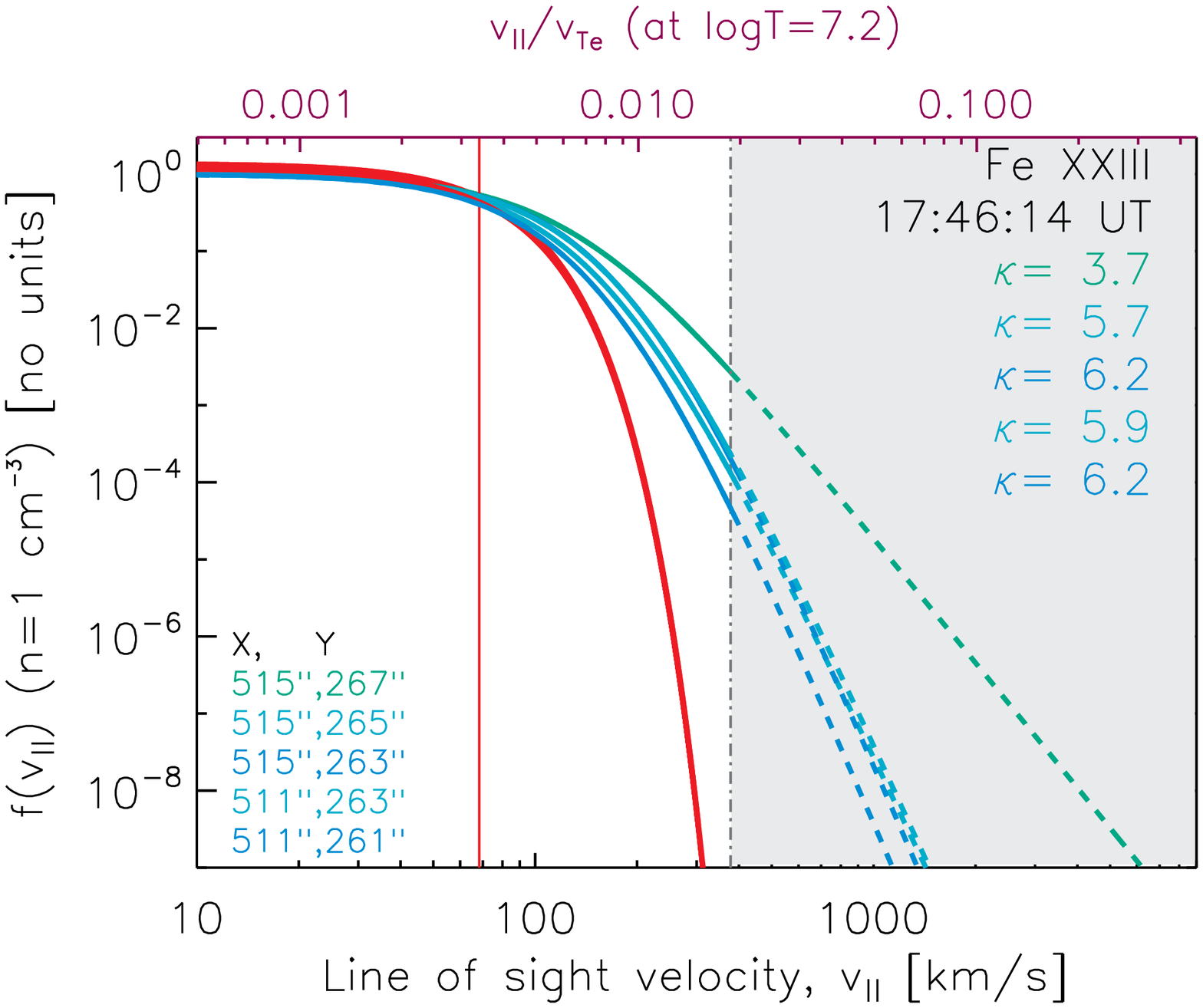}
	\caption{{\it Left:} \ion{Fe}{16} ({\it top}) and \ion{Fe}{23} ({\it bottom}) $\kappa$ index maps (re-plotted from Figure \ref{fig7}), at two different times (see maps). {\it Right:} For each region in the map ({\it left}), the form of possible ion velocity distributions are determined using Equation \ref{los_kap} and the observed fit values of $\kappa$ index and $\sigma_{\kappa}$ (corresponding colours). We are only interested in the form of $f(v_{||})$, not the actual values ($n$ is set to 1 and $f(v_{||})$ is also divided by the maximum value for plotting). The red curve denotes the expected Maxwellian velocity distribution with the red vertical line denoting the ion thermal speed. For \ion{Fe}{23}, the distribution tends more towards a Maxwellian as the regions move away from the X-ray source (red contour) and towards the centre of the \ion{Fe}{23} source (black contours). The grey dashed curve indicates the fit cut off velocity. The errors for $f(v_{||})$ are not shown for clarity (see Figure \ref{fig9}). Error values for $\kappa$ and $W$ are shown in Table \ref{table1}. $v_{Te}=$electron thermal speed at either $\log{T}=6.4$ or $\log{T}=7.2$.}
	\label{fig8}
	\end{figure*}

\begin{figure*}%[htp]
	\centering
	\hspace{40pt}\includegraphics[width=0.42\linewidth]{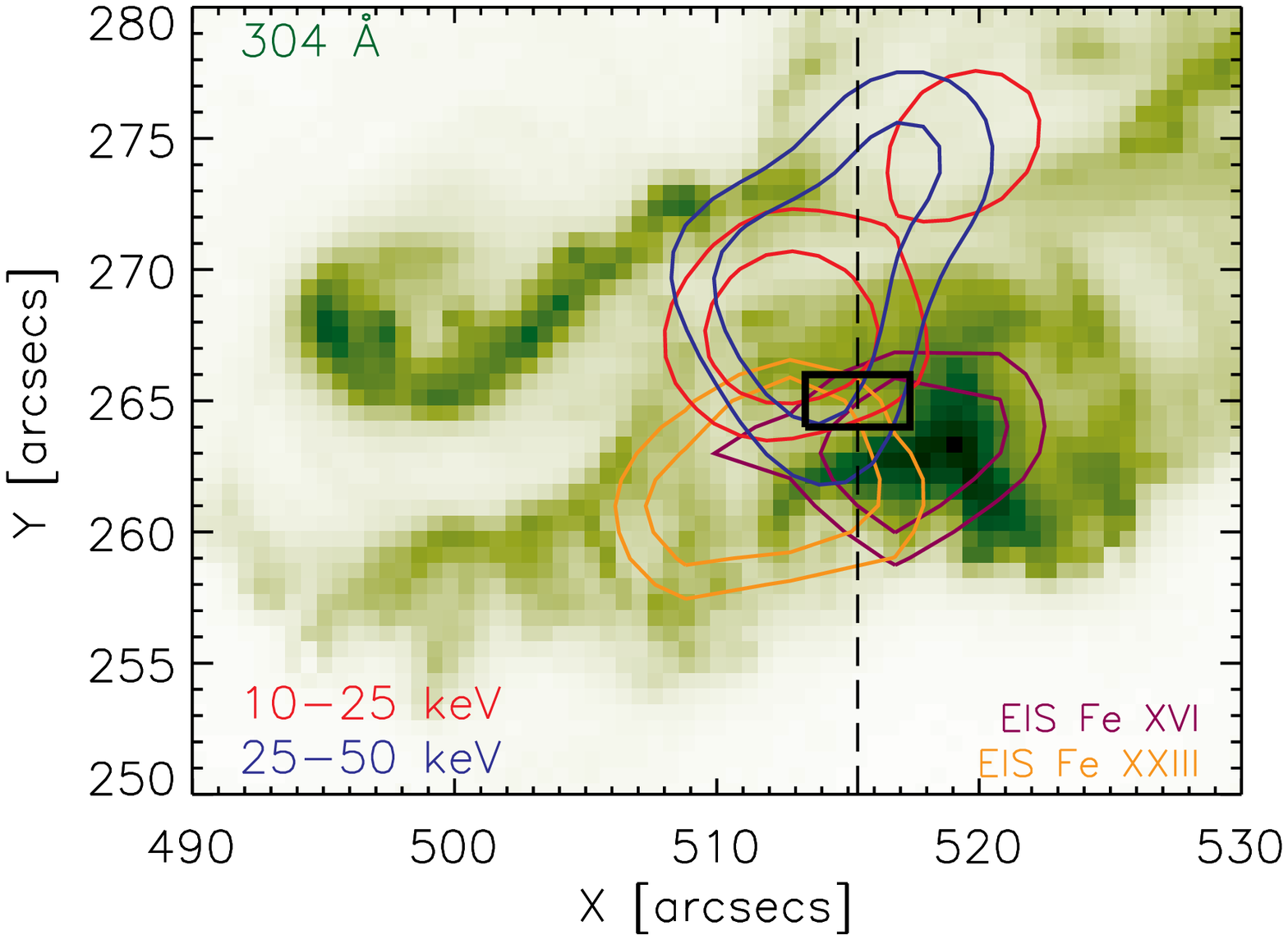}\\
	\includegraphics[width=0.35\linewidth]{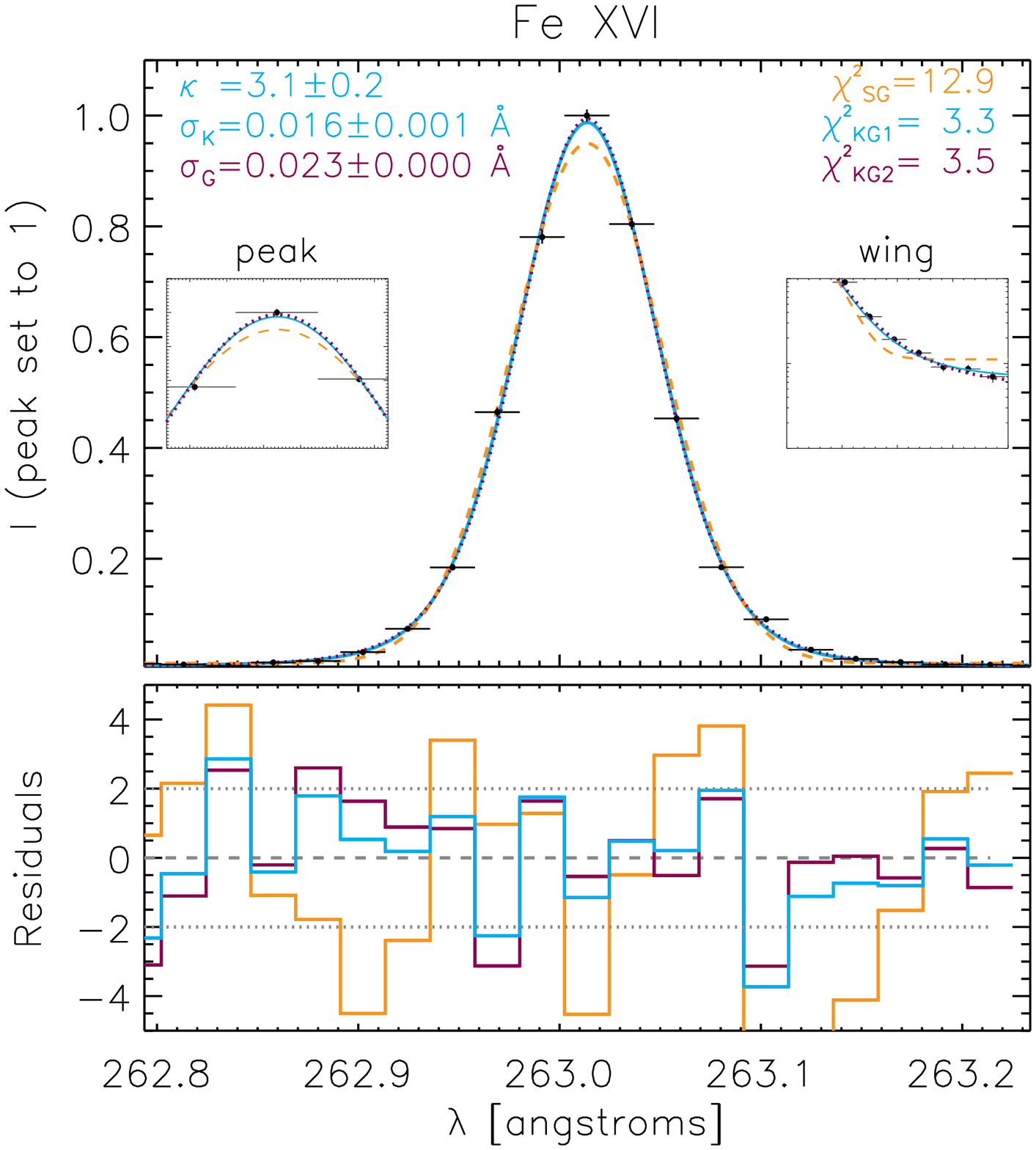}
	\includegraphics[width=0.35\linewidth]{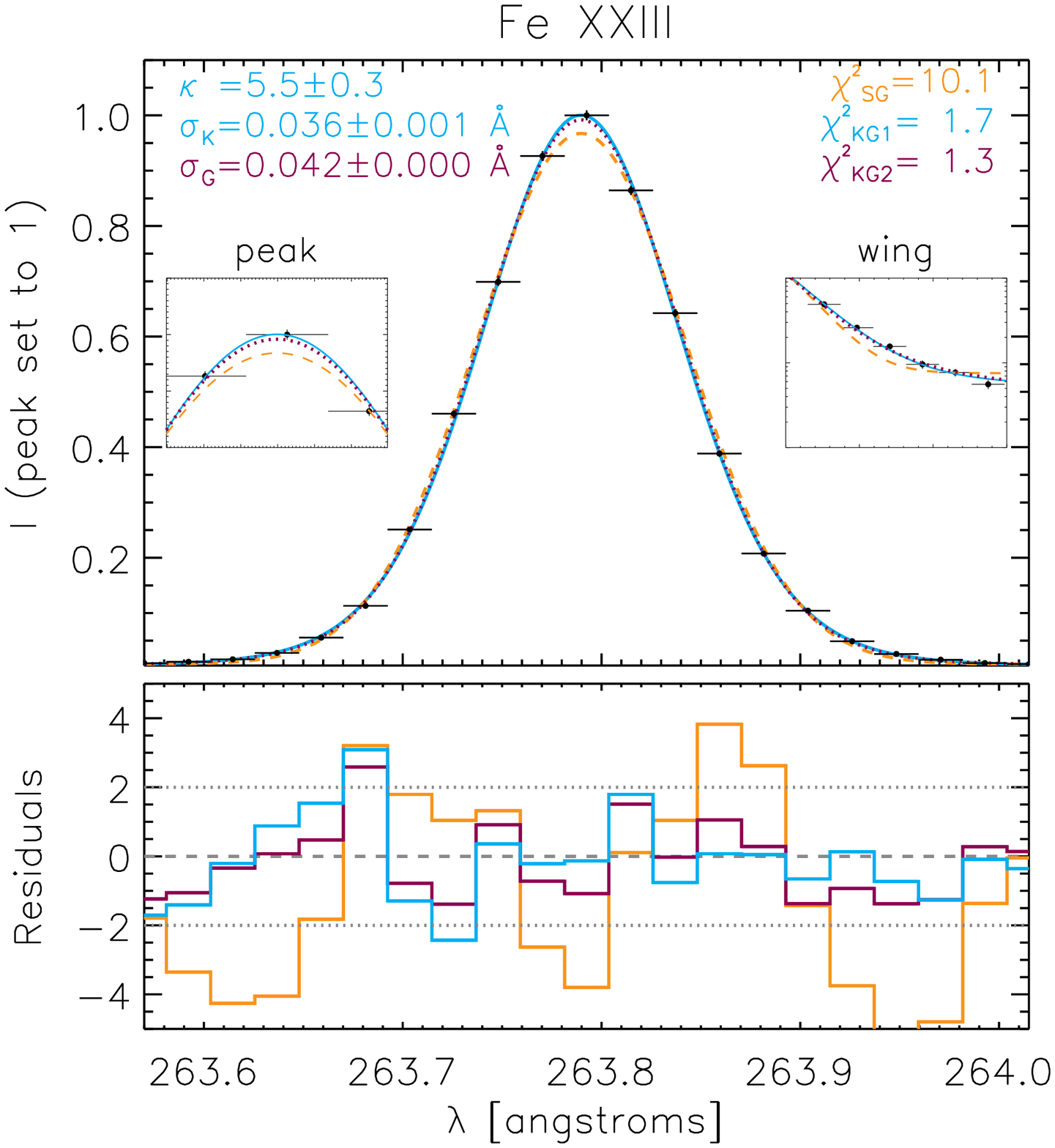}
	\includegraphics[width=0.35\linewidth]{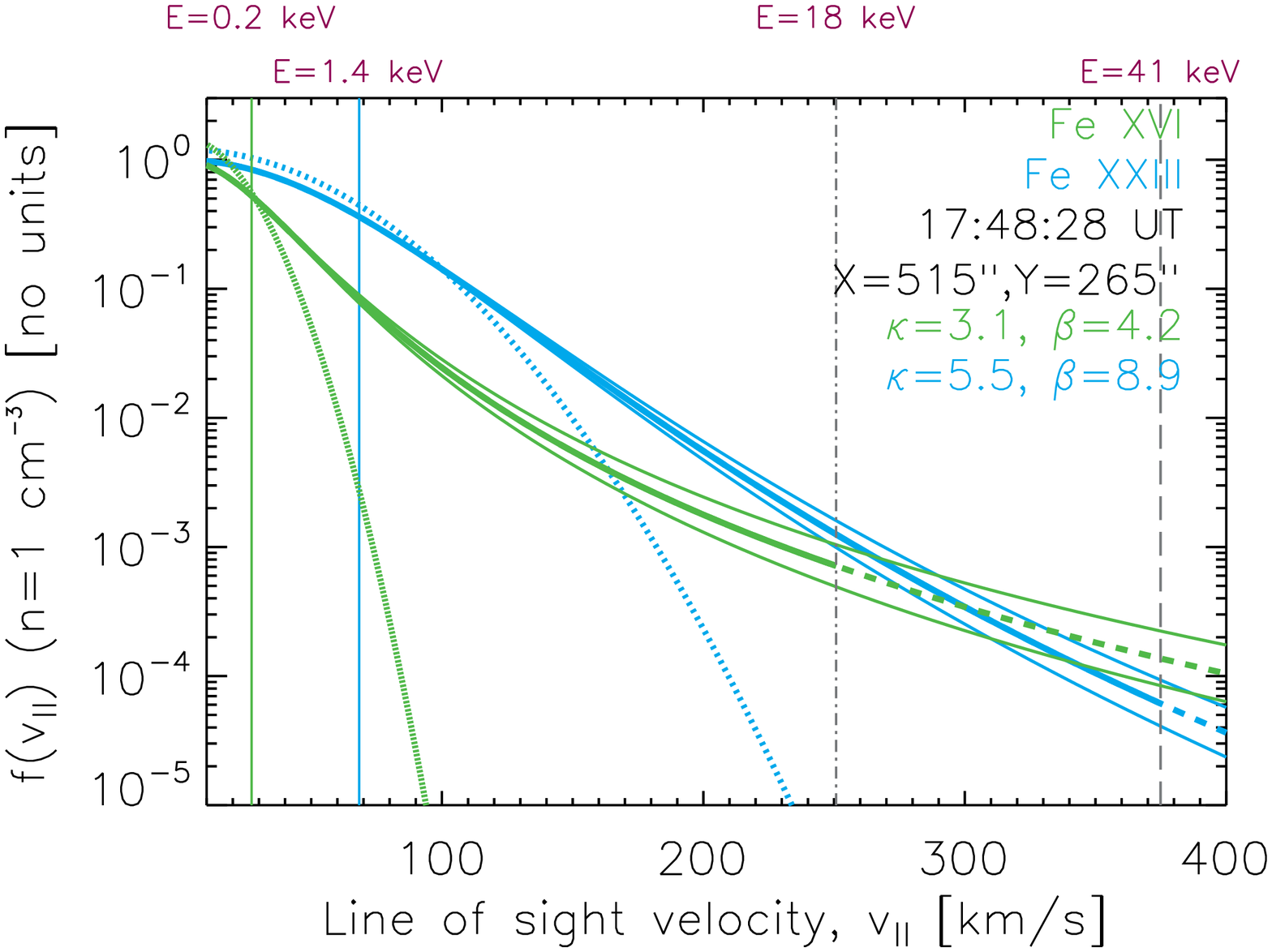}
	\includegraphics[width=0.35\linewidth]{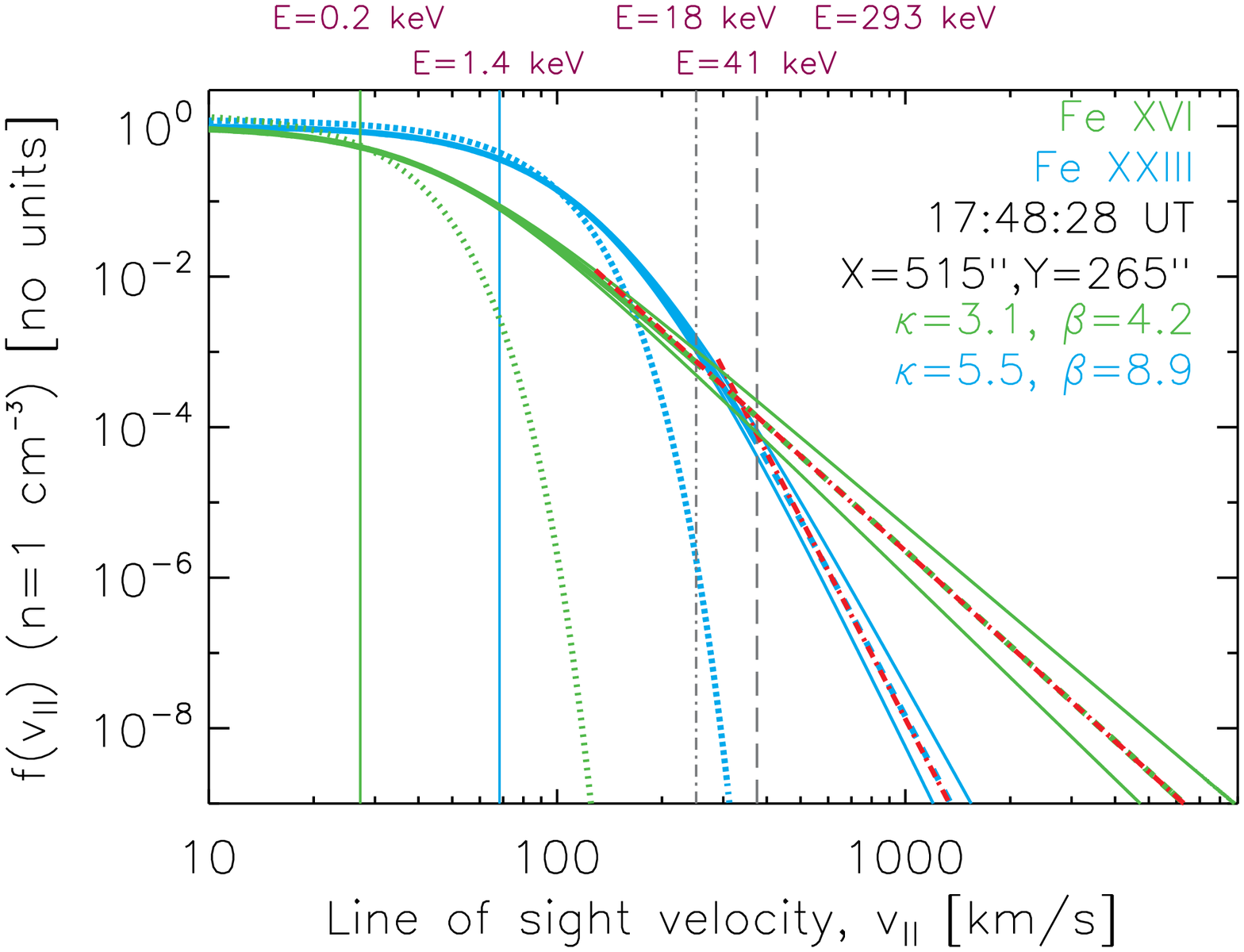}
\caption{{\it Top:} SDO AIA 304 \AA~image of SOL2014-03-29T17:44 showing RHESSI and EIS contours (see legend) and one region (at $t_3$ (17:48:28 UT) centred at [X$\sim$515'', Y$\sim$265'']) where line profiles of \ion{Fe}{23} and \ion{Fe}{16} can be fitted by KG1 using the five criteria of Section \ref{method}. {\it Middle row:} Both lines are fitted with a kappa-Gaussian and they have different values of $\kappa$ index. The residuals and $\chi^{2}$ values show that the kappa-Gaussian fits are a better model than the single Gaussian fit. {\it Bottom row:} The kappa line profiles are converted to 1-D velocity distributions $f(v_{||})$ using Equation \ref{los_kap} (and divided by the maximum value of $f(v_{||})$). The bottom left panel displays $f(v_{||})$ over the range of velocities fitted during the observation while the bottom right panel shows $f(v_{||})$ plotted over a larger range of $v_{||}$. At large $v_{||}$, a linear fit to $\log{f(v_{||})}$ versus $\log{v_{||}}$ (red lines) finds the velocity power index $\beta$, with values displayed in the legend.}
	\label{fig9}
	\end{figure*}

Though difficult to detect by other means, it is likely that protons and heavier ions are accelerated during the flare. In Figure \ref{fig8}, we interpret the results as 1-D ion velocity distributions $f(v_{||})$ at a single time for \ion{Fe}{16} ($t_3$) and \ion{Fe}{23} ($t_2$), obtained using the observed KG1 fit parameters and Equation \ref{los_kap}. The 1-D distribution is plotted against the $|$velocity$|$ in km/s and also as a fraction of the electron thermal speed ($v_{Te}=\sqrt{2k_{B}T_{e}/m_{e}}$ for $\log T_{e}=6.4$ (\ion{Fe}{16}) and $\log T_{e}=7.2$ (\ion{Fe}{23})). 
The grey region denotes ion velocities outside of the maximum fitted ion velocity (from $v_{||}=c\Delta\lambda/\lambda_{0}$), since the line fits were performed over a range of $\lambda_{0}\pm0.25$ \AA~(\ion{Fe}{16}) and $\lambda_{0}\pm0.30$~\AA\ (\ion{Fe}{23}), where $\lambda_{0}$ is the line centroid position. The maximum fitted velocities for \ion{Fe}{16} and \ion{Fe}{23} are 250 km/s and 340 km/s. The red curve denotes the expected Maxwellian ion velocity distribution at $\log T=6.4$ and $\log T=7.2$, while the solid red line denotes the expected ion thermal speed. The expected thermal speeds of \ion{Fe}{16} and \ion{Fe}{23} are $\sim30$ km/s and $\sim70$ km/s respectively. The \ion{Fe}{23} results (at $t_2$, the flare peak) are particularly interesting since we can see the $\kappa$ index increasing as we move away from the centre of the coronal X-ray source towards the centre of the \ion{Fe}{23} source. 
This shows that the velocity distribution tends towards Maxwellian further from the coronal X-ray source. In Figure \ref{fig9}, we compare $f(v_{||})$ versus $|v_{||}|$ for one \ion{Fe}{16} line and one \ion{Fe}{23} line observed in the same spatial region at the same time ($t_3$). At high velocities, the distribution tends to a power law $f(v_{||})\sim v_{||}^{-\beta}$ with power index $\beta\approx2(\kappa-1)$. The high velocity part of $f(v_{||})$ is fitted with a straight line in log-log space and values of $\beta$ are shown on the figure legend, with the $\kappa$ index values. The $\beta$ values are $\beta=4.2$ (\ion{Fe}{16}) and $\beta=8.9$ (\ion{Fe}{23}). Hence, in the same spatial region (but not necessarily the same height), the emitting \ion{Fe}{16} velocity distribution is further from Gaussian than the \ion{Fe}{23} velocity distribution. This is an interesting result since we might expect the cooler \ion{Fe}{16} to lie at a lower height in the atmosphere than the hotter \ion{Fe}{23}, and further from sites of acceleration for example.
Further, we can estimate (using the fast ion thermalization equations taken from \citet{JCallen}) whether the inferred non-thermal ion distributions can exist in a flaring plasma. Kappa distributions are routinely measured in the collisionless solar wind, but the flaring atmosphere is highly collisional with electron number densities of $10^{9}$ cm$^{-3}$ or greater. For fast ions where $v_{Tf}<<v<<v_{Te}$, (for $v_{Te}=$ electron thermal speed and $v_{Tf}=$ heavy ion thermal speed), colliding with a background electron (e) - proton (p) plasma, there are two dominant collisional regimes below and above a velocity $v_{c}$ given by
\begin{equation}\label{vel_c}
v_{c}=\left[\frac{3\sqrt{\pi}}{4}\frac{m_{e}}{m_{p}}\right]^{1/3}v_{T_{e}}.
\end{equation}
Here $v_{Te}=\sqrt{2k_{B}T_{e}/m_{e}}=1.1\times10^{4}$ km/s and $v_{Tf}=\sqrt{2k_{B}T_{f}/m_{f}}=34$ km/s for $T_{e}=T_{p}=T_{f}=4$ MK (corresponding to \ion{Fe}{16}). 
Equation \ref{vel_c} gives $v_{c}\sim989$ km/s. Above $v_{c}$, collisions with electrons are dominant but below $v_{c}$, collisions with protons are dominant. The maximum velocities determined from the line fitting are only 200-300~km/s, so only heavy ion-proton collisions are considered\footnote{Note, the negligible abundance of heavy ions means that we can ignore heavy ion - heavy ion collisions, compared to the interaction with electrons and protons.}. 

The Coulomb collisional frequencies (ion-electron f/e and ion-proton f/p) of heavy Fe ions with a background electron-proton plasma are given by
\begin{equation}
\nu_{\epsilon}^{f/e}\simeq\nu_{0}^{f/e}\frac{m_{f}}{m_{e}}\frac{8}{3\sqrt{\pi}}\left(\frac{v}{v_{Te}}\right)^{3}
=2\left(\frac{\epsilon}{\epsilon_{c}}\right)^{3/2}
\end{equation}
\begin{equation}
\nu_{\epsilon}^{f/p}\simeq\nu_{0}^{f/e}2\frac{m_{f}}{m_{p}}
\end{equation}
where $\epsilon_{c}=\frac{m_{f}v_{c}^{2}}{2}$ and $\nu_{0}$ is a reference collisional frequency (or a generalisation of the Lorentz collisional frequency) given by
\begin{equation}
\nu_{0}^{f/e}(v)=\frac{n_{e}\Gamma_{fe}}{v^{3}}
=\frac{4\pi n_{e} Z_{f}^{2} e^{4} \ln{\Lambda_{fe}}}{m_{f}^{2}v^{3}} \ ,
\end{equation}
where $Z$ the ion charge and assuming the Coulomb logarithm $\ln{\Lambda_{fe}}=\ln{\Lambda_{ee}}\sim20$ in the corona. The heavy ion collisional energy loss rate (total ion energy $\epsilon=m_{f}v^{2}/2$, not per nucleon) is then given by
\begin{equation}
\frac{d\epsilon}{dt}=-\left(\nu_{\epsilon}^{f/e} + \nu_{\epsilon}^{f/p}\right)\epsilon
\end{equation}
or in terms of ion velocity
\begin{equation}\label{dvdt}
\frac{dv}{dt}=-\frac{v}{\tau_{S}}\left[1+\frac{v_{c}^{3}}{v^{3}}\right]
\end{equation}
where
\begin{equation}
\tau_{S}=\frac{2}{\nu_{\epsilon}^{f/e}}
\end{equation}
is a characteristic fast ion slowing-down time.
	\begin{figure*}%[htp]
	\centering
	\includegraphics[width=0.43\linewidth]{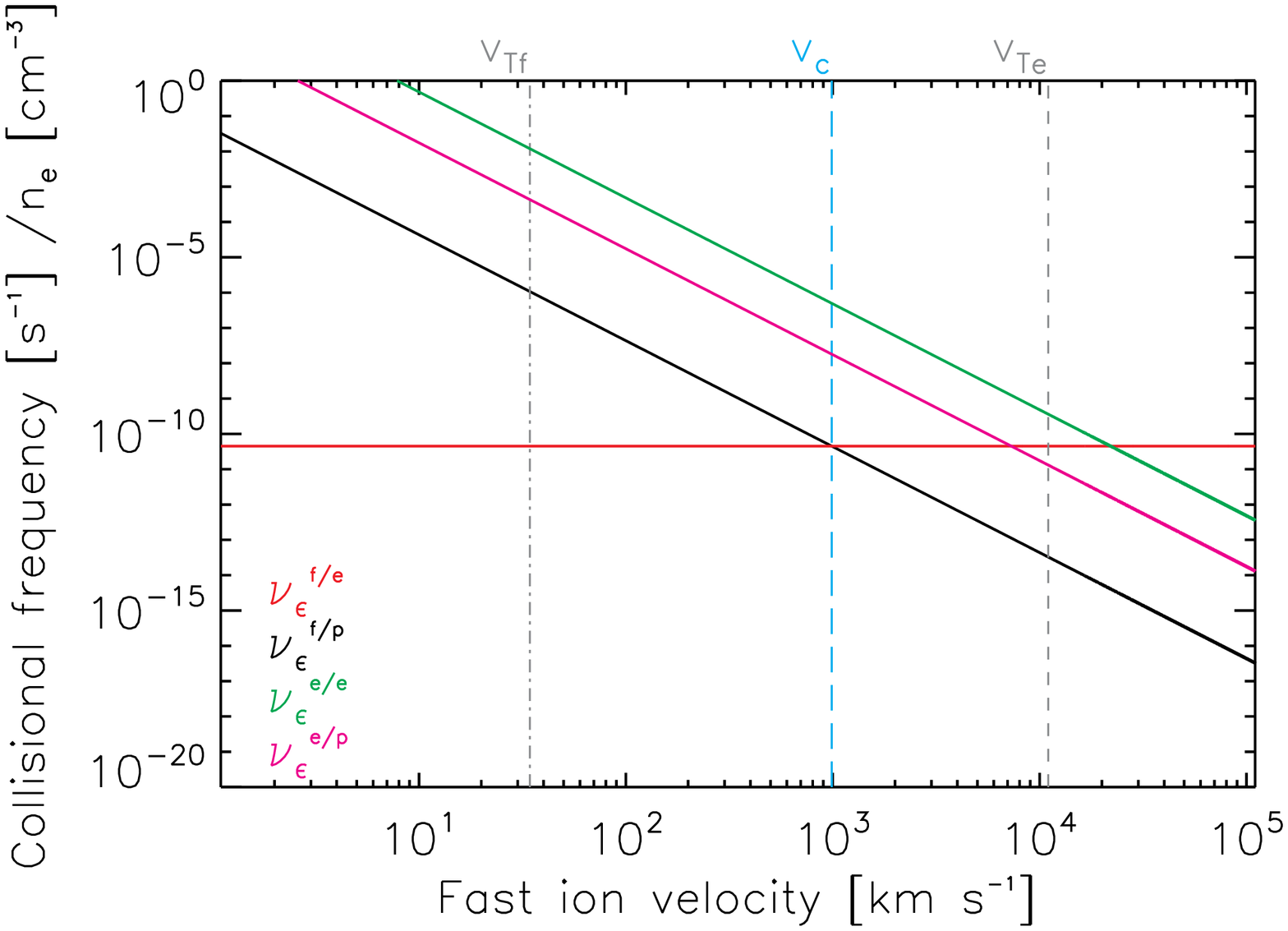}
	\includegraphics[width=0.43\linewidth]{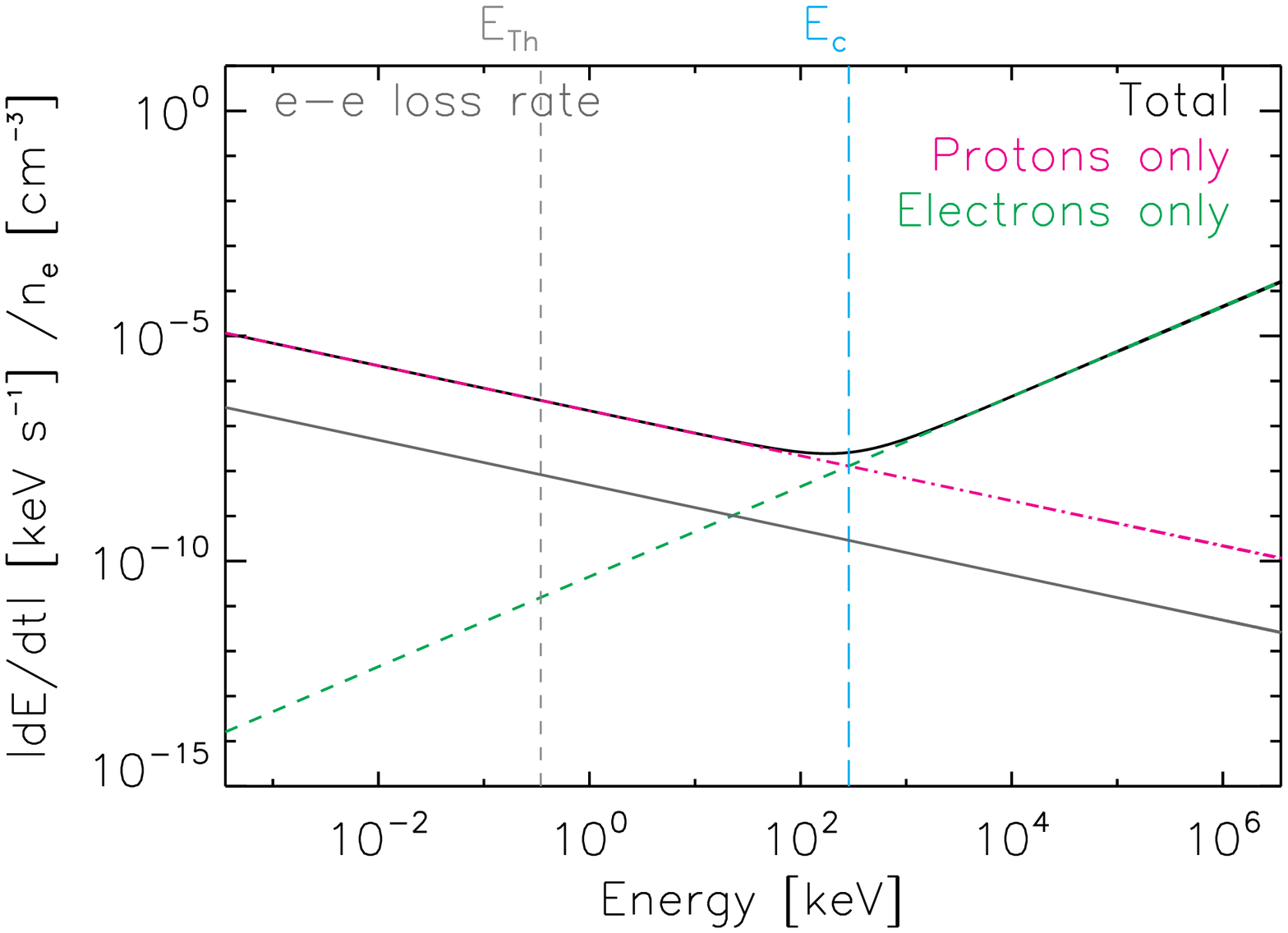}
	\includegraphics[width=0.43\linewidth]{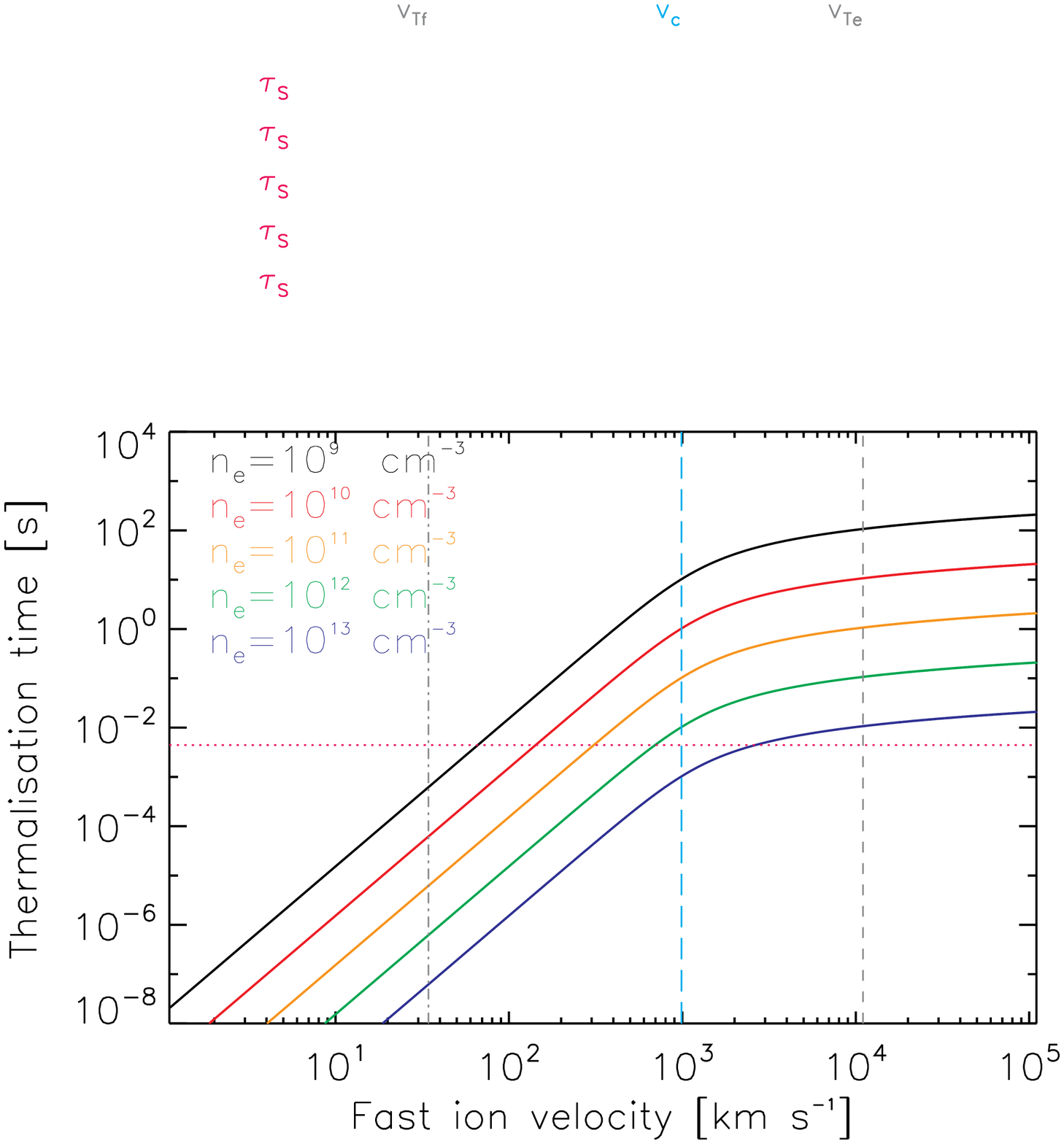}
	\includegraphics[width=0.43\linewidth]{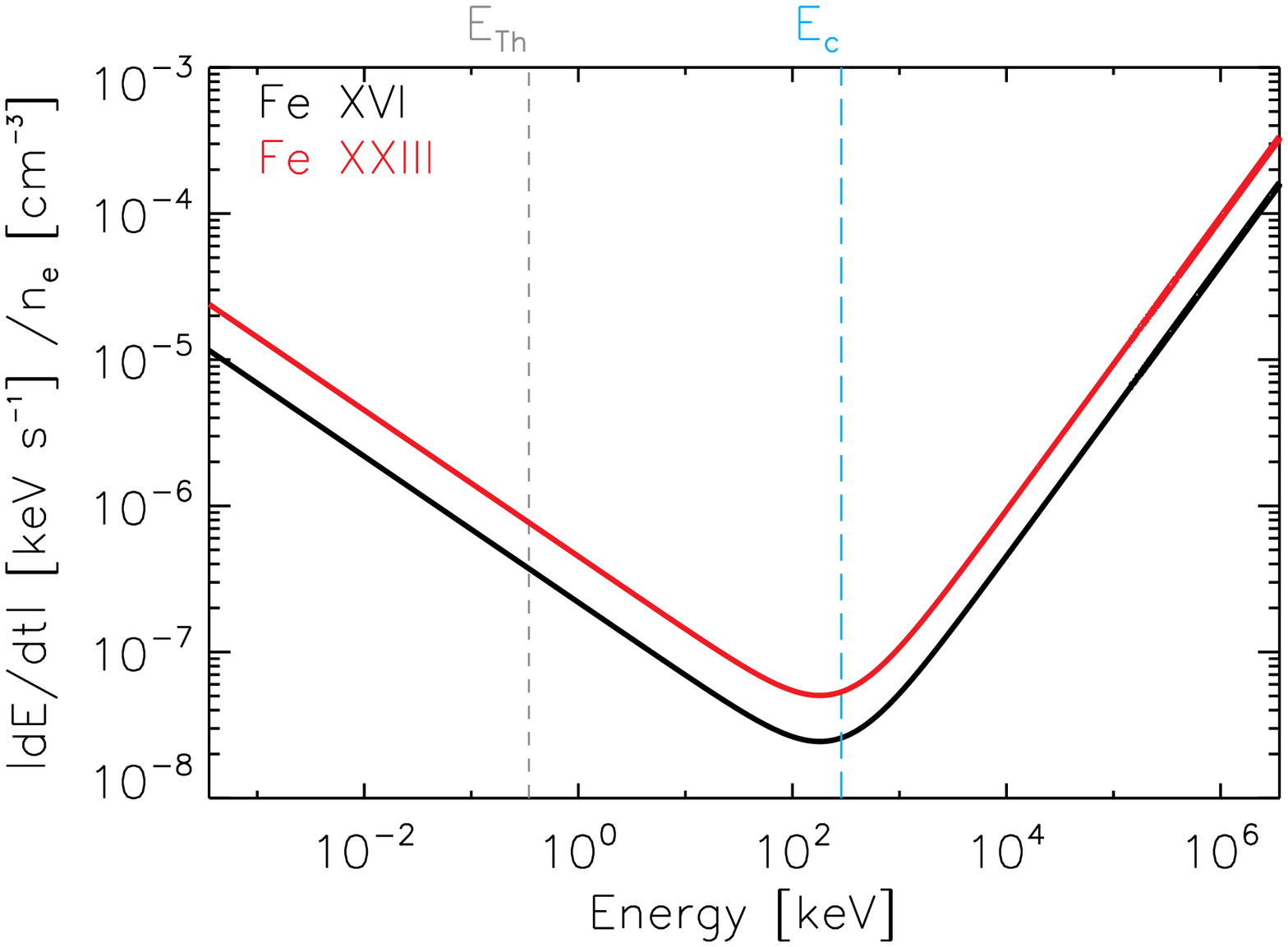}
	\includegraphics[width=0.43\linewidth]{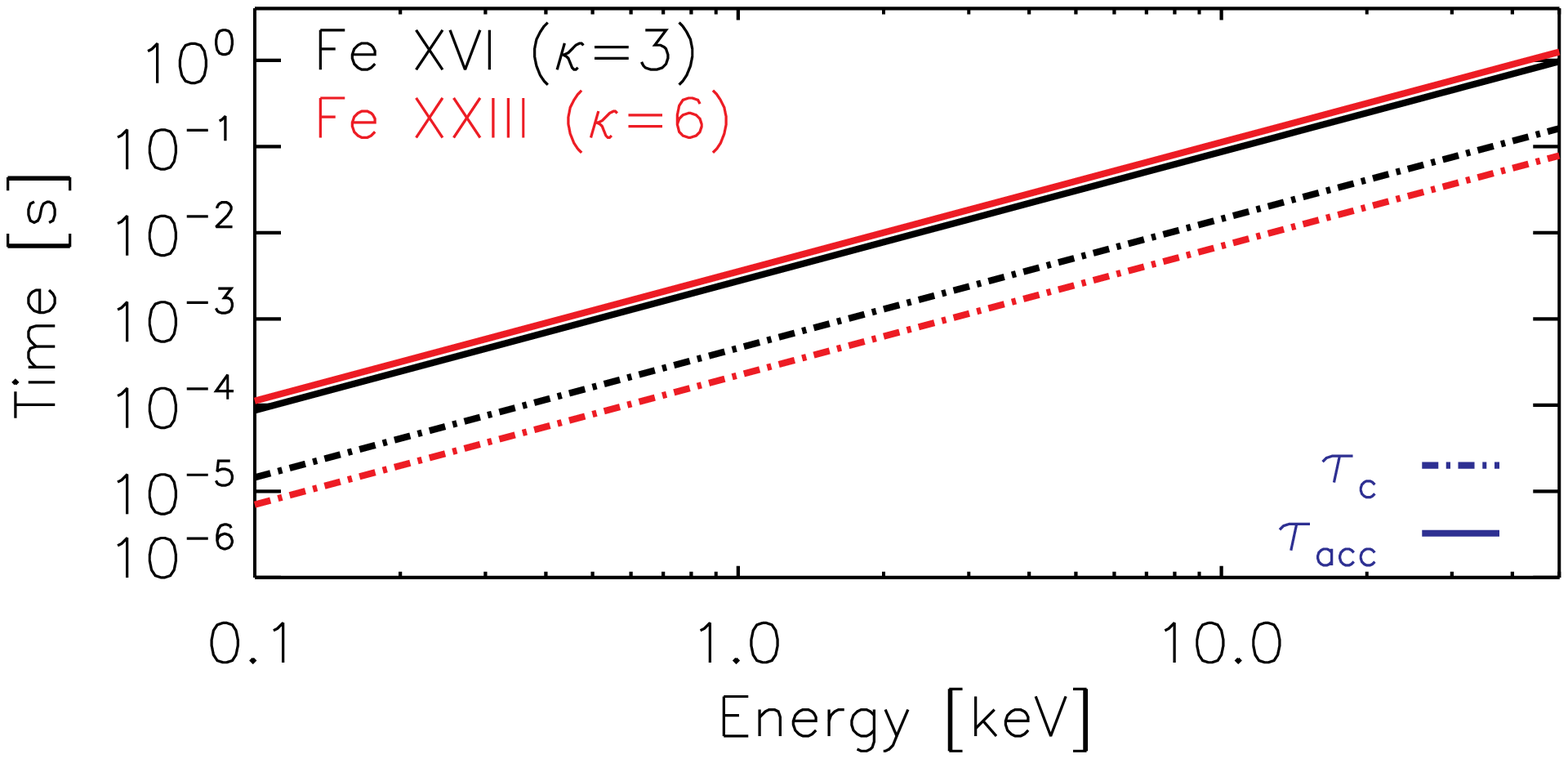}
\caption{{\it Top left:} \ion{Fe}{16} (f) collisional frequency in an electron (e) -proton (p) background plasma for f-e interactions (red) and f-p interactions (black), versus ion velocity [km/s]. {\it Top right:} Energy-loss-rate $dE/dt$ versus total ion kinetic energy (not energy per nucleon) for \ion{Fe}{16} f-e and \ion{Fe}{16} f-p collisions. The black curve is a combination of both interactions while the f-e (green) and f-p (pink) curves are also shown individually. The energy-loss-rates for e-e (grey) is also shown for comparison (note these are divided by $n_{e}$). At the observated energies of interest, f-p collisions are the dominant interaction. {\it Middle left:} The thermalization times for \ion{Fe}{16} ions due to different background electron number densities (where $n_{e}=n_{p}$). {\it Middle right:} Comparision of f-e and f-p energy-loss-rates for \ion{Fe}{16} and \ion{Fe}{23}. The energy-loss-rate for \ion{Fe}{23} is $2\times$ larger than for \ion{Fe}{16}, which might account for different $\kappa$ indices observed for both ions. {\it Bottom panel:} Comparison of the collisional (dashed) and acceleration (solid) times for \ion{Fe}{16} (black) and \ion{Fe}{23} (red) using a number density of $10^{10}$ cm$^{-3}$.}
	\label{fig10}
	\end{figure*}
Integrating Equation \ref{dvdt} allows an estimation of the ion thermalization time, $\tau_{f}$
\begin{equation}
\tau_{f}\simeq\frac{\tau_{S}}{3}\ln\left[1+\left(\frac{\epsilon}{\epsilon_{c}}\right)^{3/2}\right]
=\frac{\tau_{S}}{3}\ln\left[1+\left(\frac{v}{v_{c}}\right)^{3}\right].
\end{equation}
Figure \ref{fig10} shows the \ion{Fe}{16} collisional frequencies (divided by the electron number density), energy loss rates (divided by the electron number density), and ion thermalization times for different number densities. The bottom right panel compares the energy loss rate of \ion{Fe}{16} and \ion{Fe}{23} (using $T_{e}=T_{p}=T_{f}=15$ MK). We can see that the energy loss rate of \ion{Fe}{23} at all energies is approximately twice that for \ion{Fe}{16}. This might explain the higher observed \ion{Fe}{23} $\kappa$ index values. For a number density $n_{e}=n_{p}=10^{10}$ cm$^{-3}$ and an ion velocity of 200 km/s, the ion thermalization time is $\tau_{f}\sim0.01$ s. 

The $\kappa$ index can also be thought of as a parameter that describes the competing processes of particle acceleration and thermalization, and as a ratio of the particle acceleration time ($\tau_{acc}$) to the collisional time ($\tau_{c}$, inverse of the collisional frequency). It can be written as $\kappa=\Gamma_{c}/2D_{0}=\tau_{acc}/2\tau_{c}$, where $\Gamma_{c}$ is a collisional parameter and $D_{0}$ is a diffusion coefficient related to the acceleration mechanism. We can then estimate a local acceleration time using $\tau_{acc}=2\times\tau_{c}\kappa$. For \ion{Fe}{16}, the average $\kappa$ value is 3 while for \ion{Fe}{23} it is 6. Taking an ion velocity of 200 km/s and electron number density of $n_{e}=10^{10}$ cm$^{-3}$ with $n_{p}=n_{e}$, gives an acceleration timescale of $\tau_{acc}=0.1$ s. $\tau_{c}$ and $\tau_{acc}$ are shown for a range of ion energies in Figure \ref{fig10} (bottom panel). 

The collisional drag force on heavy ions has a minimum for ion speeds below $v_{Te}$ \citep{1995ApJ...452..451H} and a partial runaway could occur, giving a suprathermal tail of ions with velocities below $v_{Te}$. However, if we convert the estimated ion thermalization times $\tau_{f}$ to thermalization lengths using $L_{F}=v\tau_{f}$, then the thermalization of the observed ions will occur over distances of $<<1''$, for $n_{p}=10^{10}$ cm$^{-3}$. Therefore, if the line profiles are due to accelerated ions then they must undergo acceleration {\it locally and continuously} during the flare time of study. We also note that SOL2014-03-29T17:44 had no observable gamma-ray line emission, so there is no evidence for MeV ions.

We can make a rough estimate of the total energy associated with the observed kappa distributions (without separating the `thermal' and `non-thermal' components). The element and ion abundances are taken from the CHIANTI atomic database \citep{1997A&AS..125..149D,2013ApJ...763...86L}. If we assume that the range of electron number densities $n_{e}=n_{p}$ lie between $10^{9}-10^{11}$ cm$^{-3}$, then we can estimate that the number densities of \ion{Fe}{16} and \ion{Fe}{23} lie between $n_{f}=10^{3}-10^{5}$ cm$^{-3}$. Using observed \ion{Fe}{16} and \ion{Fe}{23} $\kappa$ index values of 3-6 and plugging them into $f(v)$ [cm$^{-6}$ s$^{3}$] (Equation \ref{fv_3d}), an estimate of the energy density $U$ [ergs cm$^{-3}$] above $v_{th}$ is found by numerically integrating
\begin{equation} 
U=\int_{v_{th}}^{\infty}\frac{1}{2}Mv^2f(v)d^{3}v.
\end{equation}

In order to turn this into a total energy estimate, $E$, we can multiply this by the EUV emission volume $V$. From Figure \ref{fig7}, we estimate \ion{Fe}{23} and \ion{Fe}{16} volumes (we use times $t_{2}$ and $t_{3}$ respectively and assume a spherical volume), and we calculate a volume of the order $V=10^{27}$ cm$^{3}$. We can then estimate a total energy above $v_{th}$ associated with a single ion species (\ion{Fe}{16} or \ion{Fe}{23}) using $E=VU$. Finding the values numerically gives $E\sim10^{22}-10^{24}$ ergs. We have not performed a detailed {\em RHESSI} spectroscopy analysis for this flare as it was not the purpose of the study but in comparison, the energies associated with electrons in large flares are usually of the order $10^{30}$ ergs.

An alternative scenario for producing non-Gaussian line shapes is that they originate in macroscopic velocity fields due to plasma turbulence. Support for this possibility comes from the fact that all of the observed \ion{Fe}{16} and \ion{Fe}{23} lines are slightly red-shifted, indicating small bulk downflows - even in th{}e corona - which could drive turbulence. An estimate using laboratory rest wavelengths for each line gives downflow speeds of $v_{shift}\sim+$30 to $+$70 km/s, at nearly all times and locations. It is difficult to estimate an absolute rest wavelength since these high temperature flare lines (particularly \ion{Fe}{23}) are not present in quiet Sun regions. After trying to determine an absolute wavelength scale for the cooler \ion{Fe}{16} line ($\log{T}\sim6.4$) using a `quiet Sun' region at the top of the raster at different times, we still find red-shift values of $v_{shift}\sim+$10 to $+$40 km/s, with \ion{Fe}{16} showing the larger shifts. Even taking a rather large uncertainty of $\sim10$ km/s in the inferred rest wavelength, small red-shifts are still present. This interpretation of the non-Gaussian line profiles leads to the interesting possibility of a diagnostic for localised turbulence, which could have profound consequences for theories of flare particle acceleration.

\section{Summary}
In this paper, we show that in many locations in a flare, the \ion{Fe}{16} and \ion{Fe}{23} line profiles observed by {\it Hinode} EIS are inconsistent with Gaussian spectral line shapes, and are better described by emission from a kappa distribution of ion velocities. We find that the line profile analysis of suitable unblended lines such as \ion{Fe}{16} and \ion{Fe}{23} can provide a powerful diagnostic for microscopic (non-equilibrium) or macroscopic (turbulent) ion velocities during a solar flare, that may help to constrain fundamental processes related to localised particle acceleration and/or turbulent magnetic or plasma fluctuations or flows. Straightforward estimates of ion collisional timescales suggest that the required accelerated ion distributions, with energies below $1$ MeV, can exist provided that they are accelerated close to where the EUV line emission originates. Also, the acceleration mechanism must have an acceleration time $\tau_{acc}\le0.1$ s, and must operate for the duration of the flare observations. Although, not impossible, these are stringent conditions, suggesting that the alternative possibility, line profiles due to non-Gaussian turbulent velocities is a more plausible physical explanation. This is also supported by the observation of small red-shifts at the sites of non-Gaussian \ion{Fe}{16} and \ion{Fe}{23} profiles. If broadening is due to turbulence, the physical line profile is a convolution of two physical velocities: the ion thermal velocity and a non-Gaussian spectrum of plasma velocities. Further, since SOL2014-03-29T17:44 is the second flare observed with non-Gaussian line profiles, its disk location in comparison to the close to limb location of SOL2013-05-15T01:45 studied in \citet{2016A&A...590A..99J}, is suggestive of near-isotropy more consistent with turbulent magnetic fluctuations rather than unresolved plasma flows. Lastly, it is interesting to note that in many studies of excess broadening, the highest excess broadening occurs for lines formed at the highest temperatures. In this line profile study we find that although $v_{th}$ is generally higher at early times and for hotter \ion{Fe}{23} (for all fits), the least Gaussian profiles (i.e. the smallest $\kappa$ values) are found for the cooler Fe XVI lines expected to exist at lower heights in the atmosphere.

The line profile analysis used data from {\it Hinode} EIS with an instrumental broadening of 0.059~\AA~($1''$ slit) and a spectral pixel size of 0.022~\AA, and this analysis pushes the limits of EIS. We are as certain as we can be that the non-Gaussian profile is not instrumental, however it cannot be dismissed completely for EIS. We suggest that the instrumental profile of future EUV spectrometers is measured precisely well into the line wings, so that higher moments of the line shape can be found with confidence from future solar observations. 

\acknowledgments
NLSJ, LF and NL gratefully acknowledge the financial support by the STFC Consolidated Grant ST/L000741/1. CHIANTI is a collaborative project involving George Mason University, the University of Michigan (USA) and the University of Cambridge (UK). Hinode is a Japanese mission developed and launched by ISAS/JAXA, collaborating with NAOJ as a domestic partner, NASA and UKSA as international partners. Scientific operation of the Hinode mission is conducted by the Hinode science team organized at ISAS/JAXA. This team mainly consists of scientists from institutes in the partner countries. Support for the post-launch operation is provided by JAXA and NAOJ (Japan), UKSA (U.K.), NASA, ESA, and NSC (Norway). The research leading to these results has received funding from the European Community's Seventh Framework Programme (FP7/2007-2013) under grant agreement no. 606862 (F-CHROMA). We thank Dr. David Brooks for providing the original EIS laboratory data and Dr. Eduard Kontar for informative comments.

%\bibliographystyle{apj}
%\bibliography{ms}

\appendix
\section{The EIS instrumental profile}\label{app}

\begin{figure}[ht]
	\centering
	\includegraphics[width=0.75\linewidth]{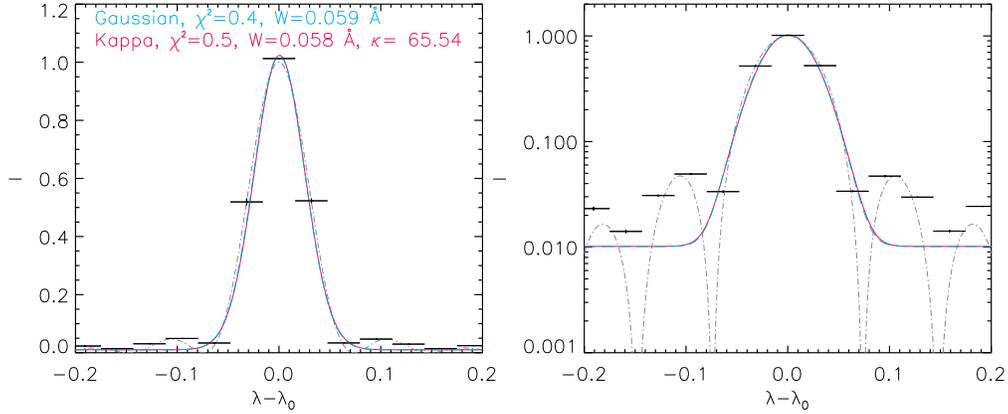}
	\caption{Modelled ${\rm sinc}^{2}{\lambda}$ line profile using the EIS spectral pixel resolution of 0.022 \AA, in a linear Y scale ({\it left}) and logarithmic Y scale ({\it right}). The same profile with a much higher resolution is displayed using a grey dash-dot curve so that the shape of the ${\rm sinc}^{2}{\lambda}$ line profile can be clearly seen. Both single Gaussian (blue) and single kappa (pink) line profiles are fitted and we can see that the central part of the profile is well-approximated by a Gaussian distribution (or a kappa distribution tending towards a Gaussian with a $\kappa$ index $\sim 66$).}
	\label{app1}
	\end{figure}

We test whether the EIS instrumental profile is responsible for the non-Gaussian line shapes observed. We find that many of the observed line profiles are well-fitted (low reduced $\chi^{2}$ values) using a convolved kappa - Gaussian function, so we fit the same function but in reverse, where the instrumental profile is approximated by a fixed kappa profile and the physical line profile is described using a Gaussian (fit KG2 in the main text). The fitting function $\mathcal{W}_{KG2}$ is then given by
\begin{equation}\label{IW}
	\mathcal{W}_{KG2}(\lambda) = \mathcal{G}(\lambda)*\mathcal{K}(\lambda)= A[0]+ 
	A[1]\sum_{\lambda^{'}}\exp{\left(-\frac{(\lambda^{'}-A[2])^{2}}{2A[3]^{2}}\right)\left(1+\frac{(\lambda-\lambda^{'}-A[2])^{2}}{2\sigma_{I}^{2}\kappa_{I}}\right)^{-\kappa_{I}+1}}
	\end{equation}
where the instrumental $\sigma_{I}$ and $\kappa_{I}$ are fixed and the Gaussian $\sigma=A[3]$, represents a physical Gaussian isothermal line width (plus excess broadening). The only constraint for $\sigma_{I}$ and $\kappa_{I}$ is that the resulting instrumental profile should be approximated by a Gaussian profile with FWHM equal to the instrumental width of $W_{inst}=0.059$ \AA, for the $1''$ slit. There is no reason for the instrumental broadening to be kappa in shape but if it is non-Gaussian, the relatively low EIS spectral pixel resolution may produce a profile well-described by such a function. It is possible that the instrumental profile is described by something closer to a ${\rm sinc}^{2}{\lambda}$ function \citep{1995ASPC...77..503J}, given by
\begin{equation}\label{eq_sinc2x}
I(\lambda)\propto{\rm sinc}^{2}{\lambda}=\left(\frac{\sin(\alpha \lambda/2)}{\alpha \lambda/2}\right)^2
\end{equation}
where $\alpha$ controls the central width of the function. This function is shown in Figure \ref{app1}. In Equation \ref{eq_sinc2x}, $\alpha\sim85$ produces a Gaussian FWHM of $W_{inst}\sim0.059$ \AA, when fitted with a single Gaussian function.

 In Figure \ref{app1}, we fitted a single kappa function and a single Gaussian function to a ${\rm sinc}^{2}{\lambda}$ profile with a 5\% Gaussian noise level. We find that both the Gaussian and the kappa functions fit the central part of ${\rm sinc}^{2}{\lambda}$ function well, with the kappa distribution tending towards a Gaussian with a kappa index larger than 20 ($\sim$66). At the given noise level, the Gaussian and kappa fits gave reduced $\chi^{2}$ values of 0.4 and 0.5 respectively, but both functions do not fit the higher order terms.

	\begin{figure}[t]
	\centering
	\includegraphics[width=0.4\linewidth]{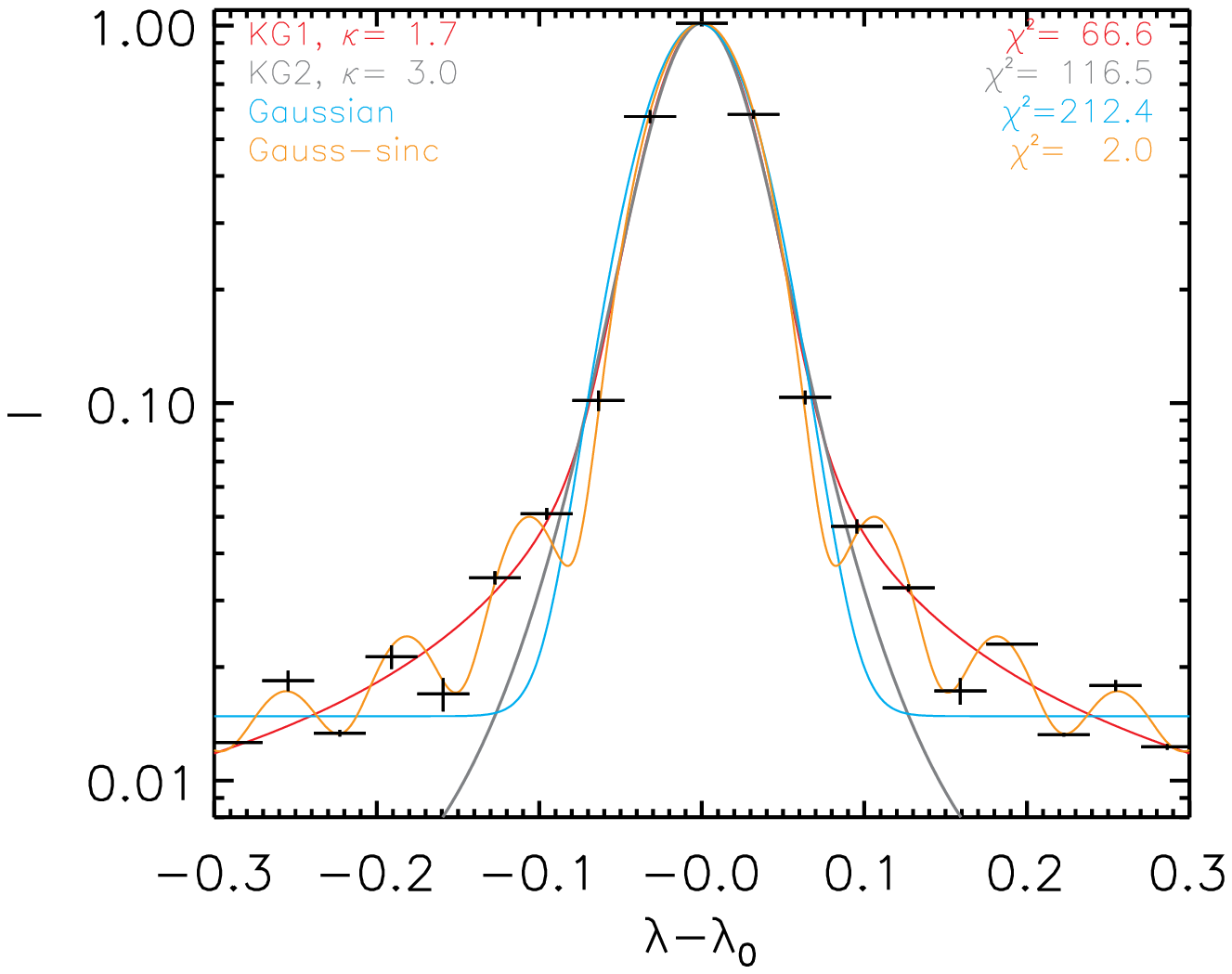}
	\includegraphics[width=0.4\linewidth]{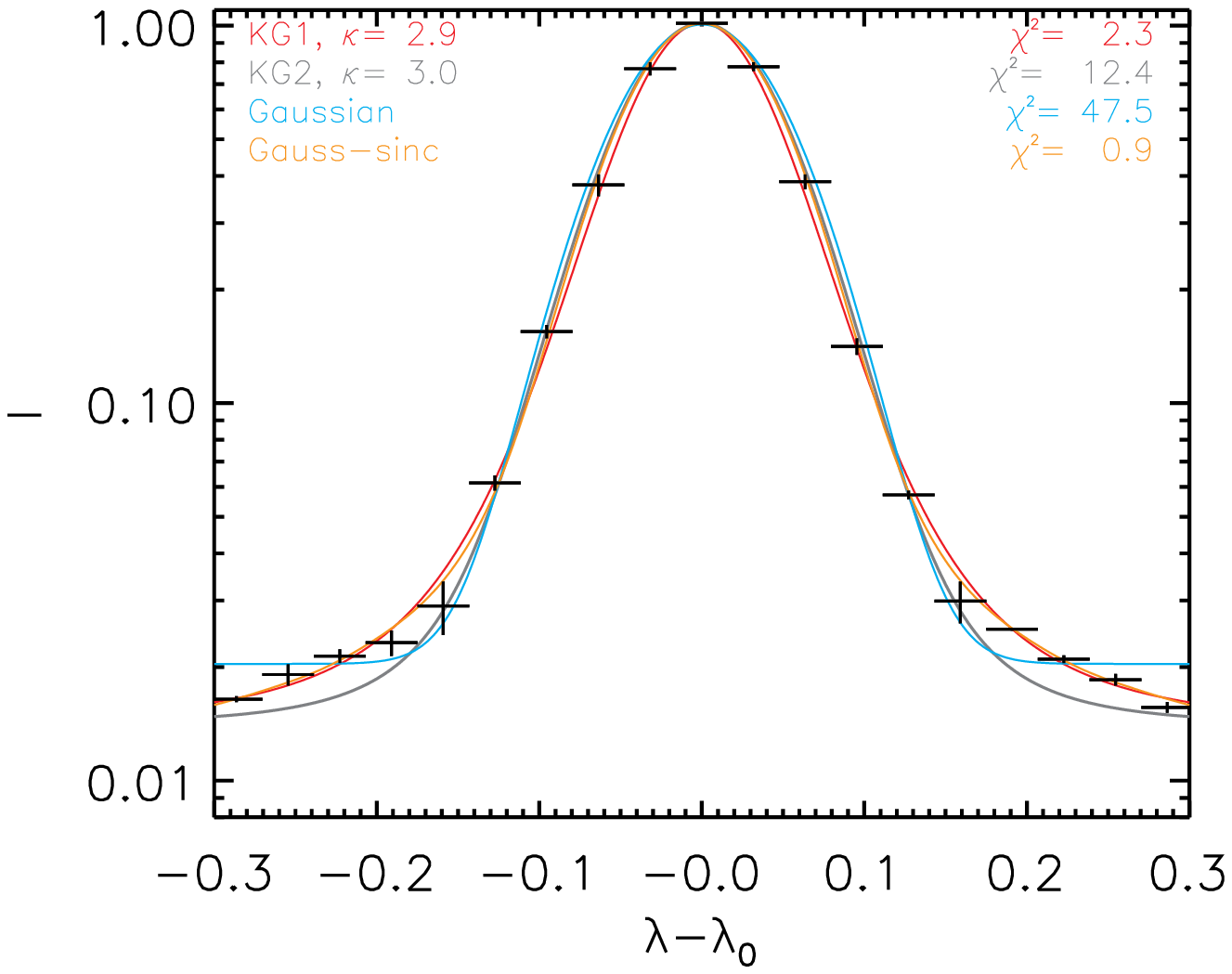}
	\caption{Modelled convolved Gaussian - ${\rm sinc}^{2}{\lambda}$ line profile using the EIS spectral pixel resolution of 0.022 \AA, for \ion{Fe}{16} ({\it left}) and \ion{Fe}{23} ({\it right}). The Gaussian part of the profile in the left column represents \ion{Fe}{16} and the right column \ion{Fe}{23} and each has a noise level of 5 \%. Each line is fitted with 1. KG1 (pink), 2. KG2 (dark grey), 3. Gaussian (blue) and 4. Gaussian - ${\rm sinc}^{2}{\lambda}$ (orange). The reduced $\chi^{2}$ values for each fit are also shown.}
	\label{app2}
	\end{figure}

For further testing, we convolve ${\rm sinc}^{2}{\lambda}$  with a Gaussian distribution (representing a physical line profile) and fit it with a number of different functions: (1.) single Gaussian, (2.) convolved kappa (physical) - Gaussian (instrumental) (main text KG1), (3.) convolved kappa (instrumental) - Gaussian (physical) (main text KG2), and (4.) a convolved Gaussian - ${\rm sinc}^{2}{\lambda}$ function. A 5\% noise level is again added and the lines and fits are shown in Figure \ref{app2}. The physical Gaussian part of the lines represent either \ion{Fe}{16} with a thermal width $W_{th}=0.039$ \AA~or \ion{Fe}{23} with $W_{th}=0.099$ \AA. The widths of the Gaussian lines are also increased to represent excess broadening. The $\kappa$ values found from the KG1 and KG2 fits and the $\chi^{2}$ values of all fits 1-4 are shown in Figure \ref{app2}.

	\begin{figure}[ht]
	\centering
	\includegraphics[width=0.49\linewidth]{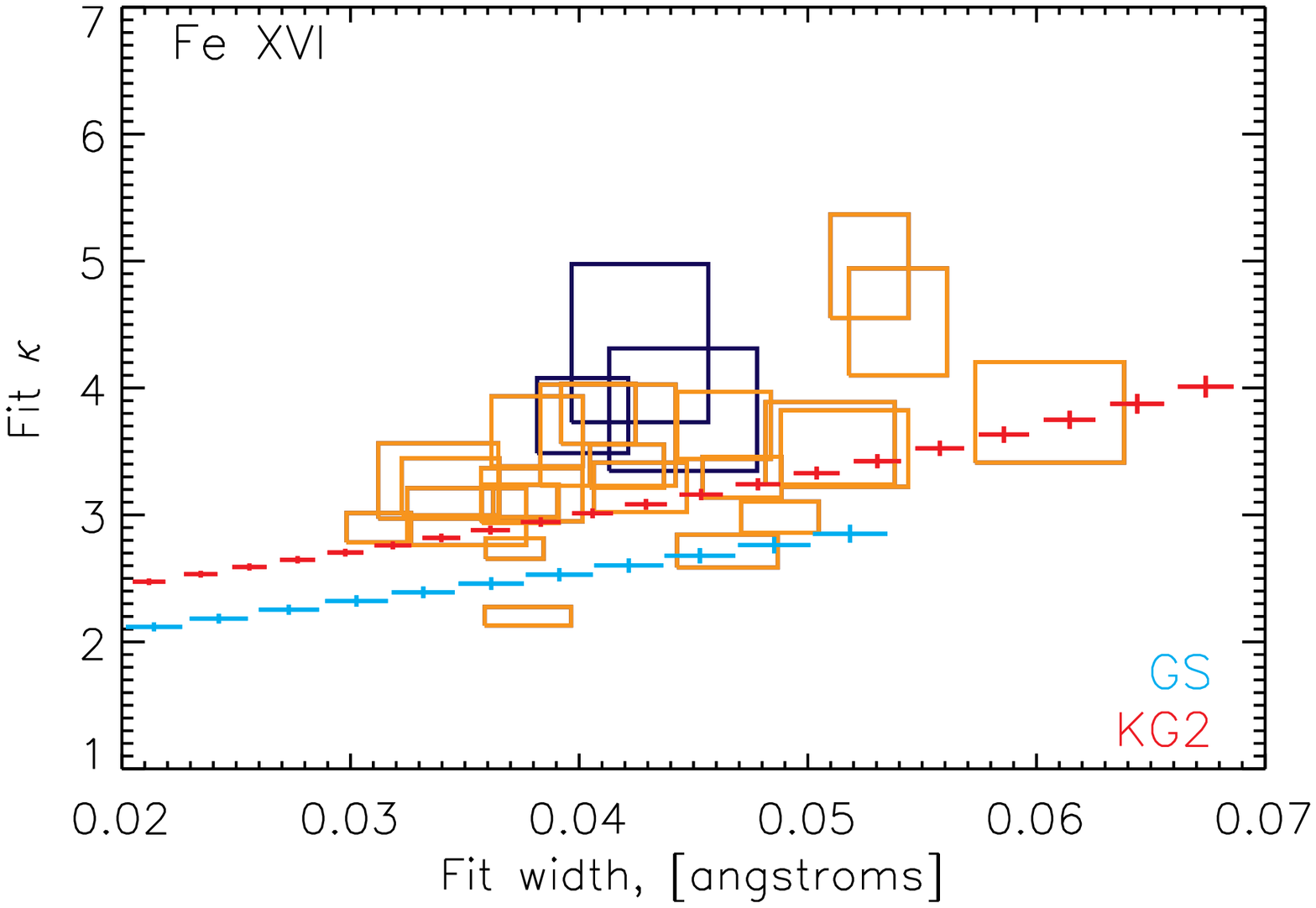}
	\includegraphics[width=0.49\linewidth]{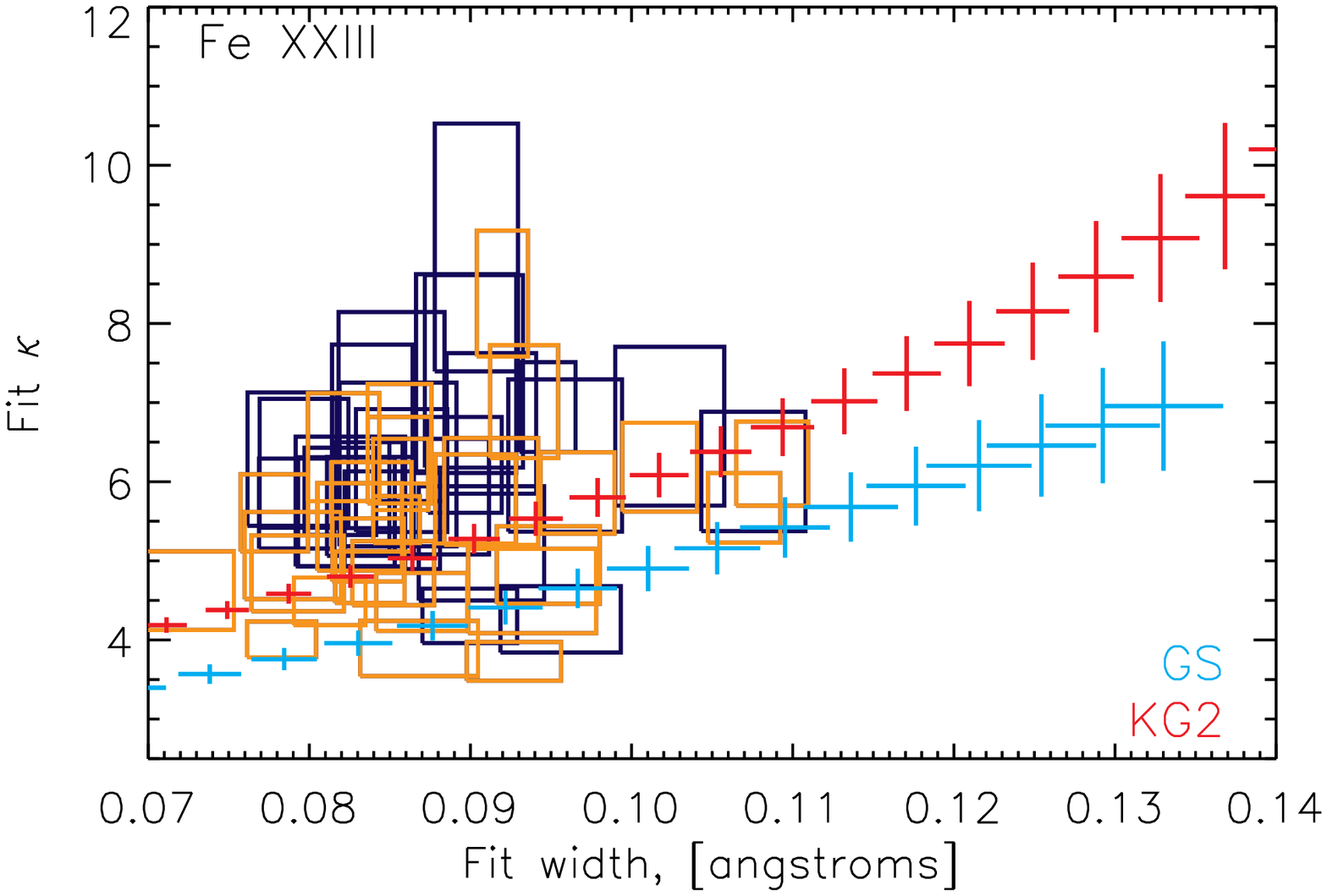}
	\caption{Lines representing \ion{Fe}{16} and \ion{Fe}{23} with Gaussian line widths ranging from 0.039~\AA~to 0.089~\AA~and from 0.089~\AA~to 0.15~\AA~respectively are convolved with either an instrumental profile of the form: (1.) a ${\rm sinc}^{2}{\lambda}$ function (as above; blue) or (2.) a kappa function with parameters $\kappa_{I}=3$ and $\sigma_{I}=0.0395$~\AA~(KG2; red). Again, a 5\% noise level is added. The KG1 function is fitted to the lines and the fitted $\kappa$ index versus the widths $W=2\sqrt{2\ln{2}}\sigma_{\kappa}$ are shown. The actual observations and fitting results from Section \ref{results} and Figure \ref{fig6} are also displayed using the orange (using the full criteria as listed in Section \ref{method}) and blue (using only criteria 1. and 2. listed in Section \ref{method}) rectangles.}
	\label{app3}
	\end{figure}
For the KG2 fit in the main text, we had to pick instrumental values for $\kappa_{I}$ index and characteristic width $\sigma_{I}$. After iteratively trying values of $\sigma_{I}$ and $\kappa_{I}$, $\kappa_{I}=3$ was chosen, matching the low $\kappa$ index values found from the observed \ion{Fe}{16} line fitting. The value of $\sigma_{I}=0.0395$ \AA~ with $\kappa_{I}=3$ produces a line profile that can be fitted with a Gaussian to produce a Gaussian FWHM equal to that of an instrumental width $W_{inst}=0.059$ \AA~for the the $1''$ slit. For the line representing \ion{Fe}{16}, Figure \ref{app2} shows that the Gaussian - sinc$^{2}$ function is the best fit ($\chi^{2}=2.1$). This is because the small physical broadening of Fe XVI does not hide the chosen instrumental form. All other functions fit the line profile poorly with high $\chi^{2}$ values of 66.6 (KG1), 116.5 (KG2) and 212.4 (Gaussian). For the line representative of \ion{Fe}{23}, we see that the Gaussian - ${\rm sinc}^{2}{\lambda}$ function is again the best fit ($\chi^{2}=0.9$), since this was the chosen form of the overall line profile. However, in this case, the KG1 fit is able to produce a low $\chi^{2}=2.3$, while KG2 gives $\chi^{2}=12.4$. The $\chi^{2}$ for the single Gaussian fit is large with a value of 47.5. Overall, the fits here are not representative of our observed EIS line profiles where a single Gaussian function and both the KG1 and KG2 functions tend to fit the line profile with much lower $\chi^{2}$ values, usually less than 10. For a final test, we vary the Gaussian widths of \ion{Fe}{16} and \ion{Fe}{23} from 0.039~\AA~to 0.089~\AA~and from 0.089~\AA~to 0.15~\AA~respectively and create two examples where the instrumental response is either represented by: (1.) a ${\rm sinc}^{2}{\lambda}$ function (as above) or (2.) a kappa function with parameters $\kappa_{I}=3$ and $\sigma_{I}=0.0395$~\AA\ (KG2). Again, a 5\% noise level is added to the lines. Each modelled line profile is fitted with a KG1 fit and we plot the resulting KG1 fitted $\kappa$ index versus fitted line width (as FWHM). For both lines representing \ion{Fe}{16} and \ion{Fe}{23}, we find that as the physical line width grows so do the KG1 fit parameters of $\kappa$ and $\sigma_{\kappa}$. In Figure \ref{app3}, the results are compared with the observations from Section \ref{results}. We can see from Figure \ref{app3} that the observed $\kappa$ indices and $W=2\sqrt{2\ln{2}}\sigma_{\kappa}$ do not follow the trends suggested by the presence of a ${\rm sinc}^{2}{\lambda}$ or KG2 instrumental response, particularly for \ion{Fe}{23} where for a given $\kappa$ index, $W$ can take multiple values, and vice versa. \ion{Fe}{16} does follow the trend expected from the presence of a KG2 instrumental response. However, again we see different values of $W$ for a single $\kappa$ index, and vice versa.
\\\\
We also examined the \ion{Fe}{16} line profiles present in other, smaller solar flares. We looked for the presence of Gaussian \ion{Fe}{16} profiles that would completely rule out an instrumental cause but we found that the lines profiles were difficult to analyse due to low intensity and high noise levels. We also obtained two sets of laboratory data when EIS was tested before launch. It was difficult to analyse the line shape with confidence due to the low EIS spectral pixel resolution, low line intensities, the presence of blends and no available intensity error values.

\end{document}